\documentclass[aps,pre,onecolumn,showpacs,showkeys,a4paper]{revtex4-1}

\usepackage{graphicx} 
\usepackage{amsmath}
\usepackage{amsfonts}
\usepackage{amscd}
\usepackage{amssymb}
\usepackage{fullpage}
\usepackage{color}

\usepackage{float}
\usepackage{empheq} 
\usepackage{fancybox}
\usepackage{pdfpages}

\usepackage{epsf}
\usepackage{epsfig}
\usepackage{epstopdf}

\begin{document}

\title{Bethe Ansatz and Q-operator for the open ASEP}
\author{Alexandre Lazarescu$^{(1)}$ and Vincent Pasquier$^{(2)}$}
\affiliation{(1) Instituut voor Theoretische Fysica, K. U. Leuven, Belgium\\
(2) Institut  de Physique Th\'eorique, C. E. A.  Saclay, France}
\begin{abstract}
In this paper, we look at the asymmetric simple exclusion process with open boundaries with a current-counting deformation. We construct a two-parameter family of transfer matrices which commute with the deformed Markov matrix of the system. We show that these transfer matrices can be factorised into two commuting matrices with one parameter each, which can be identified with Baxter's Q-operator, and that for certain values of the product of those parameters, they decompose into a sum of two commuting matrices, one of which is the usual one-parameter transfer matrix for a given dimension of the auxiliary space. Using this, we find the T-Q equation for the open ASEP, and, through functional Bethe Ansatz techniques, we obtain an exact expression for the dominant eigenvalue of the deformed Markov matrix.
\end{abstract}

\maketitle

The asymmetric simple exclusion process (ASEP), a one-dimensional discrete lattice gas model with hard core exclusion and biased diffusion, is one of the most extensively studied models of non-equilibrium statistical physics \cite{Derrida199865,1742-5468-2007-07-P07023,0034-4885-74-11-116601,Schütz20011}. There are several good reasons for this. First of all, it is a simple, physically reasonable toy model, yet its behaviour is complex enough to be interesting. Furthermore, it has the mathematical property of being integrable, which makes it a good candidate for exact calculations. It is also related to other models, like growing interfaces \cite{kardar1986dynamic}, the XXZ spin chain \cite{sandow1994partially}, or certain random matrices \cite{ferrari2010interacting}, which makes its study relevant not only for itself, but for a wide range of physical systems.

One of the fundamental quantities to describe the behaviour of the ASEP, as a particle gas driven out of equilibrium, is the current of particles that flows through it in its steady state. That current is closely related to the entropy production in the system \cite{Lebowitz99agallavotti-cohen}, which is the defining characteristic of non-equilibrium systems, and is therefore of particular importance. The statistics of that current can be described through its large deviation function \cite{Touchette20091}, which is the rescaled logarithm of its probability distribution, or, equivalently, through the generating function of its cumulants. This generating function can be expressed as the maximal eigenvalue of an integrable deformation of the Markov matrix of the system. That method was used in \cite{derrida1998exact}, in conjunction with the coordinate Bethe Ansatz \cite{Gaudin1983}, to find the generating function of the cumulants of the current in the periodic totally asymmetric simple exclusion process (TASEP), a simpler variant where the particles move only in one direction, and later in the periodic ASEP \cite{prolhac2010tree}. For the open ASEP, unfortunately, the presence of reservoirs makes it impossible to number the particles, which makes the coordinate Bethe Ansatz inapplicable. The generating function of the cumulants of the current was first obtained in \cite{PhysRevLett.107.010602} in the large size limit and in certain phases of the system. An exact general expression was then conjectured in \cite{Lazarescu2011} for the TASEP and \cite{gorissen2012exact} for the ASEP (through calculations \cite{1751-8121-46-14-145003} relying on the famous matrix Ansatz \cite{derrida1993exact} that describes the steady state of the system), and their structure was found to be extremely similar to the periodic case, but no rigorous proof was found, and no precise explanation for that similarity. This is what we propose to do in this paper.

~~~~

There is a version of the Bethe Ansatz more versatile than the coordinate one, called the algebraic Bethe Ansatz \cite{Faddeev1996}, which can in principle be used for the open ASEP. Its formulation comes from the fact that the Markov matrix of the ASEP (or the Hamiltonian of the XXZ spin chain), can be expressed in relation to the row-by-row transfer matrix of the six-vertex model \cite{Baxter1982}. That transfer matrix commutes with the Markov matrix, and depends on a spectral parameter. It has the form of a product of local tensors, (Lax matrices), traced over an auxiliary space which is usually of dimension $2$, like the physical space of a single site, although any dimension can be chosen. That product of Lax matrices, if the trace on the auxiliary space is not taken, is called the monodromy matrix of the system, and creation and annihilation operators for the particles can be extracted from it, that depend on their own spectral parameters. The algebraic Bethe Ansatz then consists in finding a trivial eigenvector of the transfer matrix (a vacuum state), on which the creation operator is then applied to obtain the other eigenvectors. Commutation relations between the transfer matrix and the creation operators, depending on their respective spectral parameters, produce the same Bethe equations as for the coordinate Bethe Ansatz, where the role of the Bethe roots is played by those spectral parameters.

In the case of a periodic system, with a fixed number of particles (or a fixed magnetisation sector), that vacuum state can be chosen as either completely empty, or completely full, which are two trivial eigenvectors of the system. In the case of an open system, there are two extra difficulties. Firstly, the transfer matrix is more complicated, and involves two rows of the vertex model instead of one, with certain reflection operators at the boundaries \cite{Kulish1992,Sklyanin1988}. This makes things harder, but not intractable. The second difficulty, however, does: for an open system, with no occupation sectors and no trivial eigenvectors, we do not know, in general, how to find a suitable vacuum state. Such states have been found for certain special boundary conditions, such as triangular boundary matrices \cite{Belliard2012, Ragoucy2012}, or full matrices with constraints on their coefficients \cite{Crampe2010,crampe2011matrix,de2005bethe,simon2009construction}, for which some pseudo-particles are conserved and the full construction can be performed. Recent progress has also been made for a semi-infinite chain \cite{Baseilhac2012,Baseilhac2007}, which has only one boundary, through a method alternative to the Bethe Ansatz. The vacuum state for completely general boundary conditions, however, remains elusive.

There is however a way to obtain the eigenvalues of the Markov matrix (or Hamiltonian, or transfer matrix) we are interested in without having to deal with the eigenvectors at all. This can be done through Baxter's so-called `Q-operator' method. It was first used as an alternative method to solve the six-vertex model \cite{Baxter1982}, but was later discovered to be, in fact, a limit of the transfer matrix with an infinite-dimensional auxiliary space \cite{Frappat2007,Nepomechie2003}. Certain algebraic relations between the Q-operator and the transfer matrix, called `T-Q' relations, allow to obtain the functional Bethe Ansatz equations for the eigenvalues directly, without need of the eigenvectors \cite{Yung1995,Bazhanov2010,Korff2005,Korff2007,Yang2006}. However, even with that method, the open case was solved only for certain constraints on the boundary parameters \cite{Nepomechie2004,Murgan2005,Yang2006} (in the case of the ASEP deformed to count the current, those constraints involve all four boundary parameters, the current-counting fugacity, and the size of the system). More recently, a variant of the T-Q equation was devised, for the XXZ spin chain \cite{Cao2013a,Cao2013,Kitanine2014} as well as in the XXX case \cite{Nepomechie2013,Cao2013b}, where an extra inhomogeneous term is added to the equation. This extra term allows to define a polynomial equivalent to the Q-operator which verifies the same relation with the eigenvalue of the corresponding Hamiltonian (the usual Q-operator cannot be polynomial itself for general values of the boundary parameters, unless the aforementioned constraints are verified). While that method applies to the situation we are considering, it is for the moment unclear how it compares with the one presented in this paper. We will come back to this issue in the conclusion.

~~

In this paper, we show that it is in fact possible to treat the most general case without introducing an inhomogeneity, by constructing explicitly the Q-operator (as well as the P-operator for the `other side of the equator' \cite{Pronko1999}) for the open ASEP with any boundary parameters and current-counting deformation, and obtaining the functional Bethe equations for the eigenvalues of the deformed Markov matrix. We construct a transfer matrix with an auxiliary space of infinite dimension and two spectral parameters instead of one (the second of which is what is usually used as a representation parameter of the U$_q$[SU(2)] algebra \cite{chaichian1996introduction} and fixed to a specific value, but we will see that it is essential to us to treat it as a free parameter). This is a natural generalisation of the transfer matrix presented in \cite{1751-8121-46-14-145003}, with the boundary vectors playing the part of the reflection operators we mentioned. We identify that transfer matrix as the product of P and Q, and we find that, taking special values of these two spectral parameters, we can recover the usual one-parameter transfer matrices with any auxiliary space dimension, and the T-Q relations for all of those matrices, as well as the corresponding fusion rules. This is the reverse of the usual construction, where Q is obtained through the fusion of an infinite number of 2-dimensional transfer matrices.

In the first section, we consider the periodic case, where everything is known from the coordinate Bethe Ansatz, as a benchmark for our construction, and we connecting this approach to that of the functional Bethe Ansatz presented in \cite{Prolhac2008a,prolhac2010tree}, noticing that the polynomials $P$ and $Q$ that are constructed there are the eigenvalues of the operators $P$ and $Q$ that we consider here. In the second section, we apply the same method to the open ASEP, and see thet the addition of the boundaries does not modify the structure of the Q-operator. This allows us to use the same method as in \cite{prolhac2010tree} to obtain the expression of the generating function of the cumulants of the current in the open ASEP that was conjectured in \cite{gorissen2012exact}. We settle, in passing, the question of how the matrix Ansatz for the steady state of the ASEP \cite{derrida1993exact} relates to the algebraic Bethe Ansatz, and we show how our construction can also be applied to the spin-1/2 XXZ chain.

\section{Periodic ASEP}
\label{V-1}

In this first section, we treat the periodic case, for which the coordinate Bethe Ansatz solution is known \cite{prolhac2010tree}. By generalising the tensors $X$ that were defined in \cite{1751-8121-46-14-145003}, as well as the algebraic relations their elements satisfy, we construct a transfer matrix with two free parameters, which commutes with the current-counting deformed Markov matrix of the periodic ASEP for any values of those parameters. We then show that, for certain special values, the transfer matrix decomposes into two independent blocks, one of which is the one-parameter transfer matrix for some dimension of the auxiliary space. We also show that our transfer matrix is in fact the product of two one-parameter operators $P$ and $Q$. Putting these results together, we are able to recover the functional Bethe equations for the periodic ASEP \cite{Prolhac2008a}.

~~

The matrix that we want to diagonalise here is the Markov matrix of the periodic ASEP, with a current-counting deformation, which is given by:
\begin{equation}
M=\sum_{i=1}^{L} M^{(i)}\nonumber
\end{equation}
with
\begin{equation}
M^{(i)}=\begin{bmatrix} 0 & 0 & 0 & 0 \\ 0 & -q & {\rm e}^{\mu_i} & 0 \\ 0 & q{\rm e}^{-\mu_i} & -1 & 0 \\ 0 & 0 & 0 & 0 \end{bmatrix}\nonumber
\end{equation}
acting on sites $i$ and $i+1$ in basis $\{00,01,10,11\}$, and where $M^{(L)}$ connects site $L$ with site $1$. If all the $\mu_i$ are taken to be $0$, this gives the Markov matrix of the open ASEP (which is stochastic : the columns sum to $0$). Deforming it with fugacities allows to keep track of the number of times a given jumping rate is used in a given realisation of the system, and hence to get the statistics of the particle currents. The largest eigenvalue of that deformed matrix can then be found to be the generating function for the cumulants of those currents, which is the quantity that we want to obtain. It can be shown \cite{Lebowitz99agallavotti-cohen} that the spectrum of that deformed matrix depends only on the sum of all the $\mu_i$, so that we can for instance choose $\mu_L=\mu$ and set all the others to $0$. We will denote that deformed matrix by $M_\mu$, and the corresponding highest eigenvalue by $E(\mu)$.

\subsection{Bulk algebra and commutation relations}
\label{V-1-1}

The starting point for the results of this paper is to realise that the matrices $d$ and $e$ that are defined, for instance, in \cite{1751-8121-40-46-R01,prolhac2009matrix}, with the algebraic relation that they satisfy, $d e-q~ed=(1-q)$, correspond to a special representation of the U$_q$[SU(2)] algebra (up to a simple gauge transformation that we present in section \ref{V-2-5}). In light of this, it seems natural to wonder whether a more general representation might be used, and produce different, and perhaps better, results.

~~

Let us therefore define:
\begin{align}
X&=\begin{bmatrix} n_0 & e\\ d & n_1\end{bmatrix}=~~~~~~~~~~~\begin{bmatrix}1+x A & e\\ d & 1+y A\end{bmatrix},\nonumber\\
\hat{X}&=\begin{bmatrix} \hat{n}_0 & \hat{e}\\ \hat{d} & \hat{n}_1\end{bmatrix}=\frac{(1-q)}{2}\begin{bmatrix} 1-x A & e\\ -d & -1+y A\end{bmatrix}.\nonumber
\end{align}
in basis $\{0,1\}$, corresponding to the occupancy of one of the sites, and where matrices $A$, $d$ and $e$ satisfy:
\begin{empheq}[box=\fbox]{align}
de -q~ed&=(1-q)(1-x y  A^2),\\ 
A e&=q~e A , \\
d A&=q~A d.
\end{empheq}

For practical purposes, we will be using a specific solution to these equations, given by:
\begin{equation}\label{V-1-A}
A=\sum\limits_{n=0}^{\infty} q^n|\!|n\rangle\!\rangle\langle\!\langle n|\!|
\end{equation}
\begin{equation}\label{V-1-d}
d=\sum\limits_{n=1}^{\infty}(1-q^n)|\!|n-1\rangle\!\rangle\langle\!\langle n|\!|=S^-(1-A)
\end{equation}
and
\begin{equation}\label{V-1-e}
e=\sum\limits_{n=0}^{\infty}(1-x y q^n)|\!|n+1\rangle\!\rangle\langle\!\langle n|\!|=S^+(1-x y A)
\end{equation}
where $S^+$ and $S^-$ are simply operators increasing or decreasing $n$ by $1$ (not to be confused with spin operators). We recover the simpler versions of these matrices simply by taking $x=y=0$.

A few remarks need to be made here. First of all, the matrix $A$ that we have just defined plays an important role in building the matrix Ansatz for the multispecies periodic ASEP \cite{prolhac2009matrix}. Secondly, we could have chosen for $d$ and $e$ their contragredient representation $\overline{e}={}^t\!d$ and $\overline{d}={}^t\!e$, which is equivalent to a gauge transformation on $d$ and $e$. We will be using this fact abundantly in the rest of the chapter. Finally, we can actually define $A$, $S^+$ and $S^-$ over $\mathbb{Z}$ rather than $\mathbb{N}$, so that $S^+$ and $S^-$ are the inverse of one another: $S^+S^-=1$ (which wouldn't work on $\mathbb{N}$ because of the cut at $-1$). Because of the term $(1-q^n)$ in $d$, which is $0$ between states $|\!|0\rangle\!\rangle$ and $|\!|-1\rangle\!\rangle$, we are assured that, if starting from a state $|\!|n\rangle\!\rangle$ with $n\geq 0$, we can never go to one with $n<0$ through any combination of $d$, $e$ and $A$. We just need to make sure that those expressions are always applied to vectors that have non-zero coefficients only for $n\geq0$, which is enforced by the matrix $A_\mu$ that we will define momentarily, on $\mathbb{N}$ alone. This fact will make many future calculations much easier.

All our matrices are now combinations of only $A$ and $S$, which satisfy a simple algebra:
\begin{equation}
A S=q~S A \label{V-1-AS+},
\end{equation}
with $S^+=S$ and $S^-=(S)^{-1}$.

We also need to define another diagonal matrix $A_\mu$, given by
\begin{equation}\label{V-1-Amu}
A_\mu=\sum\limits_{n=0}^{\infty} {\rm e}^{-n\mu}|\!|n\rangle\!\rangle\langle\!\langle n|\!|
\end{equation}
such that
\begin{equation}
A_\mu S={\rm e}^{-\mu}~S A_\mu \label{V-1-AS+}.
\end{equation}

Note that this doesn't have a well-defined trace for $\mu=0$. If we want to consider that limit, we will have to multiply $A_\mu$ by $(1-{\rm e}^{-\mu})$ first.

~~

Finally, we define the transfer matrix
\begin{equation}\label{Tmuper}\boxed{
T_\mu^{per}(x,y)={\rm Tr}[A_\mu\prod\limits_{i=1}^{L}X^{(i)}]
}\end{equation}
where the product symbol refers to a matrix product in the auxiliary space (i.e. the internal space of matrices $A$ and $S^+$) and a tensor product in configuration space, and the trace is taken only on the auxiliary space. The superscript $(i)$ only indicates to which physical site each matrix $X$ corresponds to (and their place in the product). In other terms, the weight of that matrix between configurations ${\cal C}=\{\tau_i\}$ and ${\cal C}'=\{\tau_i'\}$, where $\tau_i$ is the number of particles on site $i$ in ${\cal C}$, is given by
\begin{equation}
T_\mu^{per}(x,y)\big|_{\cal{C}'\!,\cal{C}}={\rm Tr}[A_\mu\prod\limits_{i=1}^{L}X_{\tau_i',\tau_i}]\nonumber
\end{equation}

Through a calculation which can be found in appendix \ref{App-A-1}, we show that each of the local matrices in $M_\mu$ satisfy
\begin{equation}\label{XX-XX}
[M^{(i)}, X^{(i)}X^{(i+1)}]=\hat{X}^{(i)}X^{(i+1)}-X^{(i)}\hat{X}^{(i+1)}
\end{equation}
and, for $M^{(L)}$, which contains the deformation,
\begin{equation}\label{XX-XX2}
[M^{(L)}, X^{(L)}A_\mu X^{(1)}]=\hat{X}^{(i)}A_\mu X^{(i+1)}-X^{(i)}A_\mu \hat{X}^{(i+1)}.
\end{equation}

Note that these relations are in fact the infinitesimal equivalent of the so called `RLL equation' for the commutation of matrices $X$ with different parameters, where $\hat{X}$ is the derivative of $X$ for a well chosen variable.

~~

We can now recover $M_\mu$ in its entirety by summing over $i$ in (\ref{XX-XX}) and (\ref{XX-XX2}). The hat matrices cancel out from one term to the next, and we are left with $0$, so that:
\begin{equation}\boxed{\boxed{
[M_\mu,T_\mu^{per}(x,y)]=0.
}}\end{equation}

$T_\mu^{per}(x,y)$ has therefore the same eigenvectors as $M_\mu$.

~~

Note that for a general set of fugacities $\{\mu_i\}$, the corresponding transfer matrix is the same as the one we defined here, with matrices $A_{\mu_i}$ inserted at their appropriate place in the matrix product:
\begin{equation}\label{V-1-Tmui}
T_{\{\mu_i\}}^{per}(x,y)={\rm Tr}[A_{\mu_0}\prod\limits_{i=1}^{L}X^{(i)}A_{\mu_i}].
\end{equation}

\subsection{Decomposition of the transfer matrix}
\label{V-1-2}

Considering the representation we chose for matrix $e$ in (\ref{V-1-e}), namely $S(1-x y A)$, an interesting case to consider is $xy=q^{-k+1}$ with $k\in\mathbb{N}^\star$ (which sets one coefficient to $0$ in $e$). 

Let us therefore impose $y=1/q^{k-1} x$. The four matrices in $X$ become:
\begin{equation}\label{V-1-dexqk/x}
d=\sum\limits_{n=1}^{\infty}(1-q^n)|\!|n-1\rangle\!\rangle\langle\!\langle n|\!|~~~~,~~~~e=\sum\limits_{n=0}^{\infty}(1- q^{n-k+1})|\!|n+1\rangle\!\rangle\langle\!\langle n|\!|
\end{equation}
\begin{equation}\label{V-1-nnxqk/x}
n_0=\sum\limits_{n=0}^{\infty}(1+q^{n}x)|\!|n\rangle\!\rangle\langle\!\langle n|\!|~~~~,~~~~n_1=\sum\limits_{n=0}^{\infty}(1+q^{n-k+1}/x)|\!|n\rangle\!\rangle\langle\!\langle n|\!|
\end{equation}
and the coefficient of $|\!|k\rangle\!\rangle\langle\!\langle k-1|\!|$ in $e$ vanishes. This makes all these matrices lower block-triangular ($n_0$ and $n_1$ obviously are, since they are diagonal, and $d$ already was, but not $e$ in general) with a block of size $k$ (for $n$ from $0$ to $k-1$ in the four series above) and one of infinite size (for $n$ from $k$ to $\infty$).

The coefficients of that second block happen to be the same as the coefficients of the whole matrix for $x\rightarrow q^k x$ and $y\rightarrow q/x$ and in the contragredient representation of $d$ and $e$ (i.e. exchanging and transposing them):
\begin{equation}\label{V-1-deqkx/x}
^t\!e=\sum\limits_{n=0}^{\infty}(1- q^{n+k})|\!|n-1\rangle\!\rangle\langle\!\langle n|\!|~~~~,~~~~^t\!d=\sum\limits_{n=1}^{\infty}(1-q^{n+1})|\!|n+1\rangle\!\rangle\langle\!\langle n|\!|,
\end{equation}
\begin{equation}\label{V-1-nnqkx/x}
n_0=\sum\limits_{n=0}^{\infty}(1+q^{n+k}x)|\!|n\rangle\!\rangle\langle\!\langle n|\!|~~~~,~~~~n_1=\sum\limits_{n=0}^{\infty}(1+q^{n+1}/x)|\!|n\rangle\!\rangle\langle\!\langle n|\!|.
\end{equation}
Indeed, taking $n\rightarrow n+k$ in (\ref{V-1-dexqk/x}) and (\ref{V-1-nnxqk/x}), and removing the $k$ negative indices, we get exactly (\ref{V-1-deqkx/x}) and (\ref{V-1-nnqkx/x}).

Since the trace of a product of block-diagonal matrices is the sum of the traces of the products of the blocks, this gives us an equation for $T_\mu^{per}$, which is one of the two results essential to our derivation of the functional Bethe Ansatz:
\begin{equation}\label{V-1-TtT}\boxed{\boxed{
T_\mu^{per}(x,1/q^{k-1}x)=t^{[k]}(x)+{\rm e}^{-k\mu}T_\mu^{per}(q^k x,q/x)
}}\end{equation}
where the factor ${\rm e}^{-k\mu}$ is the first coefficient of $A_\mu$ on the second block. The transfer matrix $t^{[k]}$ is the contribution coming from the first block, which can be written as:
\begin{equation}\boxed{
t^{[k]}(x)={\rm Tr}[A_\mu\prod\limits_{i=1}^{L}X_k^{(i)}(x)]
}\end{equation}
where $X_k(x)$ contains the same entries as $X(x,1/q^{k-1}x)$, but truncated at $n=k-1$ (so that the auxiliary space is $k$-dimensional).

We saw that $T_\mu^{per}$ commutes with $M_\mu$ for any values of $x$ and $y$, so it also commutes with another $T_\mu^{per}$ at different values of the parameters (this is in fact not certain, because any one of those matrices could have a degenerate eigenspace, but we will assume that it is true, for now; there is a better way to show that two matrices $T_\mu^{per}$ at different values of $x$ and $y$ commute, and we will come back to it in the next section). This tells us that those matrices also commute with $t^{[k]}(x)$ for any $k$, and that the $t^{[k]}(x)$'s with different $k$'s commute together. This matrix equation therefore implies the same relation for the eigenvalues $\Lambda_i(x,y)$ of $T_\mu^{per}(x,y)$ and the eigenvalues $\Lambda^{[k]}_i(x)$ of $t^{[k]}(x)$.

~~

To go further, we need to examine the first two cases in this last equation. Note that, to be consistent with our notation for $X$, all the matrices will be written with the physical space as their outer space and the auxiliary space as their inner space (i.e. separated in $2\times 2$ blocks of matrices acting on the auxiliary space), which is opposite to the standard convention in some cases.

For $k=1$, the first block of $X$ is given by:
\begin{equation}
X_1(x)=\left[\begin{array}{c|c} 1+x & 0\\ \hline 0 & 1+\frac{1}{x}\end{array}\right].\nonumber
\end{equation}
The matrix $t^{[1]}(x)$, which is scalar inside of a given occupancy sector, is then given by:
\begin{equation}\boxed{
t^{[1]}(x)=(1+x)^{L-N}(1+x^{-1})^{N}=h(x).
}\end{equation}
This is the same as the function $h(x)$ that is defined in \cite{prolhac2010tree}, up to a sign.

~~

For $k=2$, the $2\times 2$ blocks from $n_0$, $e$, $n_1$ and $d$ are:
\begin{equation}
X_2(x)=\left[\begin{array}{c c|c c} 1+x & 0 & 0 & 0 \\ 0 & 1+q x & 1-\frac{1}{q} & 0 \\ \hline 0 & 1-q & 1+\frac{1}{q x} & 0\\ 0 & 0 & 0 &  1+\frac{1}{x}\end{array}\right] \nonumber
\end{equation}
and
\begin{equation}\boxed{
t^{[2]}(x)={\rm Tr}[A_\mu^{[2]}\prod\limits_{i=1}^{L}X_2^{(i)}(x)]
}\end{equation}
with $A_\mu^{[2]}=\begin{bmatrix} 1 & 0\\ 0 & {\rm e}^{-\mu}\end{bmatrix}$.

\begin{figure}[ht]
\begin{center}
\includegraphics[width=0.5\textwidth]{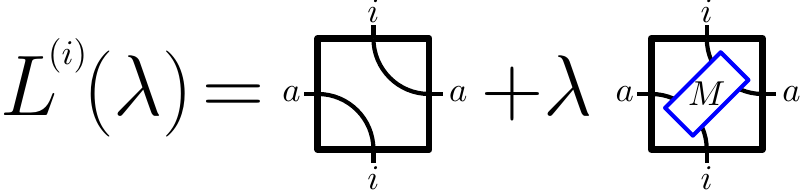}
\caption{Schematic representation of $L^{(i)}(\lambda)$. The first box represents $P^{(i)}$, which exchanges the occupancies of the auxiliary space $a$ and the physical space $i$ (the matrix is seen as acting from SE to NW). The second box is the same exchange operator, with the local matrix $M$ acting during the exchange.}
\label{fig-Li}
 \end{center}
 \end{figure}

This matrix is, in fact, the standard transfer matrix for the periodic XXZ spin chain, with a two-dimensional auxiliary space. To write it in its usual form, we need to make a few transformations and change variables. To that effect, let us consider:
\begin{equation}\label{V-1-L}
\frac{1}{1+x} \left[\begin{array}{ c | c} 1 & 0   \\ \hline 0 & x  \end{array}\right] \centerdot X_2(x)=\left[\begin{array}{c c|c c} 1 & 0 & 0 & 0 \\ 0 & \frac{1+q x}{1+x} & \frac{q-1}{q(1+x)} & 0 \\ \hline 0 &\frac{x(1-q)}{1+x} & \frac{1+q x}{q(1+x)} & 0\\ 0 & 0 & 0 &  1 \end{array}\right] = \left[\begin{array}{c c|c c} 1 & 0 & 0 & 0 \\ 0 & q \lambda &1-\lambda & 0 \\ \hline 0 &1-q \lambda & \lambda & 0\\ 0 & 0 & 0 &  1 \end{array}\right]
\end{equation}
with $\lambda = \frac{1+q x}{q(1+x)}$, i.e. $x=-\frac{1-q \lambda}{q(1-\lambda)}$. We use the symbol $\centerdot$ to signify a product in configuration space, so that it not be confused with a product in the auxiliary space (for which we use the usual product notation). This is the common form of the Lax matrix for the ASEP:
\begin{equation}
L^{(i)}(\lambda)= \left[\begin{array}{c c|c c} 1 & 0 & 0 & 0 \\ 0 & q \lambda &1-\lambda & 0 \\ \hline 0 &1-q \lambda & \lambda & 0\\ 0 & 0 & 0 &  1 \end{array}\right]=P^{(i)}(1+\lambda M^{(i)})\nonumber
\end{equation}
where $P^{(i)}$ is a permutation matrix which has the effect of exchanging the physical space at site $i$ with the auxiliary space (fig.-\ref{fig-Li}).

The matrix we applied to $X_2$ from the left in (\ref{V-1-L}) multiplies every entry by $x$ for each occupied site (and since this number is conserved between the left and right entries of the transfer matrix, this operation actually commutes with $t^{[2]}$, so that we haven't modified its eigenvectors). All in all, this operation multiplies $t^{[2]}$ by $x^N/(1+x)^L=1/h(x)$. We therefore define:
\begin{equation}\label{V-1-t2t}
\hat{t}^{[2]}(\lambda)=\frac{t^{[2]}(x)}{h(x)}={\rm Tr}[A_\mu^{[2]}\prod\limits_{i=1}^{L}L^{(i)}(\lambda)]
\end{equation}
which is the usual transfer matrix for the periodic ASEP with one marked bond.

\begin{figure}[ht]
\begin{center}
\includegraphics[width=0.8\textwidth]{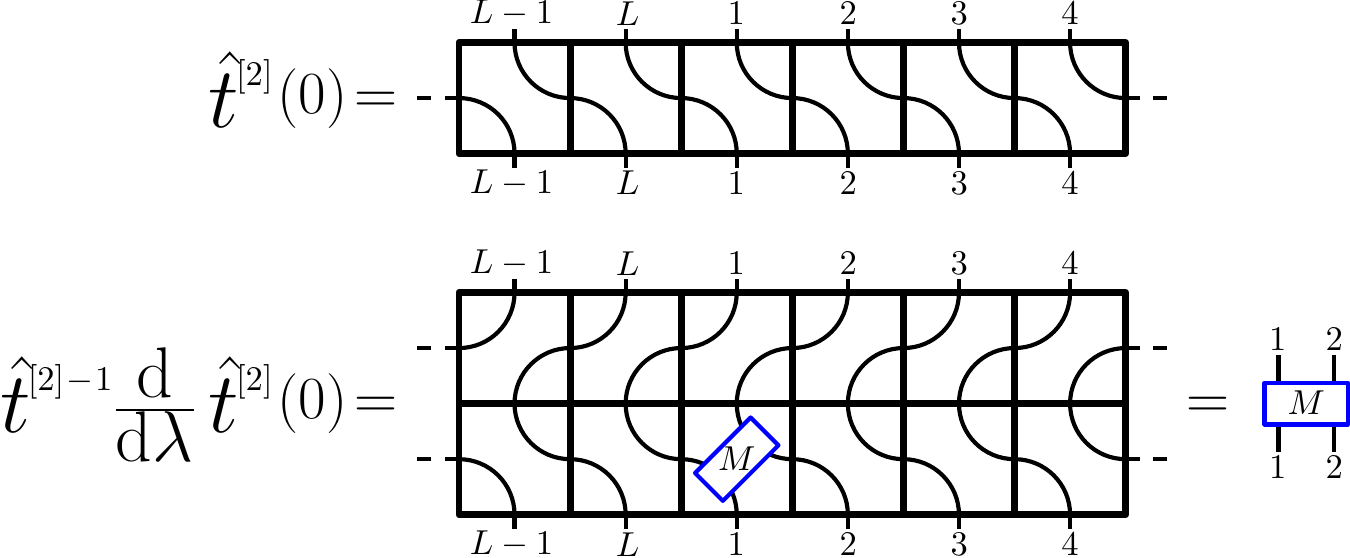}
\caption{Schematic representation of the value and first logarithmic derivative of $\hat{t}^{[2]}$ at $0$. The first is a translation matrix which takes each site $i$ to $i-1$. The second (of which only one part of the sum over sites is represented) gives the local jump matrix $M^{(i)}$.}
\label{fig-t2}
 \end{center}
 \end{figure}

Since $L^{(i)}(0)=P^{(i)}$ is a permutation matrix, and its derivative with respect to $\lambda$ at $0$ is $\frac{{\rm d}}{{\rm d}\lambda}L^{(i)}(0)=P^{(i)} M^{(i)}$, we find that $\hat{t}^{[2]}(0)$ is the matrix that transposes the whole system back one step, and that, for the whole transfer matrix $\hat{t}^{[2]}$ (fig.-\ref{fig-t2}):
\begin{equation}
M_\mu=\bigl(\hat{t}^{[2]}(\lambda)\bigr)^{-1}\frac{{\rm d}}{{\rm d}\lambda}\hat{t}^{[2]}(\lambda)\bigl|_{\lambda=0}=\frac{{\rm d}}{{\rm d}\lambda}\log\bigl(\hat{t}^{[2]}(\lambda)\bigr)\bigl|_{\lambda=0}
\end{equation}
(we have not considered how $A_\mu$ comes into play, but we can easily check that it gives the correct term in $M_\mu$).

Written in terms of $t^{[2]}(x)$, this becomes:
\begin{equation}\label{V-1-ET}\boxed{\boxed{
M_\mu=(1-1/q)\frac{d}{dx}\log\biggl(\frac{t^{[2]}(x)}{h(x)}\biggr)\biggl|_{x=-1/q}.
}}\end{equation}

Doing the same calculations at $\lambda=\infty$ instead of $0$, after a few modifications (such as taking the contragredient representation for $X$ and multiplying everything by $h(x)/h(qx)$) would have instead given us:
\begin{equation}\label{V-1-ET2}
M_\mu=(1-q)\frac{d}{dx}\log\biggl(\frac{t^{[2]}(x)}{h(qx)}\biggr)\biggl|_{x=-1}
\end{equation}
which is also what we would have obtained if we had considered a system with $L-N$ particles, exchanging $Q$ with $P$, and kept $y=1/qx$ as a variable. This identity is usually called the `crossing symmetry' in the language of quantum spin chains.

\subsection{R matrix}
\label{V-1-3}

The next step is to show that the eigenvalues $\Lambda_i(x,y)$ of $T_\mu^{per}$ are in fact a product of a function of $x$ and a function of $y$, which we will note as $P_i(x)$ and $Q_i(y)$. This factorisation property is a well known fact for the periodic XXX \cite{Derkachov2006,Derkachov2008} and XXZ \cite{Pronko1999,Korff2005} spin chains. This result will actually be derived in the next section, as there are a few preliminary calculations which need to be done first, mainly in order to find the R matrix of our system.

One way to go about this is through a method used in \cite{Bernard1990,Bazhanov1990} which consists in introducing two new Lax matrices, defined by:
\begin{align}
L_1&=L(a_1,b_1,c_1,d_1)=\begin{bmatrix} a_1 A_1 & b_1 S_1\\ -c_1  S^{-1}_1 A_1& d_1\end{bmatrix},\nonumber\\
\tilde{L}_2&=\tilde{L}(a_2,b_2,c_2,d_2)=\begin{bmatrix} a_2 A_2 & -c_2  S_2 A_2\\ b_2 S^{-1}_2 & d_2\end{bmatrix},\nonumber
\end{align}
where the operators $A_i$ and $S_j$ obey (\ref{V-1-AS+}) for $i=j$ (i.e. if they act on the same space), and commute otherwise.

We then take the product of those matrices (which is a matrix product on the physical two-dimensional space and a tensor product on the infinite-dimensional auxiliary spaces of $L_1$ and $\tilde{L}_2$):
\begin{equation}
L_1\tilde{L}_2=\begin{bmatrix} a_1 a_2  A+b_1 b_2 S_1/S_2 & b_1 d_2 S_1-a_1 c_2  S_2 A\\ d_1 b_2 S_2^{-1}-c_1 a_2  S_1^{-1} A & c_1 c_2  S_2/ S_1 A + d_1 d_2\end{bmatrix}\nonumber
\end{equation}
with $A=A_1 A_2$. We will omit the notation $\centerdot$ for the product between those new Lax matrices, since the indices are there to signify that their elements act on different spaces.

We now need to consider two special cases for the coefficients of $L_1$ and $\tilde{L}_2$. Let us first set them as follows: $a_1=x$, $c_2=y$ and the rest is $1$. We write the corresponding matrices as $L_1^+$ and $\tilde{L}_2^-$:
\begin{equation}
L_1^+(x) \tilde{L}_2^-(y)=\begin{bmatrix} x A+ S_1 / S_2 & S_1-xy  S_2 A\\  S_2^{-1}-  S_1^{-1}A & y  S_2/ S_1 A+1\end{bmatrix} \nonumber
\end{equation}
and, by projecting each element on $S_1 =S_2=S$ (i.e. by applying $\sum |\!|i,j\rangle\!\rangle\langle\!\langle i+j|\!|$ to the right and its contragredient to the left), we get:
\begin{equation}\boxed{
L_1^+(x) \tilde{L}_2^-(y)=\begin{bmatrix} x A+ 1 & S(1-xy A) \\  S^{-1}(1-  A) & y A+1\end{bmatrix} =X(x,y).
}\end{equation}
Naturally, we check that $A$ and $S$ satisfy the correct relations.

Let us now set $a_2=x$, $c_1=y/q$, $c_2=q$ and the rest to $1$. We write the corresponding matrices as $L_1^-$ and $\tilde{L}_2^+$:
\begin{equation}
L_1^-(y)\tilde{L}_2^+(x)=\begin{bmatrix} x A+ S_1/ S_2 & S_1- q S_2 A\\  S_2^{-1}- xy/q S_1^{-1}A & y  S_2/ S_1 A+1\end{bmatrix} \nonumber
\end{equation}
and, through the same operation as before, we get:
\begin{equation}\boxed{
L_1^-(y)\tilde{L}_2^+(x)=\begin{bmatrix} x A+ 1 & (1- A)S \\  (1- xy A)S^{-1} & y A+1\end{bmatrix} =\overline{X}(x,y).
}\end{equation}

In both of those special cases, one of the non-diagonal elements has a factor $(1-A)$ which allows us to truncate the representation at state $|\!|0\rangle\!\rangle$ and avoid some convergence issues. It would not be the case, however, if we were to construct $L_1^+(x) \tilde{L}_2^+(y)$ or $L_1^-(x) \tilde{L}_2^-(y)$, which we will therefore avoid at any cost.

~~

We will now use this formalism in order to find the so-called `$R$ matrix' which is such that:
\begin{equation}
X(x,y)\centerdot X(x',y')~R(x,y;x',y')=R(x,y;x',y')~X(x',y')\centerdot X(x,y)\nonumber
\end{equation}
where $R$ acts on the two auxiliary spaces of both $X$ matrices. We will comment on the use of such a matrix at the end of this section.

Considering that $X(x,y)\centerdot X(x',y')=L_1^+(x) \tilde{L}_2^-(y) L_3^+(x') \tilde{L}_4^-(y')$, we will perform this exchange of parameters in steps, exchanging the parameters of only two $L_i$'s at a time. The first thing we could try is to exchange $y$ and $x'$, but this would transform $\tilde{L}_2^-(y)$ into $\tilde{L}_2^+(x')$, so we would have $L_1^+(x) \tilde{L}_2^+(x')$ on the left, which we want to avoid. The solution is then to first exchange $x'$ and $y'$, then $y'$ and $y$, and finally $y$ and $x'$, to obtain $X(x,y')\centerdot X(x',y)$, and then do the same once more to exchange $x$ and $x'$.

~~

We first need to find $f_{1 2}(x,y)$ such that $L_1^+(x) \tilde{L}_2^-(y) f_{1 2}(x,y) = f_{1 2}(x,y) L_1^-(y)\tilde{L}_2^+(x)$, which is to say:
\begin{equation}
X(x,y)f_{1 2}(x,y)=f_{1 2}(x,y)\overline{X}(x,y).\nonumber
\end{equation}
That $f_{1 2}$ may depend on $A$ and $S$. In terms of the elements of $X(x,y)$, this writes:
\begin{align}
[1+xA,f_{1 2}]&=0,\nonumber\\
[1+yA,f_{1 2}]&=0,\nonumber\\
S(1- xy A) ~f_{1 2}&=f_{1 2}(1-A)S,\nonumber\\
S^{-1}(1-  A) ~f_{1 2}&=f_{1 2}(1-xyA)S^{-1}.\nonumber
\end{align}

The first and second equations tell us that $f_{1 2}$ commutes with $A$ (i.e. it is diagonal), so it should be a function of $A$ alone. The third or fourth equations then give us:
\begin{equation}
S(1- xy A) f_{1 2}[A]=f_{1 2}[A](1-A)S=S(1-qA)f_{1 2}[qA]\nonumber
\end{equation}
(where the second equality is due to the commutation of $S$ with $A$), which we can rewrite as
\begin{equation}
\frac{f_{1 2}[A]}{f_{1 2}[q A]}=\frac{(1- q A)}{(1-xyA)}.\nonumber
\end{equation}

Iterating this last equation, we finally find:
\begin{equation}\boxed{
f_{1 2}(xy)=\frac{(q A)_\infty}{(xyA)_\infty},
}\end{equation}
where $(x)_\infty$ is the infinite q-Pochhammer symbol
\begin{equation}
(x)_\infty=\prod\limits_{k=0}^{\infty}(1-q^kx).\nonumber
\end{equation}

To exchange the parameters back, one simply has to apply $f_{1 2}^{-1}(xy)$.

 \begin{figure}[ht]
\begin{center}
\includegraphics[width=0.6\textwidth]{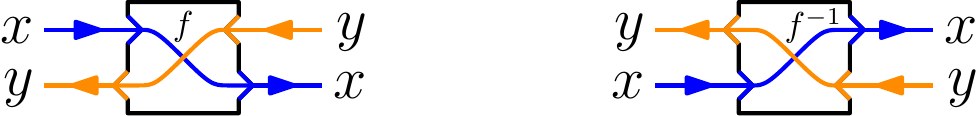}
\caption{Schematic representation of $f_{12}$ and $f_{12}^{-1}$. The exponents $+$ and $-$ of the $L$ matrices on one side and the other are represented by, respectively, right and left arrows (the top row being $L_1$, and the second $\tilde{L}_2$), and their arguments are represented by colours (blue for $x$ and orange for $y$).}
\label{fig-f12}
 \end{center}
 \end{figure}

~~

This was for the exchange of parameters between the first and second or third and fourth matrices in $L_1^+(x) \tilde{L}_2^-(y) L_3^+(x') \tilde{L}_4^-(y')$ (i.e. inside of a same $X$ matrix). We now need to exchange parameters between the second and third matrices in that product.

Let us consider $\tilde{L}_2L_3$ with $c_2=y$, $c_3=y'/q$, and the rest set to $1$:
\begin{equation}
\tilde{L}_2^-(y) L_3^-(y')=\begin{bmatrix}(1+ y y'/q S_2/ S_3) A_2 A_3 & (S_3 -y S_2)A_2  \\ (S_2^{-1}-y'/q S_3^{-1}) A_3 & S_3 /S_2+1\end{bmatrix}.\nonumber
\end{equation}

We need to find $g_{2 3}^-(y,y')$ such that $\tilde{L}_2^-(y) L_3^-(y')g_{2 3}^-(y,y')=g_{2 3}^-(y,y')\tilde{L}_2^-(y') L_3^-(y)$. As before, the commutation for each of the four elements of $\tilde{L}_2^-(y) L_3^-(y')$ is:
\begin{align}
[(1+ y y'/q S_2/ S_3) A_2 A_3,g_{2 3}^-]&=0,\nonumber\\
[1+S_3/ S_2,g_{2 3}^-]&=0,\nonumber\\
 (S_3-y S_2)A_2 ~ g_{2 3}^-&=g_{2 3}^- (S_3 -y' S_2)A_2 ,\nonumber\\
(S_2^{-1}-y'/q S_3^{-1}) A_3 ~g_{2 3}^-&=g_{2 3}^-(S_2^{-1}-y/q S_3^{-1}) A_3.\nonumber
\end{align}

The first and second equations tell us that $g_{2 3}^-$ commutes with $S_2/ S_3$ and with $A_2 A_3$. the third and fourth suggest that $g_{2 3}^-$ keeps $A_2$ and $A_3$ separated, so it should only depend on $S_2/ S_3$. The third equation then gives:
\begin{equation}
 (S_3 -y S_2)A_2 ~ g_{2 3}^-[S_2/ S_3]=g_{2 3}^-[q S_2/S_3] (S_3 -y S_2)A_2  =g_{2 3}^-[S_2/ S_3] (S_3 -y' S_2)A_2 \nonumber
\end{equation}
which can be rewritten as:
\begin{equation}
\frac{g_{2 3}^-[S_2/S_3]}{g_{2 3}^-[q S_2/ S_3]}=\frac{(1 -y S_2/S_3)}{(1 -y' S_2/S_3)}\nonumber
\end{equation}
and produces, through iteration:
\begin{equation}\boxed{
g_{2 3}^-(y,y')=\frac{(y S_2/S_3)_\infty}{(y' S_2/S_3)_\infty}.
}\end{equation}

~~

Finally, we look for $g_{2 3}^+(x,x')$ such that $\tilde{L}_2^+(x) L_3^+(x')g_{2 3}^+(x,x')=g_{2 3}^+(x,x')\tilde{L}_2^+(x') L_3^+(x)$. Setting $a_2=x$, $c_2=q$, $a_3=x'$, and the rest to $1$, in $\tilde{L}_2 L_3$, we find:
\begin{equation}
\tilde{L}_2^+(x) L_3^+(x')=\begin{bmatrix}(x x'+ q S_2/ S_3) A_2 A_3 & (x S_3 -q S_2)A_2  \\ (x' S_2^{-1}- S_3^{-1}) A_3 & S_3/S_2+1\end{bmatrix} \nonumber
\end{equation}
and the exact same operations as before produce:
\begin{equation}\boxed{
g_{2 3}^+(x,x')=\frac{(x' S_3/S_2)_\infty}{(x S_3/S_2)_\infty}.
}\end{equation}

Let us note that $g^+$ and $g^-$ commute with the projection which we perform on $L_1\tilde{L}_2$ and $L_3\tilde{L}_4$ to get the $X$ matrices.

 \begin{figure}[ht]
\begin{center}
\includegraphics[width=0.6\textwidth]{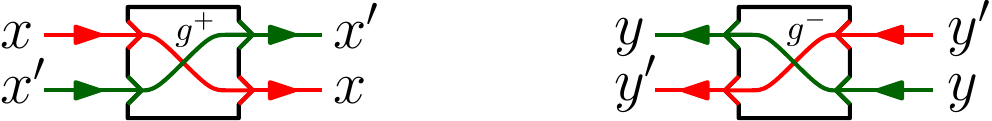}
\caption{Schematic representation of $g^+_{23}$ and $g^-_{23}$. The top row represents $\tilde{L}_2$, and the second $L_3$, and their arguments are represented by colours (red for $x$ and $y$ and green for $x'$ and $y'$).}
\label{fig-gpm}
 \end{center}
 \end{figure}

~~

We can finally forget everything about that alternative construction of $X(x,y)$, and simply define those operators $f$ and $g^\pm$ as what we found them to be. After re-indexing them so that $1$ refers to the auxiliary space of the first $X$ matrix, and $2$ to the second, we can write:
\begin{empheq}[box=\fbox]{align}
X_1(x,y)\centerdot X_2(x',y')~R_y(x,y;x',y')&=R_y(x,y;x',y')~X_1(x,y')\centerdot X_2(x',y)\label{V-1-Ry},\\
X_1(x,y)\centerdot X_2(x',y')~R_x(x,y;x',y')&=R_x(x,y;x',y')~X_1(x',y)\centerdot X_2(x,y')\label{V-1-Rx},
\end{empheq}
with
\begin{empheq}[box=\fbox]{align}
R_y(x,y;x',y')&=f_2(x' y') g^-_{1 2}(y,y') f_2^{-1}(x' y),\\
R_x(x,y;x',y')&=f_1(x y) g^+_{1 2}(x ,x') f_1^{-1}(x' y).
\end{empheq}
where we relabelled $f_{12}$ as $f_1$, $f_{34}$ as $f_2$, and $g^\pm_{23}$ as $g^\pm_{12}$, consistently with the indexes of the $X$ matrices.

\begin{figure}[ht]
\begin{center}
\includegraphics[width=0.9\textwidth]{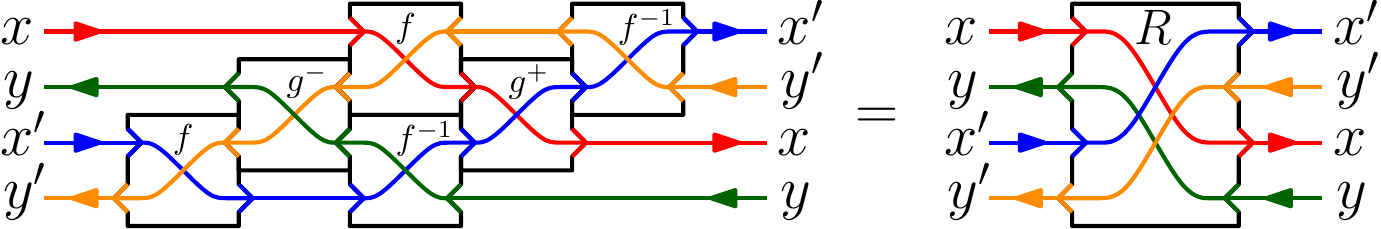}
\caption{Schematic representation of the $R$ matrix for the periodic ASEP.}
\label{fig-Rper}
 \end{center}
 \end{figure}

Applying those one after the other, we find the full $R$ matrix:
\begin{equation}\boxed{
X_1(x,y)\centerdot X_2(x',y')~R(x,y;x',y')=R(x,y;x',y')~X_1(x',y')\centerdot X_2(x,y)
}\end{equation}
with
\begin{empheq}[box=\fbox]{align}\label{V-1-Rmatrix}
R(x,y&;x',y')\!=\!R_y(x,y;x',y')R_x(x,y';x',y)\nonumber\\
&\!=\!\frac{(q A_2)_\infty}{(x'y'A_2)_\infty}\frac{(y S_1/S_2)_\infty}{(y' S_1/S_2)_\infty}\frac{(x' y A_2)_\infty}{(q A_2)_\infty}\frac{(q A_1)_\infty}{(xy'A_1)_\infty}\frac{(x' S_2/S_1)_\infty}{(x S_2/S_1)_\infty}\frac{(x'y' A_1)_\infty}{(q A_1)_\infty}
\end{empheq}
(cf. figs.-\ref{fig-f12},\ref{fig-gpm},\ref{fig-Rper}), or an equivalent expression if we apply $R_x$ to the left of $R_y$ instead.

\begin{figure}[ht]
\begin{center}
\includegraphics[width=\textwidth]{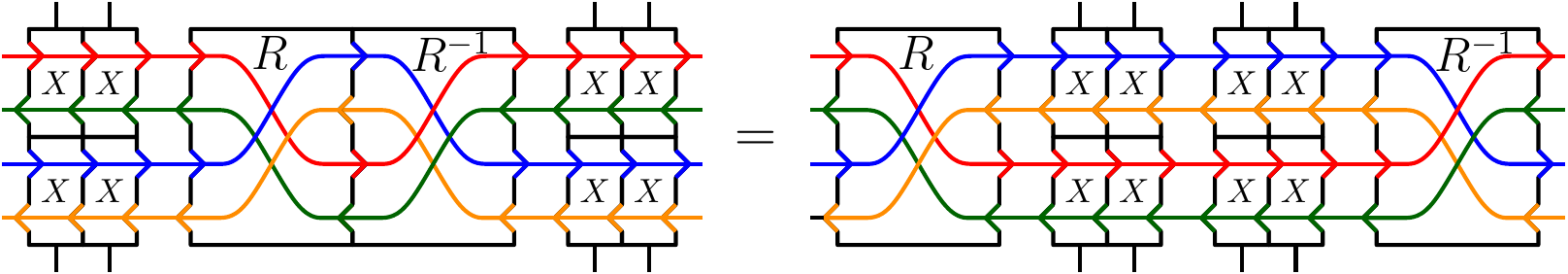}
\caption{Schematic representation of the commutation between $R$ matrices and $X$ matrices with different parameters (represented by different colours).}
\label{fig-RperComm}
 \end{center}
 \end{figure}

There are many things to be said about that $R$ matrix. First of all, it is the rigorous way to prove that $T^{per}_\mu(x,y)$ commutes with $T^{per}_\mu(x',y')$ for any value of those parameters: to do that, we insert $R(x,y;x',y')R^{-1}(x,y;x',y')$ at any point in the matrix product expression of $T^{per}_\mu(x,y)T^{per}_\mu(x',y')$, and make $R(x,y;x',y')$ commute to the left all the way around the trace (fig.-\ref{fig-RperComm}), exchanging parameters between the two rows along the way. When crossing the marked bond, we just need to note that $R$ commutes with $A_\mu\centerdot A_\mu$.

Equations (\ref{V-1-Ry}) and (\ref{V-1-Rx}) are also of interest by themselves: they tell us that $T^{per}_\mu(x,y)$ and $T^{per}_\mu(x',y')$ can actually exchange only one of their parameters and keep the other, instead of commuting altogether (by applying the procedure we just described, but with $R_x$ or $R_y$ instead of $R$). This decomposition of the R matrix is a well known property \cite{Derkachov2006,Derkachov2008}, and will be extremely useful to us in the next section.

Moreover, considering the decomposition (\ref{V-1-TtT}) which we obtained in the previous section for special values of the spectral parameters in $T^{per}_\mu(x,y)$, by taking $y=1/q^{k-1}x$ and $y'=1/q^{l-1}x'$ in $R(x,y;x',y')$, we should be able to recover the $R$ matrix between auxiliary dimensions $k$ and $l$ as an independent part of the whole matrix. In other words, this $R$ matrix of twice infinite dimension should contain all the smaller $R$ matrices for the XXZ chain. Taking $y=1/qx$, in particular, should yield the Lax matrix $X$ (possibly up to a permutation). Because of the complexity of eq.(\ref{V-1-Rmatrix}), those facts are yet to be verified.

\subsection{Q-operator and T-Q equations}
\label{V-1-4}

From the previous section, we now know that $T^{per}$ satisfies:
\begin{equation}\label{V-1-TTTT}
T^{per}(x,y)T^{per}(x',y')=T^{per}(x,y')T^{per}(x',y).
\end{equation}

Fixing $x'=y'=0$ in that equation (although any other constants would do), we can therefore write:
\begin{equation}\label{V-1-TPQ}\boxed{\boxed{
T^{per}(x,y)=P(x)Q(y)
}}\end{equation}
with:
\begin{equation}
P(x)=\Bigl[T_\mu^{per}(0,0)\Bigr]^{-1}T_\mu^{per}(x,0)~~~~,~~~~Q(y)=T_\mu^{per}(0,y).\nonumber
\end{equation}
Since the matrices $P$ and $Q$ are combinations of the matrix $T_\mu^{per}$ taken at various values of its parameters, and considering eq.(\ref{V-1-TTTT}), we immediately see that $P(x)$ and $Q(y)$ commute for any values of $x$ and $y$.

This relation is crucial to our reasoning, and, put together with (\ref{V-1-TtT}), will allow us to reach our conclusion in just a few more lines.

Using this, we can now rewrite (\ref{V-1-TtT}) as:
\begin{equation}\label{V-1-PQ}\boxed{
P(x)Q(1/q^{k-1}x)=t^{[k]}(x)+{\rm e}^{-k\mu}P(q^k x)Q(q/x).
}\end{equation}

The first and second orders of this equation give:
\begin{align}
P(x)Q(1/x)&=h(x)+{\rm e}^{-\mu}P(q x)Q(q/x)\label{V-1-PQ1},\\
P(q x)Q(1/q x)&=h(q x)+{\rm e}^{-\mu}P(q^2 x)Q(1/x)\label{V-1-PQ2},\\
P(x)Q(1/q x)&=t^{[2]}(x)+{\rm e}^{-2\mu}P(q^2 x)Q(q/x)\label{V-1-PQ3},
\end{align}
where we have written the first one twice (once at $x$, once at $qx$).

Combining those equations as $Q(1/qx)\times$(\ref{V-1-PQ1})$+{\rm e}^{-\mu}Q(q/x)\times$(\ref{V-1-PQ2})$-Q(1/x)\times$(\ref{V-1-PQ3}) yields the T-Q relation:
\begin{equation}\label{V-1-TQ}\boxed{
t^{[2]}(x)Q(1/x)=h(x)Q(1/qx)+{\rm e}^{-\mu}h(qx)Q(q/x).
}\end{equation}
Using this equation in conjunction with (\ref{V-1-ET2}) then gives $M_\mu$ in terms of $Q$:
\begin{equation}\label{V-1-ETQ}
M_\mu=(1-q)\frac{d}{dx}\log\biggl(\frac{Q(q/x)}{Q(1/x)}\biggr)\biggl|_{x=-1}.
\end{equation}

Let us note that using equs.(\ref{V-1-PQ}), we can obtain all the equations from the so-called `fusion hierarchy' \cite{Yang2006,Korff2007}, which gives equations on the decomposition of products of matrices $t^{[k]}$, as well as the T-Q equations for any $t^{[k]}$ (which involve products of $k-1$ matrices $Q$). We find for instance that
\begin{equation}
t^{[2]}(q^2x)t^{[3]}(x)=h(q^2x)t^{[4]}(x)+{\rm e}^{-\mu}h(q^3x)t^{[2]}(x),\nonumber
\end{equation}
and
\begin{align}
t^{[3]}(x)Q(1/x)Q(1/qx)=&h(x)Q(1/qx)Q(1/q^2x)+{\rm e}^{-\mu}h(qx)Q(q/x)Q(1/q^2x)\nonumber\\
&+{\rm e}^{-2\mu}h(q^2x)Q(q/x)Q(x).\nonumber
\end{align}
See appendix \ref{App-C} for more details on these relations.

~

At this point, in order to get an explicit result for the largest eigenvalue $E(\mu)$ of $M_\mu$, we need some additional information, which we get by taking $\mu$ to $0$.

\subsection{Non-deformed case}

We want to recover the expression for $E(\mu)$ which was found in \cite{prolhac2010tree}. Everything that we have shown so far is valid for all eigenspaces of $M_\mu$, but we now need to look for any information that we might have specifically on the dominant eigenvalue or eigenvectors of $M_\mu$, so that we can solve equation (\ref{V-1-ETQ}) explicitly in that eigenspace. Starting from the coordinate Bethe Ansatz, this was done, in \cite{prolhac2010tree}, using prior knowledge on the behaviour of Bethe roots for $\mu\rightarrow 0$. We cannot use that same argument here, for two reasons: we did not start from the coordinate Bethe Ansatz, so that our Q-operator is not defined in terms of Bethe roots, and even though we can, in principle, argue that the two definitions of $Q$ have to be consistent with one another, we do not expect to be able to do the same in the open case. Fortunately, we can get to the same result by purely algebraic calculations.

~

Let us go back to the definition of $T_\mu^{per}$ as given in (\ref{Tmuper}), as well as that of $X$ in terms of $S$ and $A$:
\begin{equation}
T_\mu^{per}(x,y)={\rm Tr}[A_\mu\prod\limits_{i=1}^{L}X^{(i)}]~~~~{\rm with}~~~~X=\begin{bmatrix} 1+xA & S(1-xyA)\\ S^{-1}(1-A)& 1+yA\end{bmatrix}.\nonumber
\end{equation}

Expanding this expression in powers of $A$, we see that each weight in $T_\mu^{per}$ is a finite sum of traces of the form $x^iy^jq^k {\rm Tr}[A_\mu S^l A^m]$, where $k$, $l$, and $m$ are integers, and the term $q^k$ comes from ordering the powers of $A$ and $S$. Because of the trace, those terms are $0$ if $l\neq 0$ (which accounts for the conservation of the total number of particles: the number of $d$'s and $e$'s in the trace have to be the same). Otherwise, they are simply equal to $x^iy^jq^k/(1-q^m {\rm e}^{-\mu})$. In particular, there is exactly one term with $m=0$, which is equal to $1/(1-{\rm e}^{-\mu})$. In the non-deformed limit $\mu\rightarrow 0$, this term dominates the trace in each entry of the transfer matrix, as it is the only term that diverges, so that:
\begin{equation}
T_\mu^{per}(x,y)\sim\frac{1}{1-{\rm e}^{-\mu}}|1_N\rangle\langle 1_N|\nonumber
\end{equation}
where $|1_N\rangle$ is a vector with all entries equal to $1$ if the total number of particles is $N$, and $0$ otherwise. $|1_N\rangle\langle 1_N|$ is the projector onto the dominant eigenspace (i.e. the steady state) of $M$ in the sector with $N$ particles.

From this, we draw two important conclusions. First, the prefactor in $Q(y)$ diverges as $1/\mu$. Secondly, we see that this limit does not depend on $x$ and $y$, which means that the roots of $P(x)$ and $Q(y)$, in this eigenspace, all go to infinity when $\mu$ goes to $0$. This allows us to use a contour integral in (\ref{V-1-PQ1}) to separate the contribution from $P(x)$ and $P(qx)$ (whose roots all go to infinity, and are therefore all out of the unit circle for $\mu$ small enough) and those from $Q(1/x)$ and $Q(q/x)$ (whose roots, in terms of $x$, all go to $0$, and are inside of the unit circle for $\mu$ small enough). We also find that the eigenvalue of $t^{[2]}$ in that eigenspace is
\begin{equation}
\langle 1_N|t^{[2]}|1_N\rangle=h(x)+h(qx)\nonumber
\end{equation}
which gives $E(0)=0$. Using this, we can solve (\ref{V-1-PQ1}) perturbatively in $\mu$ and find $Q$ in terms of $h$. Putting the result in (\ref{V-1-ETQ}), we finally find $E(\mu)$. Since we will be doing the same for the open ASEP in the next section with only a few differences, we will give those final steps in detail only in section \ref{V-2-4}.

\section{Open ASEP}
\label{V-2}

We will now apply the same procedure to the open ASEP. Considering the same $X$ matrix as before, we first generalise the boundary vectors found in \cite{1751-8121-46-14-145003} accordingly, and see how our construction relates to the original matrix Ansatz \cite{derrida1993exact}. We then show the PQ factorisation of the transfer matrix, which is much more straightforward than in the periodic case. Finally, we see how it decomposes into blocks, one of which is a one-parameter transfer matrix with finite auxiliary dimension, according to the same equation as in the periodic case but with a different quantum determinant. We also show, as an appendix, what this all becomes in the language of the XXZ chain with spin $\frac{1}{2}$.

The current-counting Markov matrix for the open ASEP is given by:
\begin{equation}
M_\mu=m^{(0)}+\sum_{i=1}^{L-1} M^{(i)}+m^{(L)}\nonumber
\end{equation}
with
\begin{equation}
m^{(0)}=\begin{bmatrix} -\alpha & \gamma {\rm e}^{-\mu} \\ \alpha{\rm e}^{\mu} & -\gamma  \end{bmatrix}~,~ M^{(i)}=\begin{bmatrix} 0 & 0 & 0 & 0 \\ 0 & -q & 1 & 0 \\ 0 & q & -1 & 0 \\ 0 & 0 & 0 & 0 \end{bmatrix}~,~m^{(L)}=\begin{bmatrix} -\delta & \beta \\  \delta & -\beta  \end{bmatrix}\nonumber
\end{equation}
where the matrices $m^{(0)}$ and $m^{(L)}$ represent two particle reservoirs to which the system is coupled, through the first and the last site, respectively. $m^{(0)}$ acts on site $1$ and $m^{(L)}$ on site $L$, both in the canonical occupancy basis $\{0,1\}$, and, as previously, $M^{(i)}$ acts on sites $i$ and $i+1$ in basis $\{00,01,10,11\}$.

As in the periodic case, the bond over which we count the current can be chosen arbitrarily, so for the sake of simplicity we choose the one between the left reservoir and the first site.

\subsection{Boundary algebra and commutation relations}
\label{V-2-1}

Let us first find out how the presence of boundaries make this case different from the previous one. Most of the results and calculations in this section are generalisations of those found in \cite{1751-8121-46-14-145003}. We define:
\begin{empheq}[box=\fbox]{align}
U_\mu(x)&= \langle\!\langle W|\!| A_\mu \prod_{i=1}^{L}X^{(i)}(x,x) |\!| V \rangle\!\rangle,\\
T_\mu(y)&= \langle\!\langle  \tilde{W}|\!|  A_\mu \prod_{i=1}^{L}X^{(i)}(y,y)|\!|  \tilde{V} \rangle\!\rangle,
\end{empheq}
which has the same structure as the transfer matrix from \cite{1751-8121-46-14-145003}, but with the $X$ matrices replaced by their generalisation. Note that, since we have two transfer matrices, we have, in principle, four free parameters (two in each matrix), but we must in fact put the same parameter twice in each matrix (so that $U_\mu$ depends only on $x$, and $T_\mu$ on $y$) if we want to be able to find suitable boundary vectors. To simplify notations, we will therefore rewrite $X$ and $\hat{X}$ as:
\begin{equation}
X(x)=\begin{bmatrix} n_x & e\\ d & n_x \end{bmatrix}~,~\hat{X}=\frac{(1-q)}{2}\begin{bmatrix} \tilde{n}_{x} & e\\ -d & -\tilde{n}_x\end{bmatrix}\nonumber
\end{equation}
with $n_x=1+x A$ and $\tilde{ n}_x=1-x A$.

Note that for a general set of fugacities, the generalisation (\ref{V-1-Tmui}) holds, with the matrices $A_{\mu_i}$ being inserted in both $U$ and $T$.

The conditions that these boundary vectors must satisfy is:
\begin{empheq}[box=\fbox]{align}
[\beta (d+n_x) - \delta (e+n_x) -(1-q)] ~|\!| V \rangle\!\rangle &= 0 \label{V-2-V},\\
\langle\!\langle W|\!|~ [\alpha(e+n_x)- \gamma (d+n_x)-(1-q)] &= 0\label{V-2-W},\\
[\beta (d-n_y)-\delta (e-n_y) +(1-q)y A]~|\!|  \tilde{V} \rangle\!\rangle  &=0\label{V-2-Vt}, \\
\langle\!\langle \tilde{W}|\!|~ [\alpha(e-n_y) - \gamma (d-n_y)+(1-q)y A] &=0\label{V-2-Wt},
\end{empheq}
where we notice that the first two are the same as in \cite{derrida1993exact,1751-8121-46-14-145003} with $n_x$ replacing $1$, and the next two also have an extra term $(1-q)yA$. For $x=y=0$, we naturally recover the conditions given in \cite{1751-8121-46-14-145003}. Note that those conditions were found by trial and error, so that they may not be unique.

\subsubsection{Commutation relations}

As in the periodic case, we will now see how the transfer matrices we have just defined commute with each of the individual jump matrices.

For the bulk matrices $M^{(i)}$, we can simply use eq. (\ref{XX-XX}). Summing all of them, we get
 \begin{align}   \label{MUhat}
&&\Bigl[\sum_{i=1}^{L-1}M^{(i)},U_\mu\Bigr]&=\langle\!\langle W|\!| A_\mu \hat{X}^{(1)}\prod_{i=2}^{L}X^{(i)} |\!|V\rangle\!\rangle- \langle\!\langle W|\!| A_\mu \prod_{i=1}^{L-1}X^{(i)}~\hat{X}^{(L)} |\!|V\rangle\!\rangle,\\  \label{MThat}
&&\Bigl[\sum_{i=1}^{L-1}M^{(i)},T_\mu\Bigr]&= \langle\!\langle \tilde{W}|\!| A_\mu \hat{X}^{(1)}\prod_{i=2}^{L}X^{(i)} |\!|\tilde{V}\rangle\!\rangle- \langle\!\langle \tilde{W}|\!| A_\mu \prod_{i=1}^{L-1}X^{(i)}~\hat{X}^{(L)}    |\!|\tilde{V}\rangle\!\rangle.
\end{align}

Unlike in the periodic case, where the trace over the product of matrices ensures that all the $\hat{X}$ matrices cancel one another, we here need to consider the action of the boundary matrices $m^{(0)}$ and $m^{(L)}$ as well. We find that we cannot cancel the two boundary terms in (\ref{MUhat}), and those in (\ref{MThat}), independently. Instead, we have to consider the commutation with the product $U_\mu(x)T_\mu(y)$. We show, in appendix \ref{App-A-2}, that
\begin{align}
[m^{(0)}(\mu),U_\mu T_\mu]&=- \langle\!\langle W|\!| A_\mu \hat{X}^{(1)}\prod_{i=2}^{L}X^{(i)} |\!|V\rangle\!\rangle\centerdot T_\mu-U_\mu \centerdot \langle\!\langle \tilde{W}|\!| A_\mu \hat{X}^{(1)}\prod_{i=2}^{L}X^{(i)} |\!|\tilde{V}\rangle\!\rangle\label{III-2-M0UTa},\\
[m^{(L)},U_\mu T_\mu]&= \langle\!\langle W|\!| A_\mu \prod_{i=1}^{L-1}X^{(i)}\hat{X}^{(L)} |\!|V\rangle\!\rangle\centerdot T_\mu+U_\mu\centerdot \langle\!\langle \tilde{W}|\!| A_\mu \prod_{i=1}^{L-1}X^{(i)}\hat{X}^{(L)}  |\!|\tilde{V}\rangle\!\rangle. \label{III-2-M0UTb}
\end{align}
which is exactly what is needed to compensate the terms coming from the bulk.

~~

Putting those relations together, we can conclude that, for any values of $x$ and $y$, we have:
\begin{equation}\label{V-2-MUT}\boxed{\boxed{
[M_\mu, U_\mu(x)T_\mu(y)]=0.
}}\end{equation}

\subsubsection{Explicit expressions for the boundary vectors}

We will need an expression for the boundary vectors in a few calculations later, so we determine them explicitly here.

For that, we need to define four new parameters in terms of the boundary rates:
\begin{align}
a&=\frac{1}{2\alpha}\Bigl[(1-q-\alpha+\gamma)+\sqrt{(1-q-\alpha+\gamma)^2+4\alpha\gamma}\Bigr] ,\nonumber\\
\tilde{a}&=\frac{1}{2\alpha}\Bigl[(1-q-\alpha+\gamma)-\sqrt{(1-q-\alpha+\gamma)^2+4\alpha\gamma}\Bigr],\nonumber \\
b&=\frac{1}{2\beta}\Bigl[(1-q-\beta+\delta)+\sqrt{(1-q-\beta+\delta)^2+4\beta\delta}\Bigr],\nonumber\\
\tilde{b}&=\frac{1}{2\beta}\Bigl[(1-q-\beta+\delta)-\sqrt{(1-q-\beta+\delta)^2+4\beta\delta}\Bigr] ,\nonumber
\end{align}
and reversely:
\begin{align}
\alpha&=\frac{(1-q)}{(1+a)(1+\tilde{a})}, \nonumber\\
\gamma&=-\frac{a \tilde{a}(1-q)}{(1+a)(1+\tilde{a})} ,\nonumber\\
\delta&=-\frac{b  \tilde{b}(1-q)}{(1+b)(1+ \tilde{b})} ,\nonumber\\
\beta&=\frac{(1-q)}{(1+b)(1+ \tilde{b})}.\nonumber
\end{align}

In terms of these, equations (\ref{V-2-V})-(\ref{V-2-Wt}) become:
\begin{align} 
[d+b\tilde{b}e+(1+b\tilde{b})xA-(b+\tilde{b})] ~|\!| V \rangle\!\rangle &= 0 \label{V-2-Vb},\\
\langle\!\langle W|\!|~ [e+a\tilde{a}d+(1+a\tilde{a})xA-(a+\tilde{a})] &= 0\label{V-2-Vtb},\\
[d+b\tilde{b}e+(b+\tilde{b})y A-(1+b \tilde{b})]~|\!|  \tilde{V} \rangle\!\rangle &=0 \label{V-2-Wa},\\
\langle\!\langle \tilde{W}|\!|~ [e+a\tilde{a}d+(a+\tilde{a})y A-(1+a \tilde{a})] &=0\label{V-2-Wta}.
\end{align}

We first focus on the right boundary. We saw that a possible representation for $d$ and $e$ is $e=S_1(1-x^2 A_1)$ in $U_\mu$ or $e=S_2(1-y^2 A_2)$ in $T_\mu$, and $d=S_i^{-1}(1-A_i)$ (where indices $1$ and $2$ refer to the auxiliary spaces of $U_\mu$ and $T_\mu$, respectively). Equations (\ref{V-2-Vb}) and (\ref{V-2-Vtb}) become:
\begin{align} 
[S_1^{-1}(1-A_1)+b\tilde{b}S_1(1-x^2 A_1)+(1+b\tilde{b})xA_1-(b+\tilde{b})] ~|\!| V \rangle\!\rangle &= 0 ,\nonumber\\
[S_2^{-1}(1-A_2)+b\tilde{b}S_2(1-y^2 A_2)+(b+\tilde{b})y A_2-(1+b \tilde{b})]~|\!|  \tilde{V} \rangle\!\rangle &=0,\nonumber
\end{align}
which is to say, multiplying by $S_i$ to the left:
\begin{align} 
[(1-bS_1)(1-\tilde{b}S_1v)-(1-xS_1)(1-b\tilde{b}xS_1)A_1] ~|\!| V \rangle\!\rangle &= 0,\nonumber \\
[(1-S_2)(1-b\tilde{b}S_2)-(1-byS_2)(1-\tilde{b}yS_2)A_2]~|\!|  \tilde{V} \rangle\!\rangle &=0.\nonumber
\end{align}

We can write those vectors as generating functions, which will make it easier for us to manipulate them. Let us then write $|\!| V \rangle\!\rangle =V(S_1) |\!|0\rangle\!\rangle=\sum\limits_{k=0}^\infty V_k S_1^k |\!|0\rangle\!\rangle$, which is such that $A_1|\!| V \rangle\!\rangle=V(qS_1) |\!|0\rangle\!\rangle$. Those two last equations now become:
\begin{align} 
\frac{V(S_1)}{V(qS_1)} |\!|0\rangle\!\rangle&=\frac{(1-xS_1)(1-b\tilde{b}xS_1)}{(1-bS_1)(1-\tilde{b}S_1)} |\!|0\rangle\!\rangle,\nonumber\\
\frac{ \tilde{V} (S_2)}{ \tilde{V} (qS_2)} |\!|0\rangle\!\rangle&=\frac{(1-byS_2)(1-\tilde{b}yS_2)}{(1-S_2)(1-b\tilde{b}S_2)}|\!|0\rangle\!\rangle,\nonumber
\end{align}
which we can iterate to get:
\begin{empheq}[box=\fbox]{align}
|\!| V \rangle\!\rangle&=\frac{(xS_1)_\infty(b\tilde{b}xS_1)_\infty}{(bS_1)_\infty(\tilde{b}S_1)_\infty} |\!|0\rangle\!\rangle,\\
|\!|  \tilde{V} \rangle\!\rangle&=\frac{(byS_2)_\infty(\tilde{b}yS_2)_\infty}{(S_2)_\infty(b\tilde{b}S_2)_\infty} |\!|0\rangle\!\rangle
\end{empheq}
where we recall that $(x)_\infty$ is the infinite q-Pochhammer symbol:
\begin{equation}
(x)_n=\prod\limits_{k=0}^{n-1}(1-q^kx),~~~~~~~~~~~~~(x)_\infty=\prod\limits_{k=0}^{\infty}(1-q^kx).\nonumber
\end{equation}
We will sometimes write products of those in a more compact form, with only one symbol, such as $(x,y)_\infty=(x)_\infty(y)_\infty$.

As for the left boundary, it is simpler to treat it using the contragredient representation $\overline{X}$ of $X$ on both vectors (which are then the exact symmetric of $|\!| V \rangle\!\rangle$ and $|\!|  \tilde{V} \rangle\!\rangle$, but with $a$ and $\tilde{a}$ replacing $b$ and $\tilde{b}$), and then apply the operator $f_i$ that we found in section \ref{V-1-3} in order to go back to $X$. This gives us:
\begin{empheq}[box=\fbox]{align}
\langle\!\langle W|\!|&=\langle\!\langle 0|\!|\frac{(x/S_1)_\infty(a\tilde{a}x/S_1)_\infty}{(a/S_1)_\infty(\tilde{a}/S_1)_\infty}\frac{(x^2A_1)_\infty}{(qA_1)_\infty},\\
\langle\!\langle \tilde{W}|\!|&=\langle\!\langle0|\!|\frac{(ay/S_2)_\infty(\tilde{a}y/S_2)_\infty}{(1/S_2)_\infty(a\tilde{a}/S_2)_\infty}\frac{(y^2A_2)_\infty}{(qA_2)_\infty},
\end{empheq}
to which we could add factors $\frac{(q)_\infty}{(x^2)_\infty}$ and $\frac{(q)_\infty}{(y^2)_\infty}$ for normalisation.

In fact, in most future calculations, we will use $X$ for $U_\mu$ and $\overline{X}$ for $T_\mu$, so that the factor $\frac{(y^2A_2)_\infty}{(qA_2)_\infty}$ goes to $|\!| \tilde{V} \rangle\!\rangle$ instead of $\langle\!\langle \tilde{W}|\!|$.

Note finally that scalar products involving those vectors (like, for instance, all the weights of $U_\mu$ and $T_\mu$) are well defined only for $|a|$, $|b|$, $|\tilde{a}|$ and $|\tilde{b}|$ all smaller than $1$. This corresponds to the ASEP being in its so-called `maximal current' phase. We can assume this to be the case for now, but a simple analytic continuation can be performed in order to access more generic values of the parameters \cite{1751-8121-40-46-R01}.

~~

We now need to show two things: that the transfer matrix is a product of two commuting matrices, one depending on $x$ and one on $y$, and that for special values of the parameters it decomposes into two independent blocks.

\subsection{R matrix and PQ factorisation}
\label{V-2-2}

Unlike what we did for the periodic case, we will first show that $U_\mu(x)T_\mu(y)$ factorises into two commuting matrices $P(x)$ and $Q(y)$. Note that this decomposition is not trivial: $U_\mu$ and $T_\mu$ do not commute, so $P$ is not simply equal to $U_\mu$ and $Q$ to $T_\mu$.

~~

First, we need to show that $U_\mu(x)T_\mu(y)$ commutes with $U_\mu(x')T_\mu(y')$ for any values of $x$, $y$, $x'$ and $y'$. If we want to be rigorous, the fact that both commute with $M_\mu$ is not enough ($M_\mu$ might have a degenerate eigenspace which is not degenerate for $U_\mu(x)T_\mu(y)$), so, as in the periodic case, we need to use R matrices to exchange spectral parameters. We cannot do as easily as in the periodic case, because of the boundaries: instead of taking $R$ around a trace to exchange the top and bottom matrices, we now need to show that when applied to a boundary, $R$ transfers the spectral parameters from a vector to another. We find the two following results (which are proven in appendix \ref{App-B-1} and illustrated on fig.-\ref{fig-RRvv}):
\begin{align}
R^{-1}(x,x;y,y)~|\!| V(x) \rangle\!\rangle\centerdot|\!| \tilde{V}(y) \rangle\!\rangle=|\!| V(y) \rangle\!\rangle\centerdot|\!| \tilde{V}(x) \rangle\!\rangle,\nonumber\\
R^{-1}(y,y;x,x)~|\!|  \tilde{V}(y) \rangle\!\rangle\centerdot|\!| V(x)\rangle\!\rangle=|\!|  \tilde{V}(x) \rangle\!\rangle\centerdot|\!|V(y) \rangle\!\rangle,\nonumber
\end{align}
where we use the same R matrix as defined in (\ref{V-1-Rmatrix}).

 \begin{figure}[ht]
\begin{center}
\includegraphics[width=0.8\textwidth]{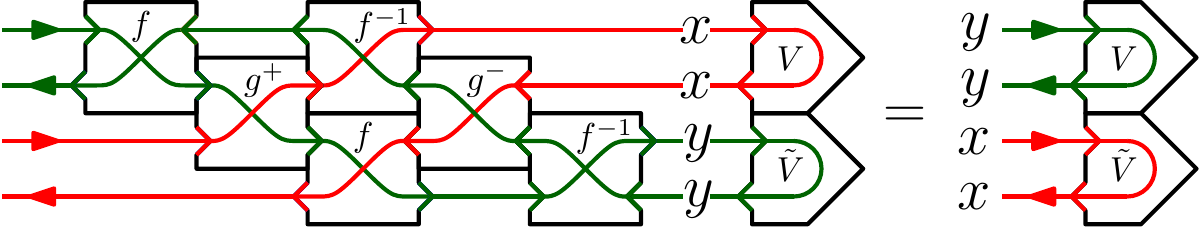}
\caption{Action of the R matrix on the right boundary.}
\label{fig-RRvv}
 \end{center}
 \end{figure}

These relations imply, at the level of the transfer matrices, that
\begin{empheq}[box=\fbox]{align}
U_\mu(x)T_\mu(y)=U_\mu(y)T_\mu(x),\label{UxTyUyTx}\\
T_\mu(y)U_\mu(x)=T_\mu(x)U_\mu(y),
\end{empheq}
which immediately gives that $U_\mu(x)T_\mu(y)$ commutes with $U_\mu(x')T_\mu(y')$, but also that
\begin{equation}
U_\mu(x)T_\mu(y)~U_\mu(0)T_\mu(0)=U_\mu(x)T_\mu(0)~U_\mu(0)T_\mu(y)\nonumber
\end{equation}
so that we can write, dividing by $U_\mu(0)T_\mu(0)$,
\begin{equation}\boxed{\boxed{
U_\mu(x)T_\mu(y)=P(x)Q(y)
}}\end{equation}
with
\begin{equation}
P(x)=U_\mu(x)\Bigl[U_\mu(0)\Bigr]^{-1}~~~~,~~~~Q(y)=U_\mu(0)T_\mu(y),\nonumber
\end{equation}
where $P$ and $Q$ commute with each other. Note that, in terms of $P$ and $Q$ and at $y=0$, eq.(\ref{UxTyUyTx}) becomes $P(x)Q(0)=P(0)Q(x)$, so that $P$ and $Q$ are the same up to some constant (since $P(0)$ is set to $1$ but not $Q(0)$). We will not need to use this identity in our calculations, so we will keep using distinct names for those two objects, to have notations that are consistent with the periodic case.

\subsection{Decomposition of the transfer matrix}
\label{V-2-3}

In this section, we investigate whether anything happens for those special values of $x$ and $y$ that we considered in section \ref{V-1-2}. The calculations involved are slightly more complicated than they were then, not only because of the boundaries, but also because $x$ and $y$ are now separated, and the factors $(1-xyA)$ in $e$ have been replaced by $(1-x^2A_1)$ and $(1-y^2A_2)$, which do not give anything useful for $xy=q^{1-k}$.

The simple solution to this problem is to put $x$ and $y$ back together, using  $g_{12}^-(x,y)$ (as defined in section \ref{V-1-3}) to go from $X(x,x)\centerdot \overline{X}(y,y)$ to $X(x,y)\centerdot \overline{X}(y,x)$. We choose to keep the second row of matrices in the contragredient representation to simplify future calculations. We recall that it corresponds to having $f(x y)$ commute through $X(x,y)$, so that
\begin{equation}
\overline{X}(x,y)=f(xy)X(x,y)f^{-1}(xy)=\begin{bmatrix} n_x & {}^t\!d\\ {}^t\!e & n_y \end{bmatrix}\nonumber
\end{equation}

\begin{figure}[ht]
\begin{center}
\includegraphics[width=0.8\textwidth]{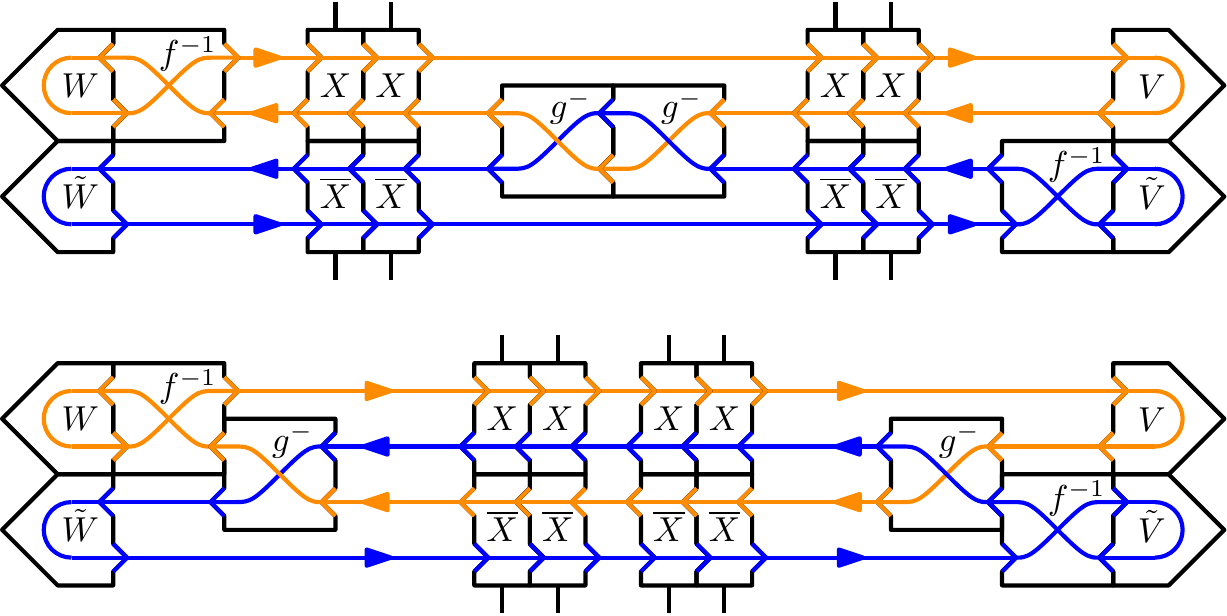}
\caption{Two equivalent representations for $U_\mu(x)T_\mu(y)$. In the first one, the two rows of $X$ matrices are independent, and the two spectral parameters $x$ (orange) and $y$ (blue) are separated. In the second, $x$ and $y$ are back together in $X$, but the two rows are now intertwined.}
\label{fig-RowsLoop}
 \end{center}
 \end{figure}

This operation makes the boundary vectors more complicated: they are not a product of two vectors acting each on one row any more (see fig.-\ref{fig-RowsLoop}). We will write those vectors as $K^+$ for the right boundary and $K^-$ for the left one:
\begin{empheq}[box=\fbox]{align}
K^-&\!=\!\langle\!\langle 0_1,0_2|\!|\frac{(x/S_1)_\infty(a\tilde{a}x/S_1)_\infty}{(a/S_1)_\infty(\tilde{a}/S_1)_\infty}\frac{(x^2A_1)_\infty}{(qA_1)_\infty}\frac{(ay/S_2)_\infty(\tilde{a}y/S_2)_\infty}{(1/S_2)_\infty(a\tilde{a}/S_2)_\infty}\frac{(x S_1/S_2)_\infty}{(y S_1/S_2)_\infty}\frac{(q)_\infty}{(x^2)_\infty},\label{Kminus}\\
K^+&\!=\!\frac{(q)_\infty}{(y^2)_\infty}\frac{(y S_1/S_2)_\infty}{(x S_1/S_2)_\infty}\frac{(xS_1)_\infty(b\tilde{b}xS_1)_\infty}{(bS_1)_\infty(\tilde{b}S_1)_\infty}\frac{(y^2A_2)_\infty}{(qA_2)_\infty}\frac{(byS_2)_\infty(\tilde{b}yS_2)_\infty}{(S_2)_\infty(b\tilde{b}S_2)_\infty} |\!|0_1,0_2\rangle\!\rangle.\label{Kplus}
\end{empheq}

We will denote by $K^+_{i,j}$ the coefficient of $ |\!|i,j\rangle\!\rangle$ in $K^+$ (i.e. the coefficient of $S_1^{i}S_2^{j}$ in the expansion of the ratios of q-Pochhammer symbols in $K^+$), and by $K^-_{j,i}$ the coefficient of $\langle\!\langle i,j|\!|$ in $K^-$. Note that, in this section, the normalisation of the boundary vectors is important. The factors $\frac{(q)_\infty}{(x^2)_\infty}$ and $\frac{(q)_\infty}{(y^2)_\infty}$ in $K^+$ and $K^-$ are there so that $K^-_{0,0}=K^+_{0,0}=1$. Also note that the transpose of $K^-$ can be obtained from $K^+$ through the transformation $\{x\leftrightarrow y,S_1\rightarrow aS_2,S_2\rightarrow aS_1, b\rightarrow 1/a,\tilde{b}\rightarrow\tilde{a}\}$, so that we only need to do calculations on $K^+$.

~~

Now that $x$ and $y$ have been reunited in the bulk of the system, the same decomposition (into block triangular matrices) happens as did in the periodic case. We have to see whether the same happens to the boundary matrices $K^\pm$.

\subsubsection{One-way boundaries}

We first focus on the simpler case where $\tilde{a}=\tilde{b}=0$, which is to say $\gamma=\delta=0$. For this simpler case, we have:
\begin{equation}
K^+=\frac{(xS_1)_\infty}{(bS_1)_\infty}\frac{(y S_1/S_2)_\infty}{(x S_1/S_2)_\infty}\frac{(y^2A_2)_\infty}{(qA_2)_\infty}\frac{(byS_2)_\infty}{(S_2)_\infty} \frac{(q)_\infty}{(y^2)_\infty}\nonumber
\end{equation}
where the factor $|\!|0_1,0_2\rangle\!\rangle$ on which this whole function acts is implicit.

This can be expanded into:
\begin{equation}
K^+=\frac{(xS_1)_\infty}{(bS_1)_\infty}\sum\limits_{k=0}^{\infty}\frac{(y/x)_k}{(q)_k}(x S_1/S_2)^k\sum\limits_{n=0}^{\infty}\frac{(b y)_n}{(y^2)_n}(S_2)^n\nonumber
\end{equation}
where the first sum comes from $\frac{(y S_1/S_2)_\infty}{(x S_1/S_2)_\infty}$, and the second from everything that is to the right of that.

~~

We will shortly be in need of one of Heine's transformation formulae for basic hypergeometric series, which we give now. For a function $_2\phi_1(a,b;c;z)$ defined as:
\begin{equation}
_2\phi_1(a,b;c;z)=\sum\limits_{n=0}^{\infty}\frac{(a)_n(b)_n}{(q)_n(c)_n}z^n
\end{equation}
we have \cite{Gasper2004}:
\begin{equation}\label{V-2-qHeine}\boxed{
_2\phi_1(a,b;c;z)=\frac{(z ab/c)_\infty}{(z)_\infty}~{}_2\phi_1(c/a,c/b;c;zab/c).
}\end{equation}

We will also need this simple identity on q-Pochhammer symbols:
\begin{equation}\label{V-2-xqjk}\boxed{
(x)_{j+k}=(x)_j(x q^j)_{k}.
}\end{equation}

~~

Coming back to $K^+$, we have:
\begin{align}
K^+&=\frac{(xS_1)_\infty}{(bS_1)_\infty}\sum\limits_{j=0}^{\infty}\frac{(b y)_j}{(y^2)_j}(S_2)^j\sum\limits_{k=0}^{\infty}\frac{(y/x)_k}{(q)_k}(x S_1/S_2)^k\frac{(b yq^j)_k}{(y^2q^j)_k}(S_2)^k\nonumber\\
&=\sum\limits_{j=0}^{\infty}\frac{(b y)_j}{(y^2)_j}(S_2)^j\frac{(xS_1)_\infty}{(bS_1)_\infty}{}_2\phi_1(y/x,byq^j;y^2q^j;xS_1)\nonumber\\
&=\sum\limits_{j=0}^{\infty}\frac{(b y)_j}{(y^2)_j}(S_2)^j{}_2\phi_1(xyq^j,y/b;y^2q^j,bS_1)\label{V-2-K+Heine}\nonumber
\end{align}
where we used (\ref{V-2-xqjk}) between the first line and the second, and (\ref{V-2-qHeine}) between the second and the third. This finally gives us:
\begin{equation}\boxed{
K^+_{i,j}(x,y)=\frac{(xyq^j)_i(y/b)_i b^i(b y)_j}{(q)_i(y^2)_{i+j}}
}\end{equation}
and, at the left boundary:
\begin{equation}\boxed{
K^-_{j,i}(x,y)=\frac{(xyq^i)_j(x/a)_i a^i(a x)_j}{(q)_j(x^2)_{i+j}}.
}\end{equation}

We need to compare those values for $(x,y)$ equal to $(1/q^{k-1}y,y)$ or $(q/y,q^k y)$ (which are the same values as before, but keeping $y$ as a variable instead of $x$). We get:
\begin{align}
K^+_{i,j}(1/q^{k-1}y,y)&=\frac{(q^{j-k+1})_i(y/b)_i b^i(b y)_j}{(q)_i(y^2)_{i+j}},\nonumber\\
K^+_{i,j}(q/y,q^ky)&=\frac{(q^{j+k+1})_i(yq^k/b)_i b^i(b yq^k)_j}{(q)_i(y^2q^{2k})_{i+j}}.\nonumber
\end{align}

Since $(q^{j-k+1})_i=0$ for $j-k+1\leq 0$ and $i+j-k\geq 0$, i.e. for $j\leq k-1$ and $i\geq k-j$, the first matrix is, as we expected, block triangular.

We now consider, for $j\geq k-1$, the ratio:
\begin{equation}
\frac{K^+_{i+k,j+k}(1/q^{k-1}y,y)}{K^+_{i,j}(q/y,q^ky)}=\frac{(y/b)_k b^k(b y)_k}{(y^2)_{2k}}\frac{(q^{j+1})_k}{(q^{i+1})_k}.\nonumber
\end{equation}

The term $\frac{(q^{j+1})_k}{(q^{i+1})_k}$ accounts for going to the contragredient representation on both lines (consistently with what happens to the $X$ matrices), and can be transferred to the left boundary, where they compensate similar terms, emerging from the same calculation:
\begin{equation}
\frac{K^-_{j+k,i+k}(x,1/q^{k-1}x)}{K^-_{j,i}(q^k x,q/x)}=\frac{(x/a)_k a^k(a x)_k}{(x^2)_{2k}}\frac{(q^{i+1})_k}{(q^{j+1})_k}.\nonumber
\end{equation}
All the other terms depend only on $k$, and factor out of the matrix product. We will now put them together.

First, we need to make a few transformations on those factors. For $y=1/q^{k-1}x$, we have:
\begin{align}
(y/b)_k b^k&=(-x)^{-k}q^{-k(k-1)/2}(bx)_k,\nonumber\\
(b y)_k&=(bq^{-k+1}/x)_k,\nonumber\\
(x/a)_k a^k&=(-x)^{k}q^{k(k-1)/2}(aq^{-k+1}/x)_k,\nonumber
\end{align}
which gives us:
\begin{align}
\frac{(x/a)_k a^k(a x)_k(y/b)_k b^k(b y)_k}{(x^2)_{2k}(y^2)_{2k}}&=\frac{(aq^{-k+1}/x)_k (bq^{-k+1}/x)_k (a x)_k (bx)_k }{(x^2)_{2k}(x^{-2}q^{-2k+2})_{2k}}\nonumber\\
&=\frac{h_b(q^k x)h_b(q/x)}{h_b(x)h_b(1/q^{k-1}x)}\nonumber
\end{align}
with
\begin{equation}\label{V-2-hb1}\boxed{
h_b(x)=\frac{(x^2)_{\infty}}{(ax,bx)_\infty}.
}\end{equation}

We see that the boundary matrices, just as the bulk matrices, are upper block triangular, with a first block of auxiliary dimension $k$, and a second of infinite auxiliary dimension, which is the same as the full matrix at different values of the spectral parameters, up to a global factor which we have just computed. Put into equations, this becomes:
\begin{equation}
P(x)Q(1/q^{k-1}x)=t^{[2]}(x)+{\rm e}^{-2k\mu}\frac{h_b(q^k x)h_b(q/x)}{h_b(x)h_b(1/q^{k-1}x)}P(q^k x)Q(q/x)\nonumber
\end{equation}
where the factor ${\rm e}^{-2k\mu}$ comes, as before, from the first coefficient of the matrices $A_\mu$ (of which there are now $2$) in the second block.

To make things simpler, we can redefine all our matrices as:
\begin{equation}\label{PQtilde}
\tilde{P}(x)=h_b(x)P(x)~~,~~\tilde{Q}(x)=h_b(x)Q(x)~~,~~\tilde{t}^{[k]}(x)=h_b(x)h_b(1/q^{k-1}x)t^{[k]}(x)
\end{equation}
which is equivalent to choosing another normalisation for $U_\mu$ and $T_\mu$. Written in terms of those new transfer matrices, this last equation takes its definitive form:
\begin{equation}\label{V-2-PQ}\boxed{\boxed{
\tilde{P}(x)\tilde{Q}(1/q^{k-1}x)=\tilde{t}^{[k]}(x)+{\rm e}^{-2k\mu}\tilde{P}(q^k x)\tilde{Q}(q/x).
}}\end{equation}

This has the exact same form as eq.(\ref{V-1-PQ}), the only difference being a factor $2\mu$ instead of $\mu$ in the right hand side. Note that, since $\tilde{P}$ and $\tilde{Q}$ are the same function up to a constant (as we saw at the end of section \ref{V-2-2}), we find that $\tilde{t}^{[k]}$ is symmetric under $x\leftrightarrow 1/q^{k-1}x$.

~~

The rest of the reasoning is the same as in the periodic case. We first consider the case where $k=1$. This gives us quite simply:
$t^{[1]}(x)=(1+x)^L(1+1/x)^L=h(x)$. From this, we get:
\begin{equation}
\tilde{t}^{[1]}(x)=h(x)h_b(x)h_b(x^{-1})=\frac{(1+x)^L(1+1/x)^L(x^2,x^{-2})_{\infty}}{(ax,bx,a/x,b/x)_\infty}=F(x).\nonumber
\end{equation}
This explains how the function $F(x)$ replaces $h(x)$ (from the periodic case) in all the expressions for the cumulants of the current that were found in \cite{gorissen2012exact}.

For $k=2$, through the same calculations as in the periodic case, we find the T-Q equation
\begin{equation}
t^{[2]}(x)Q(1/x)=h(x)Q(1/qx)+{\rm e}^{-2k\mu}\frac{h_b(qx)h_b(q/x)}{h_b(x)h_b(1/x)}h(qx)Q(q/x)\nonumber
\end{equation}
which translates into
\begin{equation}\label{V-2-t2Q}
\tilde{t}^{[2]}(x)\tilde{Q}(1/x)=F(x)\tilde{Q}(1/qx)+{\rm e}^{-2k\mu}F(qx)\tilde{Q}(q/x).
\end{equation}
Note that although the equation for $\tilde{t}^{[2]}$ is more compact that that for $t^{[2]}$, the former is not polynomial due to the presence of q-Pochhammer symbols in $F$ and in the normalisation of $\tilde{t}^{[2]}$, whereas the latter is.

~~

We now have to verify that $\tilde{t}^{[2]}(x)$ is related to $M_\mu$ through some derivative. We will do this directly in the general case (with all four boundary rates), in a few pages. For now, we just give the two-dimensional boundary matrices that we find from $K^+$ and $K^-$:
\begin{equation}
K^+_2=\begin{bmatrix} 1 & \frac{qx(qx-b)}{(q^2x^2-1)} \\  \frac{x(1-qxb)}{(q^2x^2-1)} & 0\end{bmatrix}~~~~,~~~~K^-_2=\begin{bmatrix} 1 & \frac{(a-x)}{(1-x^2)} \\  \frac{(ax-1)}{q(1-x^2)} & 0\end{bmatrix}.\nonumber
\end{equation}

\subsubsection{Two-way boundaries}

In the general case, where both boundaries have two non-zero rates, the calculations are much more involved, and can be found in appendix \ref{App-B-2}. Starting from expression (\ref{Kplus}) for $K^+(x,y)$, and taking $xy=q^{1-k}$, we find that
\begin{equation}\boxed{
\frac{K^+_{i+k,j+k}(x,y)}{K^+_{i,j}(q^kx,q^ky)}=\frac{(x/b)_kb^k(xb)_k(y/\tilde{b})_k(\tilde{b})^k(y\tilde{b})_k}{(y^2 )_{2k}}(-y)^kq^{k(k-1)/2}\frac{(q^{j+1})_k}{(q^{i+1})_k}
}
\end{equation}
and the corresponding result for the left boundary:
\begin{equation}\boxed{
\frac{K^-_{j+p,i+p}(x,y)}{K^-_{j,i}(q^px,q^py)}=\frac{(y/a)_pa^p(ya)_p(x/\tilde{a})_p(\tilde{a})^p(x\tilde{a})_p}{(x^2 )_{2p}}(-x)^pq^{p(p-1)/2}\frac{(q^{i+1})_p}{(q^{j+1})_p}.
}\end{equation}

Using this, we do the exact same operations as in the previous case, replacing $h_b$ with
\begin{equation}\label{V-2-hb2}
h_b(x)=\frac{(x^2)_{\infty}}{(ax,\tilde{a}x,bx,\tilde{b}x)_\infty},
\end{equation}
and we get the completely general version of the function $F(x)=\tilde{t}^{[1]}(x)$:
\begin{equation}\boxed{\boxed{
F(x)=\frac{(1+x)^L(1+1/x)^L(x^2,x^{-2})_{\infty}}{(ax,\tilde{a}x,bx,\tilde{b}x,a/x,\tilde{a}/x,b/x,\tilde{b}/x)_\infty}.
}}\end{equation}

~~

We finally look at the transfer matrix $\tilde{t}^{[2]}(x)$ and try to relate it to $M_\mu$, as we did in section \ref{V-1-2} for the periodic case. By analogy with eq.(\ref{V-1-t2t}), we will try to rewrite $\tilde{t}^{[2]}(h)/F(x)$ as the standard two-dimensional transfer matrix.

The two-dimensional blocks from $K^+$ and $K^-$ for $y=1/qx$ are:
\begin{equation}
K^+_2=\begin{bmatrix} 1 & \frac{qx(qx+qxb\tilde{b}-b-\tilde{b})}{(q^2x^2-1)} \\  \frac{x(1+b\tilde{b}-qxb-qx\tilde{b})}{(q^2x^2-1)} & -b\tilde{b}x\end{bmatrix}~~~~,~~~~K^-_2=\begin{bmatrix} 1 & \frac{(a+\tilde{a}-x-a\tilde{a}x)}{(1-x^2)} \\  \frac{(ax+\tilde{a}x-1-a\tilde{a})}{q(1-x^2)} & -\frac{a\tilde{a}}{q x}\end{bmatrix}\nonumber
\end{equation}
and the corresponding blocks from the bulk are:
\begin{align}
X_2(x)&=\left[\begin{array}{c c|c c} 1+x & 0 & 0 & 0 \\ 0 & 1+q x & 1-\frac{1}{q} & 0 \\ \hline 0 & 1-q & 1+\frac{1}{q x} & 0\\ 0 & 0 & 0 &  1+\frac{1}{x}\end{array}\right],\nonumber\\
\overline{X}_2(x)&=\left[\begin{array}{c c|c c}  1+\frac{1}{q x}& 0 & 0 & 0 \\ 0 &  1+\frac{1}{x}& 1-q & 0 \\ \hline 0 & 1-\frac{1}{q} & 1+x & 0\\ 0 & 0 & 0 & 1+q x \end{array}\right].\nonumber
\end{align}

Consider these transformations, with $\lambda = \frac{1+q x}{q(1+x)}$, i.e. $x=-\frac{1-q \lambda}{q(1-\lambda)}$ :
\begin{equation}
L^{(i)}(\lambda)=\frac{1}{1+x}  \Biggl(\!\begin{bmatrix} 1 &  0\\ 0 & -q \end{bmatrix}\! X_2(x)\!\begin{bmatrix} 1 &  0\\ 0 & -1/q \end{bmatrix}\!\Biggr) \!\centerdot\!\left[\begin{array}{ c | c} 1 & 0   \\ \hline 0 & x  \end{array}\right] \!=\! \left[\begin{array}{c c|c c} 1 & 0 & 0 & 0 \\ 0 & q \lambda &1-q\lambda & 0 \\ \hline 0 &1- \lambda & \lambda & 0\\ 0 & 0 & 0 &  1 \end{array}\right]\nonumber
\end{equation}
and
\begin{equation}
\overline{L}^{(i)}(\lambda)=\frac{1}{1+x}  \left[\begin{array}{ c | c} x & 0   \\ \hline 0 & 1  \end{array}\right]\!  \centerdot\!\Biggl(\!\begin{bmatrix} 0 &  1\\1 & 0 \end{bmatrix} \!\overline{X}_2(x)\!\begin{bmatrix}0 & 1\\ 1 & 0 \end{bmatrix}\!\Biggr) \! = \!\left[\begin{array}{c c|c c}1 & 0 & 0 & 1-q\lambda \\ 0 &  \lambda & 0& 0 \\ \hline 0 & 0& q \lambda & 0\\ 1- \lambda & 0 & 0 & 1 \end{array}\right]\nonumber
\end{equation}
where the matrix products inside the parentheses are done in the auxiliary space, on each element of $X_2$ or $\overline{X}_2$, and the third product is done on the physical space at each site. Notice that the inner products cancel out between one site and the next, and that the outer products are done between $X_2$ and $\overline{X}_2$ and amount to a global factor $\frac{x}{(1+x)^2}$ on each site. Taking a product of $L$ matrices $X_2\centerdot\overline{X}_2$, this transformation gives, apart from the inner products at each end of the chain, a global factor $\frac{x^L}{(1+x)^{2L}}$, which accounts for the bulk part $h(x)$ of $F(x)$. Also note that the matrices from $\overline{L}^{(i)}$ need to be transposed if multiplied from right to left.

Considering that $\tilde{t}^{[2]}$ has a factor $h_b(x)h_b(1/qx)$, and that
\begin{equation}
h(x)h_b(x)h_b(1/qx)=F(x) \frac{(1-1/q^2x^2)(1-1/qx^2)}{(1-a/qx)(1-\tilde{a}/qx)(1-b/qx)(1-\tilde{b}/qx)}\nonumber
\end{equation}
the transformations we need to do on the boundary matrices are:
\begin{align}
\hat{K}^+_2(\lambda)&= \frac{(1-1/q^2x^2)}{(1-b/qx)(1-\tilde{b}/qx)}\begin{bmatrix} 1 &  0\\ 0 & -q \end{bmatrix} K^+_2 \begin{bmatrix} 0 &  1\\1 & 0 \end{bmatrix},\nonumber\\
\hat{K}^-_2(\lambda)&=\frac{(1-1/qx^2)}{(1-a/qx)(1-\tilde{a}/qx)}\begin{bmatrix} 0 &  1\\1 & 0 \end{bmatrix}K^-_2\begin{bmatrix} 1 &  0\\ 0 & -1/q \end{bmatrix},\nonumber
\end{align}
of which we will not write the full expression in terms of $\lambda$. Their values and first derivatives at $\lambda=0$, which is all we will need, are, in terms of the original boundary parameters:
\begin{align}
\hat{K}^+_2=\begin{bmatrix} 1 &  0\\ 0 & 1 \end{bmatrix}~~&,~~\frac{d}{d\lambda}\hat{K}^+_2=\begin{bmatrix} -2\delta & 2\beta \\  2\delta &1-q-2\beta \end{bmatrix},\label{V-2-K+l}\\
\hat{K}^-_2=\begin{bmatrix} \frac{1-\alpha+\gamma}{1+q} &\frac{\gamma}{q} \\ \alpha &  \frac{\alpha-\gamma+q}{1+q}\end{bmatrix}~~&,~~\frac{d}{d\lambda}\hat{K}^-_2=\begin{bmatrix} -\alpha+\gamma+A- B  &\frac{\gamma(2q-\alpha-\gamma)}{q}\\\alpha(1+q-\alpha-\gamma) & -2\gamma-q-A+ B \end{bmatrix},
\end{align}
with $A=\frac{2+(\alpha-2)\alpha-\gamma^2}{1+q}$ and $B=\frac{2(1-\alpha+\gamma)}{(1+q)^2}$.

~~

\begin{figure}[ht]
\begin{center}
\includegraphics[width=0.7\textwidth]{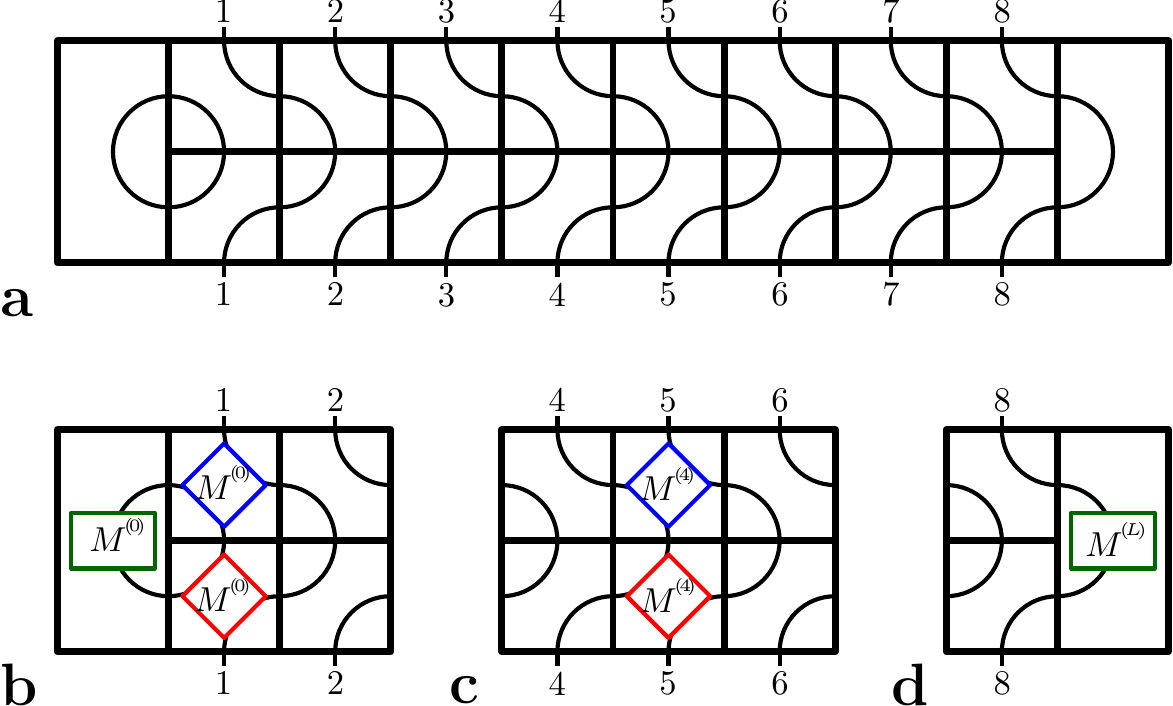}
\caption{Schematic representation of the value and first logarithmic derivative of $\hat{t}^{[2]}$ at $0$. The first is the identity matrix (a). The second is a sum of terms adding up to $M_\mu$: three terms at the left boundary (b), two for each bond in the bulk (c) and one term at the right boundary (d).}
\label{fig-t2open}
 \end{center}
 \end{figure}

We now put all these matrices together, obtaining $\tilde{t}^{[2]}(h)/F(x)$ as desired. The Lax matrices $L^{(i)}$ are not exactly the same as those we had for the periodic case. Their values at $\lambda=0$ are $L^{(i)}(0)=P^{(i)}$ and $\overline{L}^{(i)}(0)=\overline{P}^{(i)}$, where $P^{(i)}$ is the permutation matrix exchanging the auxiliary space with the physical space when applied to the right in the matrix product, and $\overline{P}v$ is the same but when applied to the left. We see that the two boundaries do not play the same role. The right boundary matrix is simply the identity at $\lambda=0$, and serves to connect the Lax matrices on the last site. The left boundary matrix is traced, at $\lambda=0$, because of the Lax matrices on the first site, and we see that its trace is $1$. All in all, at $\lambda=0$, the whole transfer matrix is simply the identity (see fig.-\ref{fig-t2open}-a).

As for its first derivative with respect to $\lambda$, we find that each pair of Lax matrices $L^{(i)}$ and $\overline{L}^{(i)}$ have a total contribution of $2M^{(i)}+(1-q)(\tau_i-\tau_{i+1})$, where $\tau_i$ is the number of particles on site $i$. The right boundary matrix, as can be seen in (\ref{V-2-K+l}), gives a term $2m^{(L)}+(1-q)\tau_L$. At the left boundary, we have three terms to consider, one involving the derivative of $K^-$ and the first Lax matrices taken at $0$, and two more where we take the derivative of the first Lax matrices, and the boundary matrix at $0$. The sum of those terms gives a total contribution of $2m^{(0)}-(1-q)\tau_1$. If we now sum all the terms that we have found, the parts proportional to $(1-q)$ all cancel out, and we are simply left with $2M_\mu$.

At the end of the day, we find that:
\begin{equation}\boxed{\boxed{
M_\mu=\frac{1}{2}\Bigl(1-\frac{1}{q}\Bigr)\frac{d}{d x} \log\biggl( \frac{\tilde{t}^{[2]}(x)}{F(x)}\biggr)\bigg|_{x=-1/q}
}} \label{MmuT2open}\end{equation}
which is the same as eq.(\ref{V-1-ET}), with $h(x)$ replaced by $F(x)$, and an extra factor $\frac{1}{2}$. We haven't mentioned the matrices $A_\mu$ in those last calculations, because their behaviour is trivial, and it is left to the reader, as an exercise, to check that adding one between two sites gives the correct deformation for $M_\mu$.

As for the periodic case, we could have done all those calculations around $\lambda=\infty$ instead of $0$, which would have given an equivalent result:
\begin{equation}
M_\mu=\frac{1}{2}(1-q)\frac{d}{d x} \log\biggl(\frac{\tilde{t}^{[2]}(x)}{F(qx)}\biggr)\bigg|_{x=-1}.
\end{equation}

~~

In order to obtain the results found in \cite{gorissen2012exact}, we need one last piece of information: how $P$ and $Q$ behave as $\mu$ goes to 0.

\subsection{Non-deformed case: matrix Ansatz}
\label{II-4}

As in the periodic case, the final step is to consider the $\mu\rightarrow 0$ limit in our transfer matrices, in order to get some information specific to the dominant eigenspace of $M_\mu$ and the corresponding eigenvalue $E(\mu)$. Once more, this is made much more difficult by the presence of the boundaries, but the principle remains the same.

We first need to consider the behaviour of $T_\mu(y)$ alone. As we recall, it is defined by:
\begin{equation}
T_\mu(y)=\langle\!\langle  \tilde{W}|\!|  A_\mu \prod_{i=1}^{L}X^{(i)}|\!|  \tilde{V} \rangle\!\rangle\nonumber
\end{equation}
where
\begin{equation}
X=\begin{bmatrix} 1+yA & S(1-y^2A)\\ S^{-1}(1-A)& 1+yA\end{bmatrix}\nonumber
\end{equation}
and
\begin{equation}
\langle\!\langle \tilde{W}|\!|=\langle\!\langle0|\!|\frac{(ay/S)_\infty(\tilde{a}y/S)_\infty}{(1/S)_\infty(a\tilde{a}/S)_\infty}\frac{(y^2A)_\infty}{(qA)_\infty}~~~~,~~~~~|\!|  \tilde{V} \rangle\!\rangle=\frac{(byS)_\infty(\tilde{b}yS)_\infty}{(S)_\infty(b\tilde{b}S)_\infty} |\!|0\rangle\!\rangle.\nonumber
\end{equation}

As in the periodic case, we can expand each entry in $T_\mu$ as a finite sum of terms of the form $y^iq^k\langle\!\langle  \tilde{W}|\!|  A_\mu S^l A^m |\!|  \tilde{V} \rangle\!\rangle$ (but this time, there is no constraint on $l$, as the number of particles is no longer conserved). In each entry, there is exactly one term with $m=0$, equal to $\langle\!\langle  \tilde{W}|\!|  A_\mu S^l |\!|  \tilde{V} \rangle\!\rangle$, where $l$ is the difference between the number of particles in the initial and final configurations. We show, in appendix \ref{App-D}, that for $\mu\rightarrow 0$, this term behaves as $1/(1-{\rm e}^{-\mu})$ times a prefactor which does not depend on $l$, and all the others remain finite, so that:
\begin{equation}
T_\mu(y)\sim \frac{1}{1-{\rm e}^{-\mu}}\frac{(ay,\tilde{a}y,by,\tilde{b}y)_\infty}{(y^2)_\infty(a\tilde{a},b\tilde{b},q)_\infty}|1\rangle\langle 1|.\nonumber
\end{equation}
Notice that the prefactor is equal to $1/h_b(y)$ as defined in (\ref{V-2-hb2}), up to a constant.

~

We can do the same calculations for $U_\mu$, and see that there is no divergence there, but it is not necessary. Instead, we just apply $U_\mu(x)$ on this last result. As we recall, $U_\mu(x)$ is defined by
\begin{equation}
U_\mu(x)= \langle\!\langle W|\!| A_\mu \prod_{i=1}^{L}X^{(i)} |\!| V \rangle\!\rangle\nonumber
\end{equation}
with
\begin{equation}
X=\begin{bmatrix} n_x & e\\ d& n_x\end{bmatrix}\nonumber
\end{equation}
and where $\langle\!\langle W|\!|$ and $ |\!| V \rangle\!\rangle$ verify:
\begin{align}
[\beta (d+n_x) - \delta (e+n_x) -(1-q)] ~|\!| V \rangle\!\rangle &= 0 ,\nonumber\\
\langle\!\langle W|\!|~ [\alpha(e+n_x)- \gamma (d+n_x)-(1-q)] &= 0.\nonumber
\end{align}
Since the vector $|1\rangle$ can be written as the tensor product of the constant vector $|1_i\rangle$ on each site $i$, applying each $X$ to $|1\rangle$ gives the vector
\begin{equation}
X|1_i\rangle=\begin{bmatrix} e+n_x \\d+n_x \end{bmatrix}~~{\rm which~we~will~note}~~\begin{bmatrix} E \\D \end{bmatrix}.\nonumber
\end{equation}
Since, as we recall, $d$, $e$, and $A$ satisfy
\begin{align}
de -q~ed&=(1-q)(1-x^2 A^2),\nonumber\\ 
A e&=q~e A , \nonumber\\
d A&=q~A d,\nonumber
\end{align}
we find that $D$ and $E$ satisfy
\begin{equation}\boxed{
DE-q~ED=(1-q)(D+E)
}\end{equation}
and the conditions on the boundary vectors simply write
\begin{empheq}[box=\fbox]{align}
[\beta D - \delta E -(1-q)] ~|\!| V \rangle\!\rangle &= 0 ,\\
\langle\!\langle W|\!|~ [\alpha E- \gamma D-(1-q)] &= 0.
\end{empheq}
The matrix $U_0(x)T_0(y)$ becomes, up to a prefactor:
\begin{equation}
U_0(x)T_0(y)\sim|P^\star\rangle\langle 1|\nonumber
\end{equation}
where, for any configuration ${\cal C}=\{\tau_i\}$,
\begin{equation}\boxed{
P^\star({\cal C})=\langle\!\langle W|\!| \prod_{i=1}^{L}[(1-\tau_i)E+\tau_i D] |\!| V \rangle\!\rangle
}\end{equation}
(where we used the fact that $A_\mu=1$ for $\mu=0$), which does not depend on $x$.

Moreover, since we know that $[M, U_0(x)T_0(y)]=0$ (by considering eq.(\ref{V-2-MUT}) at $\mu=0$), and that $\langle 1|M=0$ (because $M$ is a stochastic matrix), we find that
\begin{equation}\boxed{
M|P^\star\rangle=0
}\end{equation}
which is to say that $|P^\star\rangle$ is the steady state of the open ASEP. We therefore recover the original `matrix Ansatz', as presented in \cite{derrida1993exact}. Note that a matrix Ansatz also exists for the steady state of the periodic multi-species ASEP \cite{prolhac2009matrix}, which one should be able to recover from an appropriate transfer matrix with spectral parameters and current-counting deformations, and which would be related to the U$_q$[SU(k)] algebra (for $k$ different types of particles, counting holes as one of those types).

~

The normalisation of $U_0(x)T_0(y)$ still needs to be determined. Since we know that the only dependence in $y$ is a  prefactor $1/h_b(y)$, and that $U_\mu(x)T_\mu(y)=U_\mu(y)T_\mu(x)$ (as shown in eq.(\ref{UxTyUyTx})), we conclude that, up to a constant term which we include in $|P^\star\rangle$:
\begin{equation}
U_0(x)T_0(y)=\frac{1}{1-{\rm e}^{-\mu}} \frac{1}{h_b(x)h_b(y)}|P^\star\rangle\langle 1|\nonumber
\end{equation}
which is to say that, similarly to the periodic case, the prefactor in $\tilde{Q}(y)$ (as defined in (\ref{PQtilde})) diverges as $1/\mu$, and that $h_b(x)h_b(y)U_0(x)T_0(y)$ is independent of $x$ and $y$, so that the roots and poles of $\tilde{P}(x)$ and $\tilde{Q}(y)$ all go to infinity when $\mu$ goes to $0$.

Since, in that limit, $\tilde{Q}(y)$ is a constant, we also find, from eq.(\ref{V-2-t2Q}), that the eigenvalue of $\tilde{t}^{[2]}$ in that eigenspace is
\begin{equation}
\langle 1|\tilde{t}^{[2]}|P^\star\rangle=F(x)+F(qx).\nonumber
\end{equation}
It is straightforward to check that eq.(\ref{MmuT2open}) then gives $E(0)=0$, which we knew from the fact that $M_0$ is a stochastic matrix.

~

Starting from $\mu=0$, where, as we just saw, we know $\tilde{Q}$ explicitly, we can expand everything in series in $\mu$, and find an explicit expression for $E(\mu)$ perturbatively in $\mu$. This is what we do in the next section, where we put everything together and give a summary of the whole procedure.

\subsection{Summary - Functional Bethe Ansatz for the open ASEP}
\label{V-2-4}

In this section, we collect all the results we found for the open ASEP, and show how they lead to the expressions for the cumulants of the current obtained in \cite{gorissen2012exact}.

~~

The first step is to construct two transfer matrices $U_\mu(x)$ and $T_\mu(y)$:
\begin{align}
U_\mu(x)&=h_b(x) \langle\!\langle W|\!| A_\mu \prod_{i=1}^{L}X^{(i)}(x,x) |\!| V \rangle\!\rangle,\\
T_\mu(y)&=h_b(y)\langle\!\langle  \tilde{W}|\!|  A_\mu \prod_{i=1}^{L}X^{(i)}(y,y)|\!|  \tilde{V} \rangle\!\rangle
\end{align}
(involving the function $h_b$ defined in eq.(\ref{V-2-hb2})), such that, for any $x$ and $y$, we have:
\begin{equation}\label{V-2-MUTs}
[M_\mu, U_\mu(x)T_\mu(y)]=0.
\end{equation}
Using these, we can construct two commuting matrices $P(x)$ and $Q(y)$ as:
\begin{equation}
P(x)=U_\mu(x)\Bigl[U_\mu(0)\Bigr]^{-1}~~~~,~~~~Q(y)=U_\mu(0)T_\mu(y)
\end{equation}
such that:
\begin{equation}
U_\mu(x)T_\mu(y)=P(x)Q(y).
\end{equation}
We can show that those matrices verify, for any positive integer $k$:
\begin{equation}\label{V-2-PQs}
P(x)Q(1/q^{k-1}x)=t^{[k]}(x)+{\rm e}^{-2k\mu}P(q^k x)Q(q/x)
\end{equation}
(where we omit the tildes, since we have correctly normalised $P$ and $Q$ from the start). The matrix $t^{[k]}(x)$ is the one-parameter transfer matrix with a $k$-dimensional auxiliary space.

The first two of those relations write:
\begin{align}
P(x)Q(1/x)&=F(x)+{\rm e}^{-2\mu}P(q x)Q(q/x)\label{V-2-PQFs},\\
P(x)Q(1/qx)&=t^{[2]}(x)+{\rm e}^{-4\mu}P(q^2 x)Q(q/x)\label{V-2-PQt2s},
\end{align}
where $F(x)$ is a scalar function given by
\begin{equation}
F(x)=\frac{(1+x)^L(1+1/x)^L(x^2,x^{-2})_{\infty}}{(ax,\tilde{a}x,bx,\tilde{b}x,a/x,\tilde{a}/x,b/x,\tilde{b}/x)_\infty}
\end{equation}
and $t^{[2]}$ is such that:
\begin{equation}
M_\mu=\frac{1}{2}(1-q)\frac{d}{d x} \log\biggl(\frac{t^{[2]}(x)}{F(qx)}\biggr)\bigg|_{x=-1}.
\end{equation}
Using eq.(\ref{V-2-PQFs}) at $x$ and $qx$, and eq.(\ref{V-2-PQt2s}), we can find the T-Q equation:
\begin{equation}\label{V-2-TQs}
t^{[2]}(x)Q(1/x)=F(x)Q(1/qx)+{\rm e}^{-2\mu}F(qx)Q(q/x)
\end{equation}
which allows us to express $M_\mu$ in terms of $Q$ instead:
\begin{equation}\label{V-2-Mmut2}
M_\mu=\frac{1}{2}(1-q)\frac{d}{d x} \log\biggl(\frac{Q(q/x)}{Q(1/x)}\biggr)\bigg|_{x=-1}.
\end{equation}

We now consider:
\begin{equation}
B=-{\rm e}^{2\mu}\bigl(Q(0)\bigr)^{-1}=-{\rm e}^{2\mu}\bigl(U_\mu(0)T_\mu(0)\bigr)^{-1}.
\end{equation}
As we saw in section \ref{II-4}, the first eigenvalue of $B$ goes to $0$ for $\mu\rightarrow 0$, which is not {\it a priori} the case for the others. Moreover, all the roots and poles of $P(x)$ and $Q(y)$ go to infinity in that eigenspace.

From here on, we restrict ourselves to that specific eigenspace, so that $P$, $Q$ and $B$ refer to functions rather than matrices.

~

The next step is to rewrite eq.(\ref{V-2-PQFs}) in a different way, and see that all the information we have about $P$ and $Q$ makes it solvable. This was done in \cite{prolhac2010tree} for the periodic case, and we reproduce it here, under a slightly different form, for the open case.  Let us therefore define a function $W$ as:
\begin{equation}\label{IV-2-W}
W(x)=-\frac{1}{2}\log\biggl(\frac{P(x)Q(1/x)}{{\rm e}^{-2\mu}P(q x)Q(q/x)}\biggr),
\end{equation}
and a convolution kernel $K$, as:
\begin{equation}\label{IV-2-K}
K(x,\tilde{x})=2\sum_{k=1}^{\infty}\frac{q^k}{1-q^k}\Bigl((x/\tilde{x})^k+(x/\tilde{x})^{-k}\Bigr)
\end{equation}
along with the associated convolution operator $X$:
\begin{equation}\label{IV-2-X}
X[f](x)=\oint_{c_1}\frac{d\tilde{x}}{\imath2\pi\tilde{x}}f(\tilde{x})K(x,\tilde{x}).
\end{equation}

Using those, as we show in appendix \ref{App-E}, one can rewrite eq.(\ref{V-2-PQFs}) in terms of only one unknown function $W$:
\begin{equation}\label{IV-2-WW}\boxed{\boxed{
W(x)=-\frac{1}{2}\ln\Bigl(1-B F(x) e^{X[W](x)}\Bigr)
}}\end{equation}
which is the same as eq.(59) in \cite{prolhac2010tree}.

The last step is to take eq.(\ref{V-2-Mmut2}), and eq.(\ref{IV-2-W}) at $x=0$, to find:
\begin{equation}
E(\mu)=\frac{1}{2}(1-q)\frac{d}{dx}\log\biggl(\frac{Q(q/x)}{Q(1/x)}\biggr)\biggl|_{x=-1}~~~~,~~~~\mu=-W(0)
\end{equation}

Considering what we said before about the roots and poles of $P(x)$ being outside of the unit circle, and those of $Q(1/x)$ being inside, we can replace $\frac{1}{2}\log\bigl(\frac{Q(q/x)}{Q(1/x)}\bigr)$ by $-W(x)$ when expressing $E(\mu)$ as a contour integral over the unit circle (since $P$ will not contribute), and obtain:
\begin{equation}\label{IV-2-muB}\boxed{
\mu=-\oint_{c_1}\frac{dz}{\imath2\pi z}W(z)
}\end{equation}
and
\begin{equation}\label{IV-2-EB}\boxed{
E(\mu)=-(1-q)\oint_{c_1}\frac{dz}{\imath2\pi(1+z)^2}W(z),
}\end{equation}
in which we recognise (13) and (14) from \cite{gorissen2012exact}. All this is done for $a<1$ and $b<1$, but can then be generalised to any $a$ and $b$ through the same reasoning as in \cite{1751-8121-40-46-R01} for the mean current, replacing the unit circle $c_1$ by small contours around $\Gamma=\{0,q^k a,q^k \tilde{a},q^k b,q^k \tilde{b}\}$.

~

In this last result, $E(\mu)$ is expressed as an implicit function of $\mu$, through the variable $B$. In order to get an explicit expression, one has to invert (\ref{IV-2-muB}) to get $B$ in terms of $\mu$, and inject the result into (\ref{IV-2-EB}). Since $B$ is of order $\mu$ around $\mu=0$, this can be done perturbatively in $B$ and $\mu$, to yield the coefficients of $E(\mu)$ expanded in powers of $\mu$. Those are, up to a factorial, the cumulants of the current in our system. In the general case, they can be expressed as combinations of multiple contour integrals around powers of $F$ convolved through $K$ \cite{gorissen2012exact}. In the simpler case of the TASEP, where $q=0$, that convolution kernel vanishes, and (\ref{IV-2-muB}) and (\ref{IV-2-EB}) simplify to
\begin{equation}
\mu=-\frac{1}{2}\sum\limits_{k=1}^{\infty}\oint_{\Gamma}\frac{dz}{\imath2\pi z}F(z)^k\frac{B^k}{k}\nonumber
\end{equation}
and
\begin{equation}
E(\mu)=-\frac{(1-q)}{2}\sum\limits_{k=1}^{\infty}\oint_{\Gamma}\frac{dz}{\imath2\pi(1+z)^2}F(z)^k\frac{B^k}{k},\nonumber
\end{equation}
where $F(z)$ is also much simpler, and has at most $9$ poles. This makes calculating the cumulants of the current much easier, especially in the case where $a=b=0$ \cite{Lazarescu2011}.

\subsection{XXZ spin chain with general boundary conditions}
\label{V-2-5}

In this section, we explain how our construction for the open ASEP can be translated for the spin-$\frac{1}{2}$ XXZ chain with non-diagonal boundary conditions \cite{sandow1994partially}.

Let us first define the bulk Hamiltonian of the XXZ spin chain of length $L$:
\begin{equation}\label{HXXZ}
H_{b}=\frac{1}{2}\sum\limits_{k=1}^{L-1}h^{(i)}\nonumber
\end{equation}
with $h^{(i)}$ acting as:
\begin{equation}\label{HXXZ2}
h^{(i)}=\begin{bmatrix} \Delta & 0 & 0 & 0 \\ 0 & -\Delta & 1 & 0 \\ 0 & 1& -\Delta & 0 \\ 0 & 0 & 0 & \Delta \end{bmatrix}\nonumber
\end{equation}
on sites $i$ and $i+1$ (in basis $\{00,01,10,11\}$, as usual), and as the identity on the rest of the chain. We define $\Delta$ as $\frac{1}{2}(q^{-1/2}-q^{1/2})$, which is not the usual definition for the XXZ chain (that can be obtained simply by replacing $q$ by $q^2$).

~~

Let us also write the deformed Markov matrix $M_{\{\!\mu_i\!\}}$ for the special choice of weights defined by:
\begin{equation}
\{\mu_0=\frac{1}{2}\log{\biggl(\frac{\gamma}{\alpha}\biggr)+\nu_0},~~\mu_i=\frac{1}{2}\log{(q)},~~\mu_L=\frac{1}{2}\log{\biggl(\frac{\delta}{\beta}\biggr)+\nu_L} \}\nonumber
\end{equation}
which is on the line $\mu=\frac{1}{2} \log{\bigl(\frac{\gamma \delta}{\alpha\beta}q^{L-1}\bigr)}+\imath\mathbb{R}$ if $\nu_0$ and $\nu_L$ are imaginary numbers (in which case $M_{\{\!\mu_i\!\}}$ is Hermitian). The deformed local matrices become:
\begin{align}
m^{(0)}(\mu_0)&=\begin{bmatrix} -\alpha & \sqrt{\alpha\gamma}~{\rm e}^{-\nu_0} \\ \sqrt{\alpha\gamma}~{\rm e}^{\nu_0} & -\gamma  \end{bmatrix},\nonumber\\
M^{(i)}(\mu_i)&=~~~\begin{bmatrix} 0 & 0 & 0 & 0 \\ 0 & -q &\sqrt{q} & 0 \\ 0 & \sqrt{q}& -1 & 0 \\ 0 & 0 & 0 & 0 \end{bmatrix},\nonumber\\
m^{(L)}(\mu_l)&=\begin{bmatrix} -\delta & \sqrt{\beta\delta}{\rm e}^{\nu_L} \\   \sqrt{\beta\delta}{\rm e}^{-\nu_L} & -\beta  \end{bmatrix}.\nonumber
\end{align}

It is straightforward to check that in this case, we have $M_{\{\!\mu_i\!\}}=\sqrt{q}H+\epsilon$, where $\epsilon$ is a constant, with the boundary matrices being equal to:
\begin{align}\label{MequalH}
h^{(0)}&=\frac{1}{2\sqrt{q}}\begin{bmatrix}(1-q-\alpha+\gamma) & 2\sqrt{\alpha\gamma}~{\rm e}^{-\nu_0} \\ 2\sqrt{\alpha\gamma}~{\rm e}^{\nu_0} & (-1+q+\alpha-\gamma) \end{bmatrix}\nonumber\\
&=\frac{q^{-1/2}-q^{1/2}}{2}\frac{1}{(1+a)(1+\tilde{a})}\begin{bmatrix}a+\tilde{a}& 2\sqrt{-a\tilde{a}}{\rm e}^{\nu_0} \\ 2\sqrt{-a\tilde{a}}{\rm e}^{-\nu_0} &-a-\tilde{a} \end{bmatrix},\nonumber\\
h^{(L)}&=\frac{1}{2\sqrt{q}}\begin{bmatrix}(-1+q+\beta-\delta) & 2\sqrt{\beta\delta}{\rm e}^{\nu_L} \\ 2\sqrt{\beta\delta}{\rm e}^{-\nu_L} & (1-q-\beta+\delta)  \end{bmatrix}\nonumber\\
&=\frac{q^{-1/2}-q^{1/2}}{2}\frac{1}{(1+b)(1+\tilde{b})}\begin{bmatrix}-b-\tilde{b}& 2\sqrt{-b\tilde{b}}{\rm e}^{-\nu_L} \\ 2\sqrt{-b\tilde{b}}{\rm e}^{\nu_L} &b+\tilde{b} \end{bmatrix}.\nonumber
\end{align}

Since we have three nontrivial parameters in each of those matrices, they are completely general: we can write (without restricting ourselves to hermitian matrices)
\begin{align}
h^{(0)}=a_z\sigma_z+a_+\sigma^++a_-\sigma^-&=\begin{bmatrix}a_z& a_- \\ a_+ &-a_z \end{bmatrix},\nonumber\\
h^{(L)}=b_z\sigma_z+b_+\sigma^++b_-\sigma^-&=\begin{bmatrix}b_z& b_- \\ b_+ &-b_z \end{bmatrix},\nonumber
\end{align}
with
\begin{align}
a_z=(1-q-\alpha+\gamma)/2\sqrt{q}~~,~~&a_+=\sqrt{\alpha\gamma/q}~{\rm e}^{\nu_0}~~,~~a_-=\sqrt{\alpha\gamma/q}~{\rm e}^{-\nu_0},\nonumber\\
b_z=(-1+q+\beta-\delta)/2\sqrt{q}~~,~~&b_+=\sqrt{\beta\delta/q}~{\rm e}^{-\nu_L}~~,~~b_-=\sqrt{\beta\delta/q}~{\rm e}^{\nu_L},\nonumber
\end{align}
which is to say
\begin{align}
\nu_0&=-2\log(a_+/a_-),\nonumber\\
\alpha&=\sqrt{(\sqrt{q}a_z-(1-q)/2)^2+a_+a_-}-\sqrt{q}a_z+(1-q)/2,\nonumber\\
\gamma&=\sqrt{(\sqrt{q}a_z-(1-q)/2)^2+a_+a_-}+\sqrt{q}a_z-(1-q)/2,\nonumber
\end{align}
and
\begin{align}
\nu_L&=-2\log(b_-/b_+),\nonumber\\
\beta&=\sqrt{(\sqrt{q}b_z+(1-q)/2)^2+b_+b_-}+\sqrt{q}b_z+(1-q)/2,\nonumber\\
\delta&=\sqrt{(\sqrt{q}a_z+(1-q)/2)^2+b_+b_-}-\sqrt{q}b_z-(1-q)/2.\nonumber
\end{align}
Those are well defined for any values of $a_z$, $a_+$, $a_-$, $b_z$, $b_+$ and $b_-$.
~~

Considering expression (\ref{V-1-Tmui}), or its equivalent for an open chain, and noting that $A_{\frac{\mu_i}{2}}=A^{-1/4}$, we can rewrite $U(x)$ and $T(y)$ in a way better suited to this situation:
 \begin{align}
U(x)&= \langle\!\langle \phi |\!| \prod_{i=1}^{L}Y^{(i)}(x) |\!| \psi \rangle\!\rangle,\nonumber\\
T(y)&= \langle\! \langle \tilde{\phi}|\!|  \prod_{i=1}^{L}Y^{(i)}(y) |\!|  \tilde{\psi} \rangle\!\rangle,\nonumber
\end{align}
with
\begin{equation}\label{D0DL}
Y(x)=\begin{bmatrix} N_x & \Sigma_+\\ \Sigma_- & N_x  \end{bmatrix}\nonumber
\end{equation}
where
 \begin{align}
N_x&=A^{-1/2}-xA^{1/2}\nonumber\\
\Sigma_+&=A^{-1/4}e A^{-1/4}=q^{-1/4}S^+(A^{-1/2}-x^2A^{1/2}),\nonumber\\
\Sigma_-&=A^{-1/4}d A^{-1/4}=q^{1/4}S^-(A^{-1/2}- A^{1/2}).\nonumber
\end{align}

The boundary vectors become:
\begin{align}
\langle\!\langle \phi |\!| &= \langle\!\langle W|\!| A_{\mu_0}A^{1/4},\nonumber\\
|\!| \psi \rangle\!\rangle&=A_{\mu_L}A^{1/4}|\!|  V\rangle\!\rangle,\nonumber\\
\langle\! \langle \tilde{\phi}|\!| &= \langle\!\langle  \tilde{W}|\!| A_{\mu_0}A^{1/4},\nonumber\\
|\!|  \tilde{\psi} \rangle\!\rangle&=A_{\mu_L}A^{1/4}|\!|  \tilde{V} \rangle\!\rangle.\nonumber
\end{align}

Matrices $N_x$, $\Sigma_+$ and $\Sigma_-$ satisfy the $U_q[SU(2)]$ algebra \cite{chaichian1996introduction}:
 \begin{align}
[\Sigma_+,\Sigma_-]&=(q^{-1/2}-q^{1/2})(A^{-1}-x^2A),\nonumber\\
\Sigma_- A&=q~A \Sigma_-,\nonumber\\
A \Sigma_+&=q~\Sigma_+ A,\nonumber
\end{align}
and the conditions on the boundary vectors become:
\begin{align}
\langle\!\langle \phi |\!|~&[a_+\Sigma_+-a_-\Sigma_--2a_zN_x-(q^{-1/2}-q^{1/2})xA^{1/2}]=0,\nonumber\\
&~[b_-\Sigma_--b_+\Sigma_++2b_zN_x-(q^{-1/2}-q^{1/2})xA^{1/2}]~|\!| \psi \rangle\!\rangle=0,\nonumber\\
\langle\! \langle \tilde{\phi}|\!| ~&[a_+\Sigma_+-a_-\Sigma_--2a_zN_y+(q^{-1/2}-q^{1/2})A^{-1/2}]=0,\nonumber\\
&~[b_-\Sigma_--b_+\Sigma_++2b_zN_y+(q^{-1/2}-q^{1/2})A^{-1/2}]~|\!| \tilde{\psi} \rangle\!\rangle =0.\nonumber
\end{align}

~~

We may note that the structure of this solution bears a strong resemblance to that of the Lindblad master equation found in \cite{Prosen2013b,karevski2013exact}. In that case, the algebraic relations satisfied by the boundary vectors are different from ours, as there is only one vector per boundary but two equations per vector, which may constrain the values of the boundary parameters. The connection between integrable spin chains and certain boundary-driven Lindblad models is also investigated in \cite{Prosen2013a,Ilievski2014}, and a matrix product structure of the density matrix was observed in \cite{Prosen2014} for an open Hubbard chain, but the precise conditions for such systems to be integrable have yet to be understood.

\section{Conclusion}

In the present work, we treat the case of the asymmetric simple exclusion process with generic open boundaries and a current-counting deformation. We construct a transfer matrix with an infinite-dimensional auxiliary space and two free parameters, which commutes with the deformed Markov matrix of our system. We then derive the two essential properties of that transfer matrix: it is a factor of two matrices $P$ and $Q$ which commute and carry one of the free variables each, and for certain values of the product of those parameters it breaks into two independent matrices, one of which is the usual one-parameter transfer matrix for a given dimension of the auxiliary space. From these results, we derive the T-Q equations for the transfer matrices, which allows us to identify our $Q$ matrix with Baxter's Q-operator \cite{Baxter1982}, although it is obtained through an entirely different method. Writing the algebraic relations between the $P$ and $Q$ matrices in a given eigenstate of our system then yields the functional Bethe Ansatz equations, which we can use, through the same method as in \cite{prolhac2010tree}, to find an explicit expression for the leading eigenvalue of our deformed Markov matrix, and thus give a rigorous proof of the results conjectured in \cite{gorissen2012exact}.

We have mainly focused on one specific eigenvalue, which is of particular importance as to the physics of the ASEP, and which we were able to obtain explicitly, for any system size and boundary parameters, and perturbatively in the deformation parameter $\mu$, using specific knowledge of the corresponding eigenvector for the non-perturbed case. It is however to be noted that only a small part of our derivation depends on that specific knowledge, and that, in principle, with similar information on other eigenstates (i.e. the positions of the roots and poles of $Q$ for $\mu=0$, which can in principle be obtained through the usual Bethe equation \cite{de2005bethe,1742-5468-2006-12-P12011}), we would be able to obtain similar results. The density of the spectrum was obtained, in the thermodynamic limit, in the context of the periodic TASEP in \cite{Prolhac2013,Prolhac2014}, using the coordinate Bethe Ansatz, and it would be interesting to know whether that method can be extended to the the open case or to the ASEP, and be written in terms of the functional Bethe Ansatz (for the open ASEP, only the first excited state without deformation was analysed, in \cite{DeGier2008}). Our construction can also be applied in more detail to the XXX and the XXZ spin chain, which will be the subject of a future paper.

It should be noted that, when solving the T-Q equation for $\mu=0$, $Q$ is usually taken to have the so-called `crossing symmetry' $x\leftrightarrow 1/qx$, which is a symmetry of the equation \cite{Gaudin1983,Sklyanin1988}. In our case, $Q(y)$ is explicitly constructed as a series in powers of $y$, and the fact that it is entire and without zeros inside of the unit circle (which is impossible with the aforementioned symmetry) plays an important part in finding the explicit form of $E(\mu)$. It would be interesting to analyse further the precise relation between our construction and the usual approach to the T-Q equation.

There remains the problem of finding an expression for the eigenvectors, possibly in a form similar to that which exists for the periodic case (as a generalised determinant, or equivalently as a product of creation operators on a vacuum state). This was done for special cases where the boundary parameters satisfy certain constraints \cite{Crampe2010,crampe2011matrix,de2005bethe,simon2009construction}, and also more recently in the general symmetric case (or XXX spin chain) \cite{Belliard2013}, but no complete solution for the ASEP has yet been found.

Let us also comment on the alternative method of constructing the functional Bethe equations devised recently \cite{Cao2013a,Cao2013,Kitanine2014,Nepomechie2013,Cao2013b} in the context of the XXX and XXZ spin chains, which consists in adding an extra inhomogeneous term to the T-Q equation,  in order to make polynomial solutions possible (it is easy to see, from the T-Q equation obtained here, that it is not in general the case without that extra term), and without modifying the relation between $Q$ and the eigenvalues of the Hamiltonian. We do not know at present exactly how that method compares with ours, and whether it solves the same aspects of the problem. As far as we can tell, both methods have their merits. The alternative construction has polynomial solutions and allows for a finite set of Bethe equations, which are easier to obtain explicitly, but only when that is possible (e.g. for small system sizes), while our construction allows to use the PQ equation, which seems to be essential in obtaining a general but perhaps less explicit expression of the desired eigenvalue regardless of the system size, and which we would not be able to obtain in the same way with an inhomogeneous term. Moreover, although our construction requires some heavy calculations involving q-series, it is done explicitly and at the level of the whole matrices rather than for individual eigenvalues only, which spares us from having to make assumptions on $Q$. It would be interesting to see whether those two methods can be combined in any way.

Finally, we may note that certain steps in our derivation seem somewhat sub-optimal (in particular the proof of the block decomposition of our infinite dimensional $K$ matrices, which can be found in appendix \ref{App-B-2}), and we believe that simpler purely algebraic derivations might exist.

~~

The authors would like to thank K. Mallick, R. Blythe, E. Ragoucy, P. Di Francesco, V. Terras, G. Misguich, T. Prosen and P. Baseilhac for their helpful comments and discussions. A. L. gratefully acknowledges financial support from the Interuniversity Attraction Pole - Phase VII/18 (Dynamics, Geometry and Statistical Physics) at the K. U. Leuven.

\newpage

\appendix

\section{Commutation relations}
\label{App-A}

\subsection{Bulk}
\label{App-A-1}

The commutation relation for the periodic case or for the bulk of the open case involves a calculation almost identical to that which can be found in the appendixes of \cite{1751-8121-46-14-145003}.

Let us recall:
\begin{align}
X&=\begin{bmatrix} n_0 & e\\ d & n_1\end{bmatrix}=~~~~~~~~~~~\begin{bmatrix}1+x A & e\\ d & 1+y A\end{bmatrix},\nonumber\\
\hat{X}&=\begin{bmatrix} \hat{n}_0 & \hat{e}\\ \hat{d} & \hat{n}_1\end{bmatrix}=\frac{(1-q)}{2}\begin{bmatrix} 1-x A & e\\ -d & -1+y A\end{bmatrix}.\nonumber
\end{align}
The parts of the transfer matrix and of the Markov matrix involving sites $i$ and $i+1$ are:
 \begin{equation}
X^{(i)}X^{(i+1)}=\begin{bmatrix} n_0 n_0 & n_0 e & e n_0 & e e \\ n_0 d & n_0 n_1 & e d & e n_1 \\ d n_0 & d e & n_1 n_0 & n_1 e \\ d d & d n_1 & n_1 d & n_1 n_1 \end{bmatrix}~~,~~M_{i}=\begin{bmatrix} 0 & 0 & 0 & 0 \\ 0 & -q & ~~1 & 0 \\ 0 & ~~q & -1 & 0 \\ 0 & 0 & 0 & 0 \end{bmatrix}.\nonumber
\end{equation}

The commutator of those two gives:
\begin{align}
[M^{(i)}, X^{(i)}X^{(i+1)}]&=\begin{bmatrix} 0 & q(n_0e-en_0) & en_0-n_0e & 0  \\ dn_0-q~n_0d & de-q~ed & (1-q)ed &  n_1e-q~en_1 \\   q~n_0d-dn_0 & (q-1)de & q~ed-de & q~en_1-n_1e \\ 0 & q(dn_1-n_1d) & n_1d-dn_1 & 0 \end{bmatrix}\nonumber\\
&=\begin{bmatrix}0 & \hat{n}_0e-n_0\hat{e} & \hat{e}n_0-e\hat{n}_0 & 0  \\ \hat{n}_0d-n_0\hat{d} & \hat{n}_0n_1-n_0\hat{n}_1 & \hat{e}d-e\hat{d} &  \hat{e}n_1-e\hat{n}_1 \\  \hat{d}n_0-d\hat{n}_0& \hat{d}e-d\hat{e} & \hat{n}_1n_0-n_1\hat{n}_0 & \hat{n}_1e-n_1\hat{e} \\ 0 & \hat{d}n_1-d\hat{n}_1 & \hat{n}_1d-n_1\hat{d} & 0 \end{bmatrix}\nonumber\\
&=\boxed{\hat{X}^{(i)}X^{(i+1)}-X^{(i)}\hat{X}^{(i+1)}.
}
\end{align}
where we got from the second to the third line using:
\begin{align}
de -q ed&=(1-q)(1-x y A^2) =\hat{n}_0n_1-n_0\hat{n}_1\nonumber,\\ 
q(n_0e-en_0)&=(q-1)xA e=\hat{n}_0e-n_0\hat{e}\nonumber,\\
en_0-n_0e&=(1-q)x e A =\hat{e}n_0-e\hat{n}_0\nonumber,\\
n_1e-q~en_1&=(1-q)e=\hat{e}n_1-e\hat{n}_1\nonumber,\\
q~en_1-n_1e&=(q-1)e=\hat{n}_1e-n_1\hat{e}\nonumber,\\
q(dn_1-n_1d)&=(q-1)yd A =\hat{d}n_1-d\hat{n}_1\nonumber,\\
n_1d-dn_1&=(1-q) yA d=\hat{n}_1d-n_1\hat{d}\nonumber,\\
dn_0-q~n_0d&=(1-q)d=\hat{n}_0d-n_0\hat{d}\nonumber,\\
q~n_0d-dn_0&=(q-1)d=\hat{d}n_0-d\hat{n}_0.
\end{align}

\subsection{Boundaries}
\label{App-A-2}

We recall, for the left boundary vectors:
\begin{align}
\langle\!\langle W|\!|~ [\alpha(e+n_x)- \gamma (d+n_x)-(1-q)] &= 0,\nonumber\\
\langle\!\langle \tilde{W}|\!|~ [\alpha(e-n_y) - \gamma (d-n_y)+(1-q)y A] &=0.\nonumber
\end{align}

We consider the commutator of the left boundary matrix from $M_\mu$ with the relevant part of $U_\mu$ and $T_\mu$ (omitting the matrix $A_\mu$ here, because, as in the periodic case, its action is trivial), and use those relations. We get:
\begin{align}
\Bigl[m^{(0)},\langle\!\langle W|\!| X^{(1)}\Bigr]&=\langle\!\langle W|\!|\begin{bmatrix}\gamma d-\alpha e & -(\alpha-\gamma)e \\ (\alpha-\gamma)d & \alpha e-\gamma d \end{bmatrix}\nonumber\\
&=(\alpha-\gamma)\langle\!\langle W|\!|\begin{bmatrix} n_x & -e \\ d & -n_x \end{bmatrix}+(1-q)\langle\!\langle W|\!| \begin{bmatrix} -1 & 0 \\ 0 & 1 \end{bmatrix}\nonumber\\
\Bigl[m^{(0)},\langle\!\langle \tilde{W}|\!|  X^{(1)}\Bigr]&=\langle\!\langle \tilde{W}|\!|\begin{bmatrix}\gamma d-\alpha e & -(\alpha-\gamma)e \\ (\alpha-\gamma)d & \alpha e-\gamma d \end{bmatrix}\nonumber\\
&=(\alpha-\gamma)\langle\!\langle \tilde{W}|\!|\begin{bmatrix} -n_y & -e \\ d & n_y \end{bmatrix}+(1-q)\langle\!\langle \tilde{W}|\!| \begin{bmatrix} y A & 0 \\ 0 & -y A \end{bmatrix}.\nonumber
\end{align}

The first terms in each of those two equations cancel one another:
\begin{equation}
\begin{bmatrix} n_x & -e \\ d & -n_x \end{bmatrix}\centerdot X(y)+X(x)\centerdot\begin{bmatrix} -n_y & -e \\ d & n_y \end{bmatrix}=0\nonumber
\end{equation}
We note that if we had had two different spectral parameters for the two diagonal terms in $X(x)$ or $X(y)$, this relation wouldn't have been possible (or we would have needed two equations on each boundary vector instead of one).

As for the second terms, a straightforward calculation gives:
\begin{equation}
(1-q)\Biggl(\begin{bmatrix} -1 & 0\\ 0& 1\end{bmatrix}\centerdot X(y)+ X(x)\centerdot \begin{bmatrix} yA & 0\\ 0 & -yA\end{bmatrix}\Biggr)=-\hat{X}(x)\centerdot X(y)-X(x)\centerdot\hat{X}(y).\nonumber
\end{equation}
so that
\begin{equation}
[m^{(0)}(\mu),U_\mu T_\mu]=- \langle\!\langle W|\!| A_\mu \hat{X}^{(1)}\prod_{i=2}^{L}X^{(i)} |\!|V\rangle\!\rangle\centerdot T_\mu-U_\mu \centerdot \langle\!\langle \tilde{W}|\!| A_\mu \hat{X}^{(1)}\prod_{i=2}^{L}X^{(i)} |\!|\tilde{V}\rangle\!\rangle.\nonumber
\end{equation}

The exact same calculations can be done at the right boundary, and yields:
\begin{equation}
[m^{(L)},U_\mu T_\mu]= \langle\!\langle W|\!| A_\mu \prod_{i=1}^{L-1}X^{(i)}\hat{X}^{(L)} |\!|V\rangle\!\rangle\centerdot T_\mu+U_\mu\centerdot \langle\!\langle \tilde{W}|\!| A_\mu \prod_{i=1}^{L-1}X^{(i)}\hat{X}^{(L)}  |\!|\tilde{V}\rangle\!\rangle.\nonumber
\end{equation}

\newpage

\section{Boundary vectors: R matrix and truncation}
\label{App-B}

We start this appendix with a few definitions and formulae that we will need during our calculation. Most of those can be found in \cite{Gasper2004}.

First, the q-binomial recursion formula at order $k$:
\begin{equation}\label{binom}
\frac{(q)_{n+m}}{(q)_n (q)_m}=\sum\limits_{r+s=k}\frac{(q)_{n+m-k}}{(q)_{n-r} (q)_{m-s}}\frac{(q)_{k}}{(q)_r (q)_s}q^{s(n-r)}.
\end{equation}

Then, some useful basic hypergeometric functions:
\begin{equation}\label{2phi1}
{}_2\phi_1(\substack{a,b\\c};z)=\sum\limits_{n}\frac{(a)_n(b)_n}{(c)_n(q)_n}z^n,
\end{equation}
\begin{equation}\label{3phi2}
{}_3\phi_2(\substack{a,b,c\\d,e};z)=\sum\limits_{n}\frac{(a)_n(b)_n(c)_n}{(d)_n(e)_n(q)_n}z^n,
\end{equation}
and the q-Appell function:
\begin{equation}\label{qAppell}
\Phi_1\bigl(a,\substack{b_1\\b_2};c;\substack{z_1\\z_2})=\sum\limits_{n}\frac{(a)_{n+m}(b_1)_n(b_2)_m}{(c)_{n+m}(q)_n(q)_m}z_1^n z_2^m,
\end{equation}
along with special relations that they verify: the q-Euler formulae
\begin{equation}\label{2phi1Euler}
{}_2\phi_1(\substack{a,b\\c};z)=\frac{(abz/c)_\infty}{(z)_\infty}~{}_2\phi_1(\substack{c/a,c/b\\c};abz/c),
\end{equation}
\begin{equation}\label{qAppellEuler}
\Phi_1\bigl(a,\substack{b_1\\b_2};b_1b_2;\substack{z_1\\z_2})=\frac{(az_1/b_2)_\infty(az_2/b_1)_\infty}{(z_1)_\infty(z_2)_\infty}\Phi_1\bigl(b_1b_2/a,\substack{b_2\\b_1};b_1b_2;\substack{az_1/b_2\\az_2/b_1}),
\end{equation}
and relations that link $\Phi_1$ and ${}_3\phi_2$:
\begin{equation}\label{Phi1to3phi2}
\Phi_1\bigl(a,\substack{b_1\\b_2};b_1b_2;\substack{z_1\\z_2})=\frac{(b_1z_1)_\infty(az_2/b_1)_\infty}{(z_1)_\infty(z_2)_\infty}{}_3\phi_2(\substack{ab_1b_2,b_1,b_1z_1/z_2\\b_1b_2,b_1z_1};az_2/b_1),
\end{equation}
\begin{equation}\label{3phi2toPhi1}
{}_3\phi_2(\substack{a,b,c\\d,e};de/abc)=\frac{(e/b)_\infty(e/c)_\infty}{(e)_\infty(de/abc)_\infty}\Phi_1\bigl(d/a,\substack{b\\d/b};d;\substack{e/b\\e/c}).
\end{equation}

The integral representation of the q-Beta function:
\begin{equation}\label{qBeta}
\frac{(a)_n(b)_m}{(ab)_{n+m}}=\frac{(a)_\infty(b)_\infty}{(ab)_\infty(q)_\infty}\int\limits_0^1\frac{d_qt}{t}\frac{(qt)_\infty}{(aq^nt)_\infty}t^{\log_q(b)+m}
\end{equation}
where $q^{\log_q(x)}=x$ and $\int\limits_0^1\frac{d_qt}{t}f(t)=\sum\limits_{k=0}^\infty f(q^k)$.

The q-derivative: $D_x f(x)=\frac{f(x)-f(qx)}{x}$, and the q-Leibniz formula:
\begin{equation}\label{qDiff}
D_x^nf(x)g(x)=\sum\limits_{k=0}^n\frac{(q)_n}{(q)_k(q)_{n-k}}\bigl(D_x^kf(x)\bigr)_{x\rightarrow q^{n-k}x}~\bigr(D_x^{n-k}g(x)\bigl)
\end{equation}

Finally, a useful relation:
\begin{equation}\label{useful}
\sum\limits_{k}\frac{(a)_k(b)_{n-k}}{(q)_k(q)_{n-k}}x^k y^{n-k}=\sum\limits_{k}\frac{(ax/y)_k(by/x)_{n-k}}{(q)_k(q)_{n-k}}y^k x^{n-k}
\end{equation}
which can be found by expanding $\frac{(axt)_\infty(byt)_\infty}{(xt)_\infty(yt)_\infty}$ in two different ways.

\subsection{Action of the R matrix}
\label{App-B-1}

We start from $I_0=R^{-1}(x,x;y,y)~|\!| V(x) \rangle\!\rangle\centerdot|\!| \tilde{V}(y) \rangle\!\rangle=\frac{(x y)_\infty}{(q)_\infty}\frac{(q A_2)_\infty}{(x y A_2)_\infty}~K^+$, as defined in ...

\begin{equation}\boxed{
I_0=\frac{(xS_1)_\infty(b\tilde{b}xS_1)_\infty}{(bS_1)_\infty(\tilde{b}S_1)_\infty}~\frac{(q A_2)_\infty}{(x y A_2)_\infty}\frac{(y S_1/S_2)_\infty}{(x S_1/S_2)_\infty}\frac{(y^2A_2)_\infty}{(qA_2)_\infty}~\frac{(byS_2)_\infty(\tilde{b}yS_2)_\infty}{(S_2)_\infty(b\tilde{b}S_2)_\infty}\frac{(x y)_\infty}{(y^2)_\infty}
}\end{equation}
where the term $|\!|0_1,0_2\rangle\!\rangle$ is implicit.

We expand all but the leftmost two ratios:
\begin{equation}
I_0=\frac{(xS_1)_\infty(b\tilde{b}xS_1)_\infty}{(bS_1)_\infty(\tilde{b}S_1)_\infty}\sum\limits_{n,m,k}\frac{(x y)_k(y/x)_{n+m-k}}{(q)_k(q)_{n+m-k}}(x S_1)^{n+m-k}S_2^k\frac{(q)_{n+m}}{(y^2)_{n+m}}\frac{(y b)_n(y/b)_{m}}{(q)_n(q)_{m}}(b \tilde{b})^m.\nonumber
\end{equation}

Using equation (\ref{binom}), and relabelling $n-r$ and $m-s$ as $n$ and $m$, we get:
\begin{align}
I_0&=\frac{(xS_1)_\infty(b\tilde{b}xS_1)_\infty}{(bS_1)_\infty(\tilde{b}S_1)_\infty}\sum\limits_{n,m,r,s}\frac{(x y)_{r+s}(y/x)_{n+m}}{(y^2)_{n+m+r+s}}\frac{(y b)_{n+r}(y/b)_{m+s}}{(q)_n(q)_{m}(q)_r(q)_{s}}(q^sx S_1)^{n}(b \tilde{b}xS_1)^{m}S_2^{r}(b \tilde{b}S_2)^{s}\nonumber\\
&=\frac{(xS_1)_\infty(b\tilde{b}xS_1)_\infty}{(bS_1)_\infty(\tilde{b}S_1)_\infty}\sum\limits_{r,s}\frac{(y b)_{r}(y/b)_{s}}{(q)_r(q)_{s}}\frac{(x y)_{r+s}}{(y^2)_{r+s}}S_2^{r}(b \tilde{b}S_2)^{s}\Phi_1\bigl(y/x,\substack{q^s y/b\\q^ryb};q^{r+s}y^2;\substack{b \tilde{b}xS_1\\q^sx S_1})\nonumber
\end{align}

We then use equation (\ref{Phi1to3phi2}):
\begin{align}
I_0&=\frac{(xS_1)_\infty}{(\tilde{b}S_1)_\infty}\sum\limits_{r,s}\frac{(y b)_{r}(y/b)_{s}}{(q)_r(q)_{s}}\frac{(x y)_{r+s}}{(y^2)_{r+s}}S_2^{r}(b \tilde{b}S_2)^{s}\frac{(q^{r+s}xy)_n(q^sy/b)_n(y\tilde{b})_n}{(q^{r+s}y^2)_n(q)_n}(bS_1)^n\frac{(q^{n+s}xy\tilde{b}S_1)_\infty}{(q^sxS_1)_\infty}\nonumber\\
&=\frac{(xS_1)_\infty}{(\tilde{b}S_1)_\infty}\sum\limits_{n,s}\frac{(y/b)_{s+n}}{(q)_{s}}\frac{(x y)_{s+n}}{(y^2)_{s+n}}(b \tilde{b}S_2)^{s}\frac{(y\tilde{b})_n}{(q)_n}(bS_1)^n\frac{(q^{n+s}xy\tilde{b}S_1)_\infty}{(q^sxS_1)_\infty}{}_2\phi_1(\substack{yb,xyq^{s+n}\\y^2q^{s+n}};S_2)\nonumber
\end{align}

Then we use (\ref{2phi1Euler}) on ${}_2\phi_1$:
\begin{align}
I_0&=\frac{(xS_1)_\infty(xbS_2)_\infty}{(\tilde{b}S_1)_\infty(S_2)_\infty}\sum\limits_{n,r,s}\frac{(y/b)_{r+s+n}}{(y^2)_{r+s+n}}\frac{(y/x)_{r}}{(q)_{r}}\frac{(x y)_{s+n}(y\tilde{b})_n}{(q)_n(q)_{s}}\frac{(q^{n+s}xy\tilde{b}S_1)_\infty}{(q^sxS_1)_\infty}(xbS_2)^r(b\tilde{b}S_2)^s(bS_1)^n\nonumber\\
&=\frac{(xS_1)_\infty(xbS_2)_\infty}{(\tilde{b}S_1)_\infty(S_2)_\infty}\sum\limits_{r,s}\frac{(y/b)_{r+s}}{(y^2)_{r+s}}\frac{(y/x)_{r}}{(q)_{r}}\frac{(x y)_{s}}{(q)_{s}}(xbS_2)^r(b\tilde{b}S_2)^s\frac{(q^{s}xy\tilde{b}S_1)_\infty}{(q^sxS_1)_\infty}{}_3\phi_2(\substack{y/bq^{r+s},xyq^s,y\tilde{b}\\y^2q^{r+s},q^{s}xy\tilde{b}S_1};bS_1)\nonumber
\end{align}
and (\ref{3phi2toPhi1}) on ${}_3\phi_2$:
\begin{equation}
I_0=\frac{(xS_1)_\infty(xbS_2)_\infty}{(bS_1)_\infty(S_2)_\infty}\sum\limits_{n,m,r,s}\frac{(y/b)_{r+s}(yb)_{n+m}}{(y^2)_{n+m+r+s}}\frac{(xy)_{n+s}(y/x)_{m+r}}{(q)_n(q)_{m}(q)_r(q)_{s}}(\tilde{b} S_1)^{n}(q^s xS_1)^{m}(xbS_2)^{r}(b \tilde{b}S_2)^{s}\nonumber
\end{equation}

We then use (\ref{binom}) in reverse (and exchange $k$ and $n+m-k$ for convenience) to find:
\begin{equation}\boxed{
I_0=\frac{(xS_1)_\infty(xbS_2)_\infty}{(bS_1)_\infty(S_2)_\infty}\sum\limits_{n,m,k}\frac{(y/x)_k(xy)_{n+m-k}}{(q)_k(q)_{n+m-k}}\tilde{b}^{n+m-k} x^{k}\frac{(q)_{n+m}}{(y^2)_{n+m}}\frac{(y/b)_n(yb)_{m}}{(q)_n(q)_{m}}(xbS_2)^n(xS_1)^m\label{I0}
}\end{equation}

In this expression, we can see that, after rescaling $S_1$ to $S_1/b$, we have the symmetry $\{S_1\leftrightarrow S_2,b\leftrightarrow 1/b\}$, so that
\begin{equation}
I_0=\frac{(xbS_2)_\infty(x\tilde{b}S_2)_\infty}{(S_2)_\infty(b\tilde{b}S_2)_\infty}~\frac{(q A_1)_\infty}{(x y A_1)_\infty}\frac{(y S_2/S_1)_\infty}{(x S_2/S_1)_\infty}\frac{(y^2A_1)_\infty}{(qA_1)_\infty}~\frac{(yS_1)_\infty(b\tilde{b}yS_1)_\infty}{(bS_1)_\infty(\tilde{b}S_1)_\infty}\frac{(x y)_\infty}{(y^2)_\infty}\nonumber
\label{I02}
\end{equation}
which is to say, after rearranging a few terms:
\begin{equation}
R^{-1}(x,x;y,y)~|\!| V(x) \rangle\!\rangle\centerdot|\!| \tilde{V}(y) \rangle\!\rangle=|\!| V(y) \rangle\!\rangle\centerdot|\!| \tilde{V}(x) \rangle\!\rangle.\nonumber
\end{equation}

Through the exact same calculations, we can also show that 
\begin{equation}
R^{-1}(y,y;x,x)~|\!|  \tilde{V}(y) \rangle\!\rangle\centerdot|\!| V(x)\rangle\!\rangle=|\!|  \tilde{V}(x) \rangle\!\rangle\centerdot|\!|V(y) \rangle\!\rangle.\nonumber
\end{equation}

\subsection{Truncation at xy=q$^{{ \textbf{1-p}}}$}
\label{App-B-2}

What we want to prove here, taking $xy=q^{1-p}$, is twofold. First, we need to show that $K^+_{i,j}(x,y)=0$ for $j\leq p-1$ and $i\geq p$. Then, we need to find the relation between $K^+_{i+p,j+p}(x,y)$ and $K^+_{i,j}(q^px,q^py)$, which we will do by calculating $\overline{I_0}=D_{S_1}^p D_{S_2}^p I_0$.

~~

The first part is rather straightforward. Considering equation (\ref{I02}), and since $K^+=\frac{(q)_\infty}{(xy)_\infty}\frac{(xy A_2)_\infty}{(q A_2)_\infty}~I_0$, we see that $K^+$ has a factor $\frac{(xy A_1)_\infty}{(x y A_2)_\infty}$ which results in a factor $\frac{(xy)_i}{(x y)_j}$ in $K^+_{i,j}$. Taking $xy=q^{1-p}$, this factor is $0$ precisely when $j\leq p-1$ and $i\geq p$, and since the rest of the expression for $K^+$ has no pole for $xy=q^{1-p}$, we conclude that the corresponding terms in $K^+$ are equal to $0$. Note that, even though that prefactor is infinite for $i\leq p-1$ and $j\geq p$, the absence of poles in $K^+$ tells us that it is compensated by the rest of the expression.

~~

We now need to calculate $\overline{I_0}=D_{S_1}^p D_{S_2}^p I_0$. In order to do this, we first notice that
\begin{equation}
\frac{(q)_{n+m}}{(q)_{n+m-k}}\tilde{b}^{n+m-k}=D_{\tilde{b}}^k \tilde{b}^{n+m}\nonumber
\end{equation}
and we use equation (\ref{qBeta}) on $\frac{(y/x)_k(xy)_{n+m-k}}{(y^2)_{n+m}}$ in (\ref{I0}) to write:
\begin{align}
I_1&=\frac{(xS_1)_\infty(xbS_2)_\infty}{(bS_1)_\infty(S_2)_\infty}\sum\limits_{n,m,k}\int\limits_0^1\frac{d_qt}{t}\frac{(qt)_\infty}{(y/xq^kt)_\infty}t^{\log_q(xy)+n+m-k}\frac{x^k}{(q)_k}D_{\tilde{b}}^k \tilde{b}^{n+m}\frac{(y/b)_n(yb)_{m}}{(q)_n(q)_{m}}(bS_2)^n(S_1)^m\nonumber\\
&=\sum\limits_{k}\int\limits_0^1\frac{d_qt}{t}\frac{(qt)_\infty}{(y/xq^kt)_\infty}t^{\log_q(xy)}\frac{(x/t)^k}{(q)_k}D_{\tilde{b}}^k\frac{(xS_1)_\infty(y b \tilde{b}tS_1)_\infty}{(bS_1)_\infty(\tilde{b}tS_1)_\infty}\frac{(xbS_2)_\infty(y  \tilde{b}tS_2)_\infty}{(S_2)_\infty(b \tilde{b}tS_2)_\infty}\nonumber
\end{align}
where $I_0=I_1~\frac{(xy)_\infty(y/x)_\infty}{(y^2)_\infty(q)_\infty}$. That prefactor will be put back in later when we re-sum the integral. Note that the term $(xy)_\infty$ gives $0$ when we take $xy=q^{1-p}$, but since that term is part of $I_0$ and not of $K^+$, it merely serves to simplify the calculation, and will be taken away at the end. For that reason, we will not apply $xy=q^{1-p}$ on that term, but keep it in its generic form.

We will now differentiate with respect to $S_1$ and $S_2$. Consider the following relation, which is easy to prove by recursion: if $ab/cd=q^{1-p}$, then
\begin{equation}
D_{X}^p\frac{(a)_\infty(b)_\infty}{(c)_\infty(d)_\infty}=(a/c)_p(b/c)_pc^p\frac{(q^pa)_\infty(q^pb)_\infty}{(c)_\infty(d)_\infty}.\nonumber
\end{equation}

This relation applies to both groups of q-Pochhammer symbols at the right in the integral (remember that $xy=q^{1-p}$), and gives:
\begin{align}
D_{S_1}^p\frac{(xS_1)_\infty(y b \tilde{b}tS_1)_\infty}{(bS_1)_\infty(\tilde{b}tS_1)_\infty}&=(x/b)_p(y\tilde{b}t)_pb^p\frac{(q^pxS_1)_\infty(q^py b \tilde{b}tS_1)_\infty}{(bS_1)_\infty(\tilde{b}tS_1)_\infty}\nonumber\\
D_{S_2}^p\frac{(xbS_2)_\infty(y  \tilde{b}tS_2)_\infty}{(S_2)_\infty(b \tilde{b}tS_2)_\infty}&=(xb)_p(y\tilde{b}t)_p\frac{(q^pxbS_2)_\infty(q^py\tilde{b}tS_2)_\infty}{(S_2)_\infty(b \tilde{b}tS_2)_\infty}\nonumber
\end{align}

We then have:
\begin{align}
\overline{I_1}&=(x/b)_pb^p(xb)_p\frac{(q^pxS_1)_\infty(q^pxbS_2)_\infty}{(bS_1)_\infty(S_2)_\infty}\nonumber\\
&~~~~~~~~~~~~~~\sum\limits_{k}\int\limits_0^1\frac{d_qt}{t}\frac{(qt)_\infty}{(y/xq^kt)_\infty}t^{\log_q(xy)}\frac{(x/t)^k}{(q)_k}D_{\tilde{b}}^k(q^py\tilde{b}t)_p(q^py\tilde{b}t)_p\frac{(y b \tilde{b}tS_1)_\infty(y  \tilde{b}tS_2)_\infty}{(\tilde{b}tS_1)_\infty(b \tilde{b}tS_2)_\infty}\nonumber
\end{align}

We use the formula for the q-derivative of a product (\ref{qDiff}) on the derivative with respect to $\tilde{b}$, with $f(\tilde{b})=(y\tilde{b}t)_p$ and the rest in $g$. Considering that $D_{\tilde{b}}^l(y\tilde{b}t)_p=(q^pyt)^l(q^{-p})_l(y\tilde{b}tq^l)_{p-l}$, we find:
\begin{align}
&D_{\tilde{b}}^k(y\tilde{b}t)_p(y\tilde{b}t)_p\frac{(y b \tilde{b}tS_1)_\infty(y  \tilde{b}tS_2)_\infty}{(\tilde{b}tS_1)_\infty(b \tilde{b}tS_2)_\infty}=\nonumber\\
&~~~~~~~~~~\sum\limits_{l=0}^{k}\frac{(q)_k}{(q)_l(q)_{k-l}}(q^pyt)^l(q^{-p})_l(y\tilde{b}tq^k)_{p-l}D_{\tilde{b}}^{k-l}(y\tilde{b}t)_p\frac{(q^py b \tilde{b}tS_1)_\infty(q^py  \tilde{b}tS_2)_\infty}{(\tilde{b}tS_1)_\infty(b \tilde{b}tS_2)_\infty}\nonumber
\end{align}
so that, renaming $k-l$ as $k$, we have:
\begin{align}
\overline{I_1}&=(x/b)_pb^p(xb)_p\frac{(q^pxS_1)_\infty(q^pxbS_2)_\infty}{(bS_1)_\infty(S_2)_\infty}\nonumber\\
&~~~~~~~~~~~~~~\sum\limits_{k,l}\int\limits_0^1\frac{d_qt}{t}\frac{(qt)_\infty}{(y/xq^{k+l}t)_\infty}t^{\log_q(xy)}(q^pxy)^l\frac{(q^{-p})_l}{(q)_l}(y\tilde{b}tq^{k+l})_{p-l}\nonumber\\
&~~~~~~~~~~~~~~~~~~~~~~~~~~~~\frac{(x/t)^k}{(q)_k}D_{\tilde{b}}^k(y\tilde{b}t)_p\frac{(q^py b \tilde{b}tS_1)_\infty(q^py  \tilde{b}tS_2)_\infty}{(\tilde{b}tS_1)_\infty(b \tilde{b}tS_2)_\infty}\nonumber
\end{align}

Consider the part that depends on $l$ (writing $\frac{1}{(y/xq^{k+l}t)_\infty}=\frac{(y/xq^{k}t)_l}{(y/xq^{k}t)_\infty}$):
\begin{align}
&~~~\sum\limits_{l}(y/xq^{k}t)_l(q^pxy)^l\frac{(q^{-p})_l}{(q)_l}(y\tilde{b}tq^{k+l})_{p-l}\nonumber\\
&=\sum\limits_{l}(y/xq^{k}t)_l((y\tilde{b}tq^{k+p-1})^{-1})_{p-l}\frac{(q)_p}{(q)_l(q)_{p-l}}(xy)^l(y\tilde{b}tq^{k})^{p-l}(-1)^pq^{p(p-1)/2}\nonumber
\end{align}
which, through (\ref{useful}), is equal to:
\begin{align}
&~~~\sum\limits_{l}\frac{(y/\tilde{b})_l(q^{1-p}/xy)_{p-l}}{(q)_l(q)_{p-l}}(y\tilde{b}tq^{k})^l(xy)^{p-l}(-1)^pq^{p(p-1)/2}(q)_p\nonumber\\
&=\sum\limits_{l}\frac{(y/\tilde{b})_l(xyq^l)_{p-l}}{(q)_l(q)_{p-l}}(y\tilde{b}tq^{k})^l(-1)^lq^{l(l-1)/2}(q)_p\nonumber
\end{align}

Using that $xy=q^{1-p}$, only the term for $l=p$ survives, namely $(y/\tilde{b})_p(y\tilde{b}tq^{k})^p(-1)^pq^{p(p-1)/2}$, so that:
\begin{align}
\overline{I_1}&=(x/b)_pb^p(xb)_p(y/\tilde{b})_p(y\tilde{b})^p(-1)^pq^{p(p-1)/2}\frac{(q^pxS_1)_\infty(q^pxbS_2)_\infty}{(bS_1)_\infty(S_2)_\infty}\nonumber\\
&~~~~~~~~~~\sum\limits_{k,l}\int\limits_0^1\frac{d_qt}{t}\frac{(qt)_\infty}{(y/xq^{k}t)_\infty}t^{\log_q(xy)+p}\frac{(q^px/t)^k}{(q)_k}D_{\tilde{b}}^k(y\tilde{b}t)_p\frac{(q^py b \tilde{b}tS_1)_\infty(q^py  \tilde{b}tS_2)_\infty}{(\tilde{b}tS_1)_\infty(b \tilde{b}tS_2)_\infty}\nonumber
\end{align}

Using again the formula for the derivative of a product and relabelling $k-l$ as $k$ again, we find:
\begin{align}
\overline{I_1}&=(x/b)_pb^p(xb)_p(y/\tilde{b})_p(y\tilde{b})^p(-1)^pq^{p(p-1)/2}\frac{(q^pxS_1)_\infty(q^pxbS_2)_\infty}{(bS_1)_\infty(S_2)_\infty}\nonumber\\
&~~~~~~~~~~~~~~\sum\limits_{k,l}\int\limits_0^1\frac{d_qt}{t}\frac{(qt)_\infty}{(y/xq^{k+l}t)_\infty}t^{\log_q(xy)+p}(q^{2p}xy)^l\frac{(q^{-p})_l}{(q)_l}(y\tilde{b}tq^{k+l})_{p-l}\nonumber\\
&~~~~~~~~~~~~~~~~~~~~~~~~~~~~\frac{(q^px/t)^k}{(q)_k}D_{\tilde{b}}^k\frac{(q^py b \tilde{b}tS_1)_\infty(q^py  \tilde{b}tS_2)_\infty}{(\tilde{b}tS_1)_\infty(b \tilde{b}tS_2)_\infty}\nonumber
\end{align}

We have almost the same sum as before over $l$, only with an extra term $q^{pl}$ which changes the result to:
\begin{align}
&~~~\sum\limits_{l}(y/xq^{k}t)_l(q^{2p}xy)^l\frac{(q^{-p})_l}{(q)_l}(y\tilde{b}tq^{k+l})_{p-l}\nonumber\\
&=\sum\limits_{l}\frac{(q^py/\tilde{b})_l(xyq^{p+l})_{p-l}}{(q)_l(q)_{p-l}}(y\tilde{b}tq^{k})^l(-1)^lq^{l(l-1)/2}(q)_p\nonumber
\end{align}
so that, this time, the whole sum survives to taking $xy=q^{1-p}$.

We now have to re-sum the integral over $t$. First, we expand the two ratios of q-Pochhammer symbols on the right, and apply the derivative with respect to $\tilde{b}$:
\begin{equation}
D_{\tilde{b}}^k\frac{(q^py b \tilde{b}tS_1)_\infty(q^py  \tilde{b}tS_2)_\infty}{(\tilde{b}tS_1)_\infty(b \tilde{b}tS_2)_\infty}=t^{n+m}\sum\limits_{n,m}\frac{(q)_{n+m}}{(q)_{n+m-k}}\tilde{b}^{n+m-k}\frac{(q^py/b)_n(q^pyb)_{m}}{(q)_n(q)_{m}}(bS_2)^n(S_1)^m\nonumber
\end{equation}
so that
\begin{align}
\overline{I_1}&=(x/b)_pb^p(xb)_p(y/\tilde{b})_p(y\tilde{b})^p(-1)^pq^{p(p-1)/2}(q)_p\frac{(q^pxS_1)_\infty(q^pxbS_2)_\infty}{(bS_1)_\infty(S_2)_\infty}\nonumber\\
&~~~~~~~~~~~~~~\sum\limits_{n,m,k,l}\int\limits_0^1\frac{d_qt}{t}\frac{(qt)_\infty}{(y/xq^{k}t)_\infty}t^{\log_q(xy)+p+n+m+l-k}\frac{(q^py/\tilde{b})_l(xyq^{p+l})_{p-l}}{(q)_l(q)_{p-l}}(y\tilde{b}q^{k})^l(-1)^lq^{l(l-1)/2}\nonumber\\
&~~~~~~~~~~~~~~~~~~~~~~~~~~~~\frac{(q^px)^k}{(q)_k}\frac{(q)_{n+m}}{(q)_{n+m-k}}\tilde{b}^{n+m-k}\frac{(q^py/b)_n(q^pyb)_{m}}{(q)_n(q)_{m}}(bS_2)^n(S_1)^m\nonumber
\end{align}

We only have to consider the part that depends on $k$ and $l$, as we need to keep the terms that depend on $n$ and $m$ intact. We re-sum the integral, going back from $I_1$ to $I_0$:
\begin{equation}
\sum\limits_{k,l}\frac{(y/x)_k(x y )_{n+m-k+l+p}}{(y^2)_{n+m+l+p}}\frac{(q^py/\tilde{b})_l(xyq^{p+l})_{p-l}}{(q)_l(q)_{p-l}}(y\tilde{b}q^{k})^l(-1)^lq^{l(l-1)/2}\frac{(q^px)^k}{(q)_k}\frac{(q)_{n+m}}{(q)_{n+m-k}}\tilde{b}^{n+m-k}\nonumber
\end{equation}

Noticing that $(x y )_{n+m-k+l+p}(xyq^{p+l})_{p-l}=(x y q^{p+l})_{n+m-k}(xy)_{2p}$, we find:
\begin{equation}
\tilde{b}^{n+m}(q)_{n+m}(xy)_{2p}\sum\limits_{k,l}\frac{(y/x)_k(x y q^{p+l})_{n+m-k}}{(q)_k(q)_{n+m-k}}(q^{p+l}x/\tilde{b})^k\frac{(q^py/\tilde{b})_l}{(q)_l(q)_{p-l}}\frac{(y\tilde{b})^l}{(y^2)_{n+m+l+p}}(-1)^lq^{l(l-1)/2}\nonumber
\end{equation}

Using (\ref{useful}) on the sum over $k$, we get:
\begin{equation}\label{52}
\tilde{b}^{n+m}(q)_{n+m}(xy)_{2p}\sum\limits_{k,l}\frac{(y\tilde{b})_k(q^{p+l}y/\tilde{b})_{n+m-k}}{(q)_k(q)_{n+m-k}}(q^{p+l}x/\tilde{b})^k\frac{(q^py/\tilde{b})_l}{(q)_l(q)_{p-l}}\frac{(y\tilde{b})^l}{(y^2)_{n+m+l+p}}(-1)^lq^{l(l-1)/2}
\end{equation}

We now write:
\begin{equation}
(q^{p+l}y/\tilde{b})_{n+m-k}\frac{(q^py/\tilde{b})_l}{(y^2)_{n+m+l+p}}=\frac{(q^py/\tilde{b})_{n+m-k+l}}{(y^2)_{n+m+l+p}}=\frac{(q^py/\tilde{b})_\infty}{(y^2)_\infty}\frac{(y^2q^{n+m+l+p})_\infty}{(q^{n+m-k+l+p}y/\tilde{b})_\infty}\nonumber
\end{equation}
and
\begin{equation}
\frac{(y^2q^{n+m+l+p})_\infty}{(q^{n+m-k+l+p}y/\tilde{b})_\infty}=\sum\limits_{r}\frac{(y \tilde{b}q^k)_r}{(q)_r}(q^{p+n+m-k+l}y/\tilde{b})^r\nonumber
\end{equation}

Putting this into (\ref{52}), we get:
\begin{align}
&\tilde{b}^{n+m}(q)_{n+m}\frac{(xy)_{2p}}{(q)_p}\frac{(q^py/\tilde{b})_\infty}{(y^2)_\infty}\nonumber\\
&~~~~~~~~~~~~\sum\limits_{k,l,r}\frac{(y\tilde{b})_{k+r}}{(q)_k(q)_{n+m-k}(q)_r}(q^{p}x/\tilde{b})^k(q^{p+n+m-k}y/\tilde{b})^r\frac{(q)_p}{(q)_l(q)_{p-l}}q^{l(l-1)/2}(-q^{k+r}y\tilde{b})^l\nonumber
\end{align}

Since $(a)_p=\sum\limits_{l}\frac{(q)_p}{(q)_l(q)_{p-l}}q^{l(l-1)/2}(-a)^l$, this simplifies to
\begin{align}
&~~~\tilde{b}^{n+m}(q)_{n+m}\frac{(xy)_{2p}}{(q)_p}\frac{(q^py/\tilde{b})_\infty}{(y^2)_\infty}\sum\limits_{k,r}\frac{(y\tilde{b})_{k+r}}{(q)_k(q)_{n+m-k}(q)_r}(q^{p}x/\tilde{b})^k(q^{p+n+m-k}y/\tilde{b})^r(q^{k+r}y\tilde{b})_p\nonumber\\
&=\tilde{b}^{n+m}(q)_{n+m}\frac{(xy)_{2p}}{(q)_p}\frac{(q^py/\tilde{b})_\infty}{(y^2)_\infty}\sum\limits_{k,r}\frac{(q^py\tilde{b})_{k}(q^{p+k}y\tilde{b})_{r}}{(q)_k(q)_{n+m-k}(q)_r}(q^{p}x/\tilde{b})^k(q^{p+n+m-k}y/\tilde{b})^r(y\tilde{b})_p\nonumber
\end{align}

We can now get rid of the sum over $r$. We have:
\begin{equation}
\frac{(q^py/\tilde{b})_\infty}{(y^2)_\infty}\sum\limits_{r}\frac{(q^{p+k}y\tilde{b})_{r}}{(q)_r}(q^{p+n+m-k}y/\tilde{b})^r=\frac{(q^py/\tilde{b})_{n+m-k}}{(y^2)_{n+m+2p}}\nonumber
\end{equation}
which gives
\begin{equation}
(y\tilde{b})_p\tilde{b}^{n+m}\frac{(xy)_{2p}}{(q)_p}\frac{(q)_{n+m}}{(y^2)_{n+m+2p}}\sum\limits_{k}\frac{(q^py\tilde{b})_{k}(q^py/\tilde{b})_{n+m-k}}{(q)_k(q)_{n+m-k}}(q^{p}x/\tilde{b})^k\nonumber
\end{equation}

Using (\ref{useful}) one final time on the sum over $k$ gives:
\begin{equation}
(y\tilde{b})_p\tilde{b}^{n+m}\frac{(xy)_{2p}}{(q)_p}\frac{(q)_{n+m}}{(y^2)_{n+m+2p}}\sum\limits_{k}\frac{(y/x)_{k}(q^{2p}xy)_{n+m-k}}{(q)_k(q)_{n+m-k}}(q^{p}x/\tilde{b})^k\nonumber
\end{equation}

We can at last put this back into $\overline{K^+}$. We get:
\begin{align}
\overline{I_0}&=(x/b)_pb^p(xb)_p(y/\tilde{b})_p(\tilde{b})^p(y\tilde{b})_p(-y)^pq^{p(p-1)/2}\frac{(xy)_{2p}}{(y^2 )_{2p}}\nonumber\\
&~~~~~~\frac{(q^pxS_1)_\infty(q^pxbS_2)_\infty}{(bS_1)_\infty(S_2)_\infty}\sum\limits_{n,m,k}\frac{(y/x)_{k}(q^{2p}xy)_{n+m-k}}{(q)_k(q)_{n+m-k}}(q^{p}x)^k\tilde{b}^{n+m-k}\nonumber\\
&~~~~~~~~~~~~~~~~~~~~~~~~~~~~\frac{(q)_{n+m}}{(y^2q^{2p})_{n+m}}\frac{(q^py/b)_n(q^pyb)_{m}}{(q)_n(q)_{m}}(bS_2)^n(S_1)^m\nonumber
\end{align}
which is to say
\begin{equation}
\overline{I_0}(x,y)=\frac{(x/b)_pb^p(xb)_p(y/\tilde{b})_p(\tilde{b})^p(y\tilde{b})_p}{(y^2)_{2p}}(-y)^pq^{p(p-1)/2}(xy)_{2p}~I_0(q^px,q^py).\nonumber
\end{equation}

In terms of $K^+$, this gives us, for $xy=q^{1-p}$:
\begin{equation}\boxed{
\frac{K^+_{i+p,j+p}(x,y)}{K^+_{i,j}(q^px,q^py)}=\frac{(x/b)_pb^p(xb)_p(y/\tilde{b})_p(\tilde{b})^p(y\tilde{b})_p}{(y^2 )_{2p}}(-y)^pq^{p(p-1)/2}\frac{(q^{j+1})_p}{(q^{i+1})_p}.
}\end{equation}

The equivalent result for $K^-$ is:
\begin{equation}\boxed{
\frac{K^-_{j+p,i+p}(x,y)}{K^-_{j,i}(q^px,q^py)}=\frac{(y/a)_pa^p(ya)_p(x/\tilde{a})_p(\tilde{a})^p(x\tilde{a})_p}{(x^2 )_{2p}}(-x)^pq^{p(p-1)/2}\frac{(q^{i+1})_p}{(q^{j+1})_p}.
}\end{equation}

\newpage

\section{T-Q equations and fusion hierarchy}
\label{App-C}

In this appendix, we show how the set of PQ equations
\begin{equation}
P(x)Q(1/q^{k-1}x)=t^{[k]}(x)+{\rm e}^{-k\mu}P(q^k x)Q(q/x)\nonumber
\end{equation}
allows to recover all the T-Q equation for each $t^{[k]}$, as well as fusion equations, where products of $t^{[k]}$'s are expressed as linear combinations of $t^{[l]}$'s with different dimensions $l$. We will consider the periodic case, where $t^{[1]}(x)=h(x)$. For the open case, one simply has to replace $h(x)$ by $F(x)$ and $\mu$ by $2\mu$.

To make our notations more compact, we will make the dependence in $x$ implicit, and write $P(q^kx)=P_k$, $Q(1/q^kx)=Q_k$, $t^{[k]}(q^lx)=t^{[k]}_l$ and $h(q^kx)=h_k$. In a given equation, one can shift all the indices by any integer $k$, which corresponds to taking the equation at $q^kx$ instead of $x$.

\subsection{T-Q equations}
\label{App-C-1}

This first calculation is rather straightforward. Consider the PQ of order $1$
\begin{equation}\label{P0Q0h}
P_0Q_0=h_0+{\rm e}^{-k\mu}P_1Q_{-1}
\end{equation}
and multiply the PQ equation of order $k$
\begin{equation}
P_0Q_{k-1}=t^{[k]}_0+{\rm e}^{-k\mu}P_kQ_{-1}
\end{equation}
by the product of all $Q_l$'s for $l$ between $0$ and $k-2$:
\begin{equation}
t^{[k]}_0\prod\limits_{l=0}^{k-2}Q_l=P_0\prod\limits_{l=0}^{k-1}Q_l-{\rm e}^{-k\mu}P_k\prod\limits_{l=-1}^{k-2}Q_l.\nonumber
\end{equation}
The right-hand side of the equation can then be rewritten to give
\begin{equation}
t^{[k]}_0\prod\limits_{l=0}^{k-2}Q_l=\sum\limits_{n=0}^{k-1}{\rm e}^{-n\mu}(P_nQ_n-{\rm e}^{-\mu}P_{n+1}Q_{n-1})\prod\limits_{l=-1}^{n-2}Q_l\prod\limits_{l=n+1}^{k-1}Q_l\nonumber
\end{equation}
in which we can simply use eq.(\ref{P0Q0h}) shifted by $n$ to obtain the T-Q equation:
\begin{equation}\boxed{
t^{[k]}_0\prod\limits_{l=0}^{k-2}Q_l=\sum\limits_{n=0}^{k-1}{\rm e}^{-n\mu}h_n\prod\limits_{l=-1}^{n-2}Q_l\prod\limits_{l=n+1}^{k-1}Q_l.
}\end{equation}
As we see, this equation involves products of $k-1$ matrices $Q$. By transferring all the $Q$s to the right-hand side, we find an alternative expression:
\begin{equation}
t^{[k]}_0=\sum\limits_{n=0}^{k-1}{\rm e}^{-n\mu}h_n\frac{Q_{-1}Q_{k-1}}{Q_{n-1}Q_n}.
\end{equation}

\subsection{Fusion equations}
\label{App-C-2}

Let us consider $t^{[2]}_0t^{[k]}_1$. Using the PQ equations of order $2$ and $k$, we find:
\begin{align}
t^{[2]}_0t^{[k]}_1=&(P_0Q_1-{\rm e}^{-2\mu}P_2Q_{-1})(P_1Q_k-{\rm e}^{-k\mu}P_{k+1}Q_0)\nonumber\\
=&P_0Q_kP_1Q_1\!+\!{\rm e}^{-(k+2)\mu}P_2Q_0P_{k+1}Q_{-1}\!-\!{\rm e}^{-2\mu}P_2Q_kP_1Q_{-1}\!-\!{\rm e}^{-k\mu}P_0Q_0P_{k+1}Q_1.\nonumber
\end{align}
The first two terms of this sum can also be obtained by considering $t^{[k+1]}_0h_1$:
\begin{align}
t^{[k+1]}_0h_1=&(P_0Q_k-{\rm e}^{-(k+1)\mu}P_{k+1}Q_{-1})(P_1Q_1-{\rm e}^{-\mu}P_{2}Q_0)\nonumber\\
=&P_0Q_kP_1Q_1\!+\!{\rm e}^{-(k+2)\mu}P_2Q_0P_{k\!+\!1}Q_{-1}\!-\!{\rm e}^{-(k\!+\!1)\mu}P_{k\!+\!1}Q_1P_1Q_{1}\!-\!{\rm e}^{-\mu}P_0Q_kP_{2}Q_{-1}\nonumber
\end{align}
so that
\begin{align}
t^{[2]}_0t^{[k]}_1\!-\!t^{[k\!+\!1]}_0h_1&={\rm e}^{-(k\!+\!1)\mu}P_{k\!+\!1}Q_1P_1Q_{1}\!+\!{\rm e}^{-\mu}P_0Q_kP_{2}Q_{-1}\!-\!{\rm e}^{-2\mu}P_2Q_kP_1Q_{-1}\!-\!{\rm e}^{-k\mu}P_0Q_0P_{k\!+\!1}Q_1\nonumber\\
&={\rm e}^{-\mu}(P_0Q_0-{\rm e}^{-\mu}P_1Q_{-1})(P_2Q_k-{\rm e}^{-(k-1)\mu}P_{k+1}Q_1)\nonumber\\
&={\rm e}^{-\mu}t^{[k-1]}_2h_0.\nonumber
\end{align}
We obtain the fusion equations:
\begin{equation}\boxed{
t^{[2]}_0t^{[k]}_1=h_1t^{[k+1]}_0+{\rm e}^{-\mu}h_0t^{[k-1]}_2
}\end{equation}
which, for $k\rightarrow\infty$, allows to recover the T-Q equation (\ref{V-1-TQ}).

Through the same reasoning, we also find
\begin{equation}
t^{[k]}_0t^{[2]}_{k-1}=h_{k-1}t^{[k+1]}_0+{\rm e}^{-\mu}h_kt^{[k-1]}_0,
\end{equation}
and combining the two, we find
\begin{equation}
t^{[2]}_0t^{[k]}_1t^{[2]}_{k}=h_1h_kt^{[k+2]}_0+{\rm e}^{-\mu}h_0h_kt^{[k]}_2+{\rm e}^{-\mu}h_1h_{k+1}t^{[k]}_0+{\rm e}^{-2\mu}h_0h_{k+1}t^{[k-2]}_2.
\end{equation}

\newpage

\section{Non-deformed limit: asymptotic calculations}
\label{App-D}

We want to estimate terms of the form $I_{ij}=\langle\!\langle  \tilde{W}|\!|  A_\mu A^i S^j  |\!|  \tilde{V} \rangle\!\rangle$, with
\begin{equation}
\langle\!\langle \tilde{W}|\!|=\langle\!\langle0|\!|\frac{(ay/S)_\infty(\tilde{a}y/S)_\infty}{(1/S)_\infty(a\tilde{a}/S)_\infty}\frac{(y^2A)_\infty}{(qA)_\infty}~~~~,~~~~~|\!|  \tilde{V} \rangle\!\rangle=\frac{(byS)_\infty(\tilde{b}yS)_\infty}{(S)_\infty(b\tilde{b}S)_\infty} |\!|0\rangle\!\rangle,\nonumber
\end{equation}
for $\mu\rightarrow 0$, assuming that $|a|$, $|\tilde{a}|$, $|b|$ and $|\tilde{b}|$ are all smaller than $1$. In particular, we want to show that it diverges if $i=0$, and is finite otherwise.

Let us assume that $j\geq 0$ (the calculations for $j<0$ are similar). We have:
\begin{equation}
I_{ij}=\sum\limits_{n+m=r+s+j}\frac{(ya)_n(y/a)_{m}}{(q)_n(q)_{m}}(a\tilde{a})^{m}\frac{(q)_{n+m}}{(y^2)_{n+m}}(q^i{\rm e}^{-\mu})^{n+m}\frac{(y b)_r(y/b)_{s}}{(q)_r(q)_{s}}(b \tilde{b})^s.\nonumber
\end{equation}
Since we are looking for possible divergences when ${\rm e}^{-\mu}$ goes to $1$, we should focus on the asymptotic behaviour of the terms of the series, as $n+m$ becomes large. As we will need to approximate q-Pochhammer symbols, let us first remark that, for $0\leq x\leq 1$,
\begin{equation}
1-\frac{x}{1-q}\leq (x)_\infty\leq 1\nonumber
\end{equation}
which can be proven by recursion by considering that $(xq^k)_\infty=(1-xq^k)(xq^{k+1})_\infty$ for $k\in {\mathbb N}$ and that $(0)_\infty=1$. From this, we deduce that
\begin{equation}
|(x)_N-(x)_\infty|\leq\frac{xq^N}{1-q}(x)_N\nonumber
\end{equation}
so that $(x)_\infty$ is a good approximation for $(x)_N$ if $N$ is large enough.

~

Consider now, for $N$ large, the term $\sum\limits_{n+m=N}\frac{(ya)_n(y/a)_{m}}{(q)_n(q)_{m}}(a\tilde{a})^{m}$. It can be approximated, for $0\leq n \leq N/2$, by $\sum\limits_{n=0}^{N/2}\frac{(ya)_n(y/a)_{\infty}}{(q)_n(q)_{\infty}}(a\tilde{a})^{N-n}$ with
\begin{equation}
0\leq\sum\limits_{n=0}^{N/2}\frac{(ya)_n}{(q)_n}(a\tilde{a})^{N-n}\leq(a)^{N/2}\sum\limits_{n=0}^{\infty}\frac{(ya)_n}{(q)_n}(\tilde{a})^{n}=(a)^{N/2}\frac{(ya\tilde{a})_\infty}{(\tilde{a})_\infty}\nonumber
\end{equation}
and, for $N/2<n\leq N$, by 
\begin{equation}
\sum\limits_{m=0}^{N/2}\frac{(ya)_\infty(y/a)_{m}}{(q)_\infty(q)_{m}}(a\tilde{a})^{m}\sim\sum\limits_{m=0}^{\infty}\frac{(ya)_\infty(y/a)_{m}}{(q)_\infty(q)_{m}}(a\tilde{a})^{m}=\frac{(ya)_\infty(y\tilde{a})_{\infty}}{(a\tilde{a})_\infty(q)_{\infty}}\nonumber
\end{equation}

The same can be done with $\sum\limits_{r+s=N-j}\frac{(y b)_r(y/b)_{s}}{(q)_r(q)_{s}}(b \tilde{b})^s\sim\frac{(yb)_\infty(y\tilde{b})_{\infty}}{(b\tilde{b})_\infty(q)_{\infty}}$, which leaves us with
\begin{equation}
I_{ij}\sim\sum\limits_{N}\frac{(ya)_\infty(y\tilde{a})_{\infty}}{(a\tilde{a})_\infty(q)_{\infty}}\frac{(q)_{\infty}}{(y^2)_{\infty}}(q^i{\rm e}^{-\mu})^{n+m}\frac{(yb)_\infty(y\tilde{b})_{\infty}}{(b\tilde{b})_\infty(q)_{\infty}}=\frac{1}{1-q^i{\rm e}^{-\mu}}\frac{(ya,y\tilde{a},yb,y\tilde{b})_{\infty}}{(y^2,a\tilde{a},b\tilde{b},q)_\infty}.\nonumber
\end{equation}

As we can see, if $i=0$, $I_{ij}$ diverges as $1/\mu$, with the prefactor that is given here, and is otherwise finite. We can check that, in all cases, the error in this estimation is finite.

\newpage

\section{Self-consistent integral representation of the PQ equation}
\label{App-E}

We consider eq.(\ref{V-2-PQFs}):
\begin{equation}
P(x)Q(1/x)=F(x)+{\rm e}^{-2\mu}P(q x)Q(q/x)\nonumber
\end{equation}
and define a function $W(x)$ as in (\ref{IV-2-W}):
\begin{equation}
W(x)=-\frac{1}{2}\log\biggl(\frac{P(x)Q(1/x)}{{\rm e}^{-2\mu}P(q x)Q(q/x)}\biggr).\nonumber
\end{equation}
Combining them, along with $B=-{\rm e}^{2\mu}\bigl(Q(0)\bigr)^{-1}$, we see that we also have
\begin{equation}
W(x)=-\frac{1}{2}\log\biggl(1-\frac{BF(x)Q(0)}{P(q x)Q(q/x)}\biggr).\nonumber
\end{equation}

Since for $\mu$ small enough, all the roots and poles of $P(x)$ and $Q(y)$ are outside of the unit circle, we can expand their logarithms on and inside of the unit circle in powers of $x$ and $y$, respectively. We write:
\begin{align}
\log(P(x))=&f(x)=\sum\limits_{k=1}^{\infty}a_k x^k,\nonumber\\
\log(Q(y)/Q(0))=&g(y)=\sum\limits_{k=1}^{\infty}b_k y^k,\nonumber
\end{align}
where we recall that $P(0)$ was set to $1$.

In those terms, the two expressions for $W(x)$ become
\begin{align}
W(x)&=-\frac{1}{2}\bigl(2\mu+f(x)-f(qx)+g(1/x)-g(q/x)\bigr)\nonumber\\
W(x)&=-\frac{1}{2}\log\biggl(1-BF(x){\rm e}^{-f(qx)-g(q/x)}\biggr).\nonumber
\end{align}
Notice that in the first equation, each of the coefficients $a_k$ and $b_k$ is connected to a different power of $x$, so that $W(x)$ determines those completely. We can therefore, in principle, write the argument of the exponential in the second equation in terms of $W(x)$ alone. In this case, we can even do it explicitly: consider that the coefficients of $f(x)-f(qx)$ are of the form $a_k(1-q^k)$, and those of $f(qx)$ are $a_kq^k$. To express the latter in terms of the former, each $x^k$ has to be replaced by $\frac{q^k}{1-q^k}x^k$, which can be done through a simple convolution:
\begin{equation}
f(qx)=\oint_{c_1}\frac{d\tilde{x}}{\imath2\pi\tilde{x}}\sum_{k=1}^{\infty}\frac{q^k}{1-q^k}(x/\tilde{x})^k (f(\tilde{x})-f(q\tilde{x})).\nonumber
\end{equation}
The same can be done for $g$ with negative powers of $x$, and, putting everything together, we find:
\begin{equation}
-f(qx)-g(q/x)=X[W](x)\nonumber
\end{equation}
where
\begin{equation}
X[f](x)=\oint_{c_1}\frac{d\tilde{x}}{\imath2\pi\tilde{x}}f(\tilde{x})K(x,\tilde{x}).\nonumber
\end{equation}
with
\begin{equation}
K(x,\tilde{x})=2\sum_{k=1}^{\infty}\frac{q^k}{1-q^k}\Bigl((x/\tilde{x})^k+(x/\tilde{x})^{-k}\Bigr).\nonumber
\end{equation}
Notice that the constant term $2\mu$ in $W$ disappears in the convolution.

This allows us to finally write:
\begin{equation}\boxed{
W(x)=-\frac{1}{2}\ln\Bigl(1-B F(x) e^{X[W](x)}\Bigr).
}\end{equation}

This same method was used in \cite{prolhac2010tree} for the periodic ASEP.

\newpage

\bibliographystyle{mybibstyle}

\bibliography{Biblio}{}

\end{document}